\documentclass[12pt,a4,sans]{article}
\usepackage[latin1]{inputenc}
\usepackage{amssymb}
\usepackage{amsmath}
\usepackage{amsthm}
\usepackage{amsfonts}
\usepackage[pdftex]{graphicx}
\usepackage{geometry}
 \usepackage{bbold}
\usepackage{braket}
\usepackage{enumerate}
\usepackage[usenames,dvipsnames]{xcolor}
\usepackage{setspace}
\usepackage[backgroundcolor=white,linecolor=black]{todonotes}
\usepackage[in]{fullpage}
\usepackage{hyperref}
\usepackage{array}
\usepackage{wrapfig}
\usepackage[font=small,labelfont=bf]{caption}
\usepackage{anysize}
\usepackage{float}
\usepackage{xcolor}
\usepackage{mathtools}
\usepackage{comment}

\numberwithin{equation}{section}

\marginsize{2.1cm}{2.1cm}{2cm}{2cm}

\hypersetup{
colorlinks=true,
linktoc=page,
citecolor=PineGreen,
linkcolor=BlueViolet,
urlcolor=MidnightBlue} 

\newcommand{\comments}[1]{}

\newcommand{\C}{\mathcal{C}}

\newcommand{\T}{\mathcal{T}}

\newcommand{\Z}{\mathcal{Z}}
\newcommand{\N}{\mathcal{N}}

\newcommand{\V}{\mathcal{V}}
\newcommand{\K}{\mathcal{K}}

\newcommand{\Tr}{\text{Tr}}

\newcommand{\nn}{\nonumber\\[3mm]}

\newcommand{\vv}{\mathbb V}

\newcommand{\bea}{\begin{eqnarray}} 
\newcommand{\eea}{\end{eqnarray}} 
\newcommand{\mId}{\mathbb{1} }

\begin{document}

\makeatletter
\@addtoreset{equation}{section}
\makeatother
\renewcommand{\theequation}{\thesection.\arabic{equation}}

\rightline{QMUL-PH-14-20}
\vspace{1.8truecm}

\vspace{15pt}


{\LARGE{ 
\centerline{\bf  Quivers, Words and Fundamentals } 
}}  

\vskip.5cm 

\thispagestyle{empty} \centerline{
    {\large \bf Paolo Mattioli
 \footnote{ {\tt  p.mattioli@qmul.ac.uk}}   }
   {\large \bf and Sanjaye Ramgoolam
               \footnote{ {\tt s.ramgoolam@qmul.ac.uk}}   }}
               
\setcounter{footnote}{0}               
               
\vspace{.4cm}
\centerline{{\it Centre for Research in String Theory, School of Physics and Astronomy},}
\centerline{{ \it Queen Mary University of London},} \centerline{{\it
    Mile End Road, London E1 4NS, UK}}

\vspace{1truecm}

\thispagestyle{empty}

\centerline{\bf ABSTRACT}

\vskip.2cm 
 
A systematic study of holomorphic gauge invariant operators in general $\N=1$ quiver gauge theories, with unitary gauge groups and bifundamental matter fields, was recently presented in \cite{quivcalc}. For large ranks a simple counting formula in terms of an infinite product was given. We extend this study to quiver gauge theories with fundamental matter fields, deriving an infinite product form for the refined counting in these cases. 
The infinite products are found to be obtained from substitutions in a simple building block expressed in terms of the weighted adjacency matrix of the quiver. In the case without fundamentals, it is a determinant which itself is found to have a counting interpretation in terms of words formed from partially commuting letters associated with simple closed loops in the quiver. This is a new relation between counting problems in gauge theory and the Cartier-Foata monoid. For finite ranks of the unitary gauge groups, the refined counting is given in terms of expressions involving Littlewood-Richardson coefficients.

\newpage
\tableofcontents

\pagebreak

\section{Introduction}

The study of giant gravitons \cite{mst0003} in the context of AdS/CFT \cite{malda} has instigated detailed investigations of BPS operators in four dimensional $\N=4$ super-Yang-Mills theory (SYM) with $U(N)$ gauge group \cite{CJR01,CR02,KR1,BHR1,BHR2,BCK08,BKS08,KR2}. These studies focused on the counting of gauge-invariant operators, an inner product related to 2-point functions and higher point functions for large $N$ as well as at finite $N$. The connection between $U(N)$ gauge invariants and permutations was a central theme as well as representation theory of the permutation groups. The studies were extended beyond $\N=4$ SYM to gauge theories such as ABJM \cite{Aharony:2008ug} and the conifold \cite{Dey2011,DMMP1205,M1207,CM1210}. In \cite{quivcalc} these problems on counting and inner product were considered for general quiver gauge theories. 
These theories, often arising in the context of 3-branes transverse to 6-dimensional singular Calabi-Yau, are associated with directed graphs, i.e. collections of nodes with directed edges between them \cite{DM}. The gauge group of the theory is a product of unitary groups, one unitary group for each node. The directed edges correspond to bi-fundamental matter fields, which transform according to the anti-fundamental representation of the gauge group corresponding to the starting node and the fundamental of the ending node. 
 
Also of interest in AdS/CFT are gauge theories where some of the matter fields are in fundamental representations, rather than bi-fundamentals, see for example \cite{KarKat02,EF1012,AFR1312, Ouyang:2003df, Levi:2005hh}. 
These theories can be associated to directed graphs, where we distinguish two types of nodes: gauge nodes and global symmetry nodes. The global symmetry nodes - which will be drawn as squares - are associated with unitary groups, which are global symmetry groups rather than gauge groups. In this case the local operators of interest are not required to be invariant under the global symmetry groups. Rather they are states in specified representations of these groups. They are still invariant with respect to unitary gauge groups associated to the gauge nodes, which will be drawn as circles. Each of these gauge nodes will have a number of outgoing arrows ending on square (global symmetry) nodes and transforming as fundamentals of the global symmetry. Each gauge node will also have a number of incoming arrows which start from square (global symmetry) nodes and transform as anti-fundamentals of the global symmetry. Since the global symmetry nodes correspond to fundamental or anti-fundamental fields, they are linked to a single gauge node each. 

The main result of the present paper is to extend the results of \cite{quivcalc} 
to the case of quiver gauge theories with fundamental matter fields. These are also called 
flavoured quiver gauge theories. In order to set up this extension, we have completed the proof of a key formula in \cite{quivcalc}. We have also found a new connection between the counting of quiver gauge theory operators and a word counting problem associated with the quiver graph. This uncovers a new link between gauge invariant operators of quiver theories and the mathematics of Cartier-Foata monoids \cite{CarFoa,Krat}. The latter is expressed here in terms of a word counting problem where the letters correspond to loops on a graph, with partial commutation relations.

The central insight in \cite{quivcalc} is that the quiver, aside from encoding the gauge group and matter content of a gauge theory, is a powerful calculational tool in the application of permutation group techniques to the enumeration of gauge invariant operators, and to the calculation of their correlators. The use of the quiver as calculator starts with the process of splitting all of the gauge nodes into two pieces, one of which has all the incoming edges of the original quiver and the other of which has all the outgoing edges of the original quiver. For each node, a new edge connecting the split copies of each node is introduced. All the edges of this split-node quiver are equipped with Young diagram labels and the nodes are associated with weights which are Littlewood-Richardson coefficients. A sum over all the Young diagram labels, with weights associated to the nodes, gives the counting of gauge invariant operators for finite ranks $N$ of the gauge groups. Here we establish a similar result in the case of quivers with fundamentals. The starting point is the group integral formula for counting gauge invariant operators \cite{Sundborg:2000wp, Aharony:2003sx}. The group integrals over $U(N)$ are done by using character expansions. These character expansions introduce characters of permutation groups, because of the Schur-Weyl duality \cite{FulHar,SWrev0804} link between unitary and symmetric groups.

The finite $N$ counting formulae admit significant simplifications in the limit of large $N$. 
At finite $N$, the counting involves sums over Young diagram labels. The sizes of the Young diagrams are related to the sizes of the local operators. When these sizes are small compared to the ranks, the Young diagram sums run over complete sets of representations of symmetric groups. This allows the use of formulae from Fourier transformation over finite groups such as 
\bea 
\delta (\sigma ) = { 1 \over m! } \sum_{ R  } d_R \chi_R ( \sigma ) \,.
\eea
The delta function is $1$ if $ \sigma $ is the identity permutation in $S_m$ - symmetric group of all permutations of $m$ objects - and zero otherwise. The result is that the counting formulae can be expressed in terms of sums over multiple permutations, related by delta function constraints. These sums over permutations can be converted into sums over partitions, described by an infinite sequence of integers $p_1 , p_2 , \cdots $. This sequence is related to cycle lengths in the cycle decomposition of permutations. The upshot is that the counting of gauge invariant operators at large rank can be given in terms of a sum over the infinite sequence of integers $p_i$. The general formula takes the form of an infinite product over $i$, where $i$ is related to the cycle lengths in the above description
\bea\label{infprod1} 
\prod_{i=1}^{\infty}  F_0^{[n]}  (\{ x_{ab} \rightarrow \sum_{ \alpha } x_{ab; \alpha } ^i  \})\,.
\eea
 Each factor in the product is built from a basic function $ F_0^{[n]}  ( \{ x_{ab}  \}) $.  
The integer $n$ is the number of gauge nodes and the subscript denotes the unflavoured case. 
The index $ \alpha $ runs over the different edges with the same starting gauge node $a$ and 
the same ending gauge node $b$. If there is no edge from $a$ to $b$, we substitute $ x_{ab} \rightarrow 0$. This structure was derived in \cite{quivcalc} for the case without flavour. The function $ F_0^{[n]}  ( \{x_{ab} \}) $ was explicitly computed for the case of quivers with small numbers of nodes and a simple general formula was guessed. A general formula for $F_0^{[n]}  ( \{x_{ab}\} ) $ was also derived in terms of contour integrals. However, the proof that the contour integrals really give the guessed simple form for the $ F_0^{[n]}  ( \{x_{ab}\} ) $ was not given. This missing step is completed in this paper. We also find that this function can be written in terms of a determinant: 
\bea\label{theMatterFreeanswer}  
F_0^{[n]}  (\{ x_{ab} \}) = \frac{1}{\det\left(\mathbb 1_n- X_n\right)}\,.
\eea
The matrix $X_n$ is defined to have variables $x_{ab}$ as the entry in the $a$-th row and $b$-th column. We may think of $X_n $ as a {\it weighted adjacency matrix} associated with the quiver graph which has $n$ nodes and a single directed edge for every specified starting point $a$ and end-point $b$. We refer to this latter quiver graph as the {\it complete $n$-node quiver graph}. The notion of adjacency matrix, and weighted versions thereof, are commonly used in the context of graph theory \cite{terras2010zeta,wallis2010beginner}. The $(a,b)$ entry of the adjacency matrix of a directed graph is equal to the number of oriented edges $M_{ab}$ from node $a$ to node $b$. In the present studies, it is natural 
to associate $ \sum_{ \alpha } x_{ab , \alpha } $ as the weight for a given pair of nodes, which reduces to 
$ M_{ab}$, the entry of the adjacency matrix, when the $x_{ab , \alpha }$ are set to $1$.

While the infinite product \eqref{infprod1} counts gauge invariant operators, the building block $ F_0^{[n]} ( \{ x_{ab} \} ) $ itself \eqref{theMatterFreeanswer} has no obvious counting interpretation in terms of the original gauge theory problem. Nevertheless, after applying a well-known identity, the determinant formula \eqref{theMatterFreeanswer} makes it clear that the expansion coefficients of this building block are positive, which suggests a counting interpretation. We give such an interpretation. It is in terms of a word counting problem involving letters corresponding to simple closed loops on the complete quiver graph. Two letters commute if the loops do not share a node but they do not commute if the loops do share a node. This, we describe as the {\it 
closed string word counting} problem. There is an equivalent word counting problem in terms of {\it charge conserving open string words}. Here open string words are made of string bits - which are edges of the quiver. Two different string bits do not commute if they have the same starting point. They commute if they do not share a starting point. Charge conserving open string words have the same number of open string bits leaving any vertex as arriving at that vertex. 
This charge-conserving open string word counting is actually directly related to the formulae in our derivations leading to the result. Its equivalence to the closed string word counting is a highly non-trivial fact, which is the content of a theorem of Cartier-Foata \cite{CarFoa} from the sixties! This type of word-counting is of interest in pure mathematics and 
theoretical computer science, where it is known under the heading of Cartier-Foata monoids \cite{CarFoa,Krat,heaps}. The monoid structure arises because the words can be composed to form other words, thus giving a product which turns the set of words into a monoid.
This new connection between the counting of gauge invariant operators and Cartier-Foata monoids is the first main result of this paper.

The infinite product form and the explicit formula for the building block, for the case of flavoured quivers, is derived using contour integrals in this paper. We find that the building block for the case of flavoured quivers is closely related to the unflavoured case
(see equations \eqref{hat final2 result} \eqref{PExp like relation hat improved copy}). 
It is worth emphasizing that the contour integrals we deal with for the large $N$ limit
are significantly simpler than the original integrals over the $U(N)$ groups. 
The contour integrals we use involve $n$ complex variables $z_a$, where $n$ is the number of 
nodes in the quiver. The equations \eqref{hat final2 result}\eqref{PExp like relation hat improved copy}, which form the second main result of this paper, are derived after finding the correct pole prescription for these $n$ integrals and uncovering the structure of the residues arising when the $z_a$ are evaluated at the poles.

We stress that, even though the motivation of this work is to study 4 dimensional \(\N=1\) gauge theories,
focusing on the holomorphic gauge invariant operators made from chiral super-fields which have a complex scalar 
as the lowest component of the superspace expansion, the counting techniques we developed do not depend on either the spacetime dimension or on the amount of supersymmetry. The results apply equally to holomorphic gauge invariants of a matrix quantum mechanics, or of a matrix model of multiple complex matrices transforming as bifundamentals.

The paper is organized as follows. Section 2 gives a summary of the main results. Section 3 starts from an integral over a product of unitary groups $ \prod_a U(N_a )$,
which gives the generating function for the counting of gauge-invariant operators \cite{Sundborg:2000wp, Aharony:2003sx}. This generating function depends on chemical potentials, one for each of the bifundamental fields in the theory, i.e. one for each edge in the quiver joining gauge nodes. In addition, there are chemical potentials for the global charges under the Cartan of the global symmetry groups. The integrand is expanded in terms of characters of the (gauge and global) unitary groups along with characters of permutation groups. The gauge unitary group characters can be integrated using orthogonality of the 
irreducible characters. The resulting expressions contain sums involving Young diagrams and group theoretic multiplicities called Littlewood-Richardson coefficients \cite{FulHar}. These sums are done in Appendix \ref{Derivation of the Gen Fun} and the outcome is an infinite product parameterised by an integer $i$. For each $i$ there are sums over integers, one for each edge of the quiver. We call these edge variables $p_{ab , \alpha } ,\, p_{a , \beta } ,\, \bar p_{ a , \gamma }$. These sums are constrained by Kronecker delta functions, one for each gauge node of the quiver. The structures of the sums in each factor of the $i$-product are closely related. Once these sums are performed for $i=1$, the expressions for the factor at each $i$ can be written down. The $i=1$ factor is the building block function $ F^{[n]} (\{ x_{ab }\} ,\{ t_a\} , \{\bar t_a\} )$ which can be viewed as the generalization of $ F_0^{[n]} (\{ x_{ab}\} ) $ for unflavoured quivers to flavoured quivers. The Kronecker delta constraints on the edge variables are expressed by introducing complex variables $z_a$, giving a product of $n$ contour integrals. 

Section 4 evaluates the contour integrals for the case without fundamental matter, recovering the result written down in \cite{quivcalc}. This involves finding the right prescription for picking up poles. The prescription is simple and intuitively very plausible. 
It is derived from the inequalities which ensure the applicability of the summation formulae leading to the contour integral formula obtained in Section 3. 
The derivation is presented in Appendix \ref{r&c}. With the specified pole prescription in hand, we describe the calculation of the integral. The integrand involves $n$ factors and there are $n$ integration variables $z_1, z_2 , \cdots , z_n$. The recursive evaluation of the integral leads to a formula \eqref{allpoles same z} for the poles encountered at each stage. 
The pole coefficients in this formula can be expressed neatly in terms of paths in the complete quiver graph. 
This expression is equation \eqref{guess} and is proved in Appendix \ref{attempt}. Using this expression we are able to prove the formula for $F_0^{[n]} $, 
an inverse of a signed sum over permutations of subsets of $n$ nodes, guessed in \cite{quivcalc}. 
We then recognise that the denominator is a determinant $ \det ( \mId_n - X_n ) $, which leads to \eqref{theMatterFreeanswer}. Section \ref{subsec:WordCounting} gives the combinatoric meaning of the basic building block in terms of word counting problems. Appendix \ref{sec:FandCSW} illustrates this interpretation in the case of 2-node and 3-node quivers. 
Section 5 evaluates the $n$ countour integrals for the building block function $ F^{[n]} ( \{x_{ab}\} ,\{ t_a \}, \{\bar t_a \})$ and expresses it in terms of determinants and minors of the matrix $ ( \mathbb 1_n - X_n )$. This gives a neat formula \eqref{final2}
for $ F^{[n]} ( \{x_{ab}\} ,\{ t_a \},\{ \bar t_a\} )$ in terms of $F^{[n]}_0 ( \{x_{ab}\} )$. Appendix \ref{using guess in matter} derives this formula, following a similar strategy to the unflavoured case, namely finding expressions for pole coefficients in terms of paths in a complete $n$-node quiver. Section 6 gives applications of the general counting formulae by considering explicit quiver gauge theories with fundamental matter.

\section{Basic definitions and summary of results}
In this paper we consider quiver gauge theories with gauge group \(\prod_{a=1}^nU(N_a)\), and 
 flavour symmetry of the schematic form \(\prod SU(F)\times SU(\bar F) \times U(1) \). 
The quivers have round nodes corresponding to gauge groups, and square nodes corresponding to global symmetries.

Fields leaving gauge node \(a\) and arriving at gauge node \(b\) will be denoted by \(\Phi_{ab,\alpha}\), and will transform in the antifundamental representation of \(U(N_a)\) and the fundamental of \(U(N_b)\). The third label $\alpha $ taking values in $ \{ 1,...,M_{ab}\}$ distinguishes between \(M_{ab}\) different fields with the same transformation properties under the gauge group. At every gauge node \(a\) we allow \(M_a\) different families of quarks $\{  Q_{a,\beta} , \beta=1,...,M_a \} $ transforming in the antifundamental of \(U(N)_a\) and \(\bar M_a\) different families of antiquarks $\{  \bar Q_{a,\gamma} , \gamma=1,...,\bar M_a \} $, transforming in the fundamental of \(U(N)_a\). Here \(\beta\) and \(\gamma\) are the multiplicities of the quarks and antiquarks respectively. The flavour group of the quark \(Q_{a,\beta}\) is denoted by \(U(F_{a,\beta})\), while the one for the antiquark \(\bar Q_{a,\gamma}\) is \(U(F_{a,\gamma})\). This configuration is pictorially represented in the figure below, while the gauge and flavour groups representations carried by every field in the quiver are summarised in table \ref{transf props}. Note that if we consider the free $\N=1$ theory with chiral (and anti-chiral) multiplets, then all the global symmetry will contain factors $ SU ( F_{a} ) \times SU(\bar F_a) \times U(1)_a $ for each gauge group, 
where $F_a = \sum_{ \beta } F_{ a , \beta } = \sum_{ \gamma } \bar F_{ a , \gamma } = \bar F_a $, and the equality $ F_a = \bar F_a$ is required by cancellation of the chiral gauge anomaly. When interactions are turned on, one may be interested in a subgroup $ \times_{ \beta } U ( F_{a , \beta } ) \times_{ \gamma } U( \bar F_{ a , \gamma}) $. Our calculations 
work without any significant modification for this case of product global symmetry, hence we will work in this generality. 
To recover the results for $ SU( F_a ) \times SU ( \bar F_a ) \times U(1)_a  $ global symmetry, we just drop the $ \beta , \gamma $ indices. 
Strictly speaking the global symmetry $ SU ( F_{a} ) \times SU(\bar F_a) \times U(1)_a $ of the free theory contains only the determinant one part $ S ( U(F_{a, 1} ) \times U( F_{a,2} ) \times \cdots U (F_{a,M_a } ) \times U ( \bar F_{a,1} ) \times \cdots \times U ( \bar F_{a,\bar M_a} ) )$. This means that, although for simplicity we write $ \times_{ \beta } U ( F_{a , \beta } ) \times_{ \gamma } U( \bar F_{ a , \gamma})  $ as the global symmetry, 
all the states we count are neutral under the $ U(1) $ which acts with a phase on all of the chiral fields and the opposite phase on all of the anti-chiral fields. This $U(1)$ is part of the $U(N_a)$ gauge symmetry. 
\begin{figure}[H]
\begin{center}\includegraphics[scale=0.9]{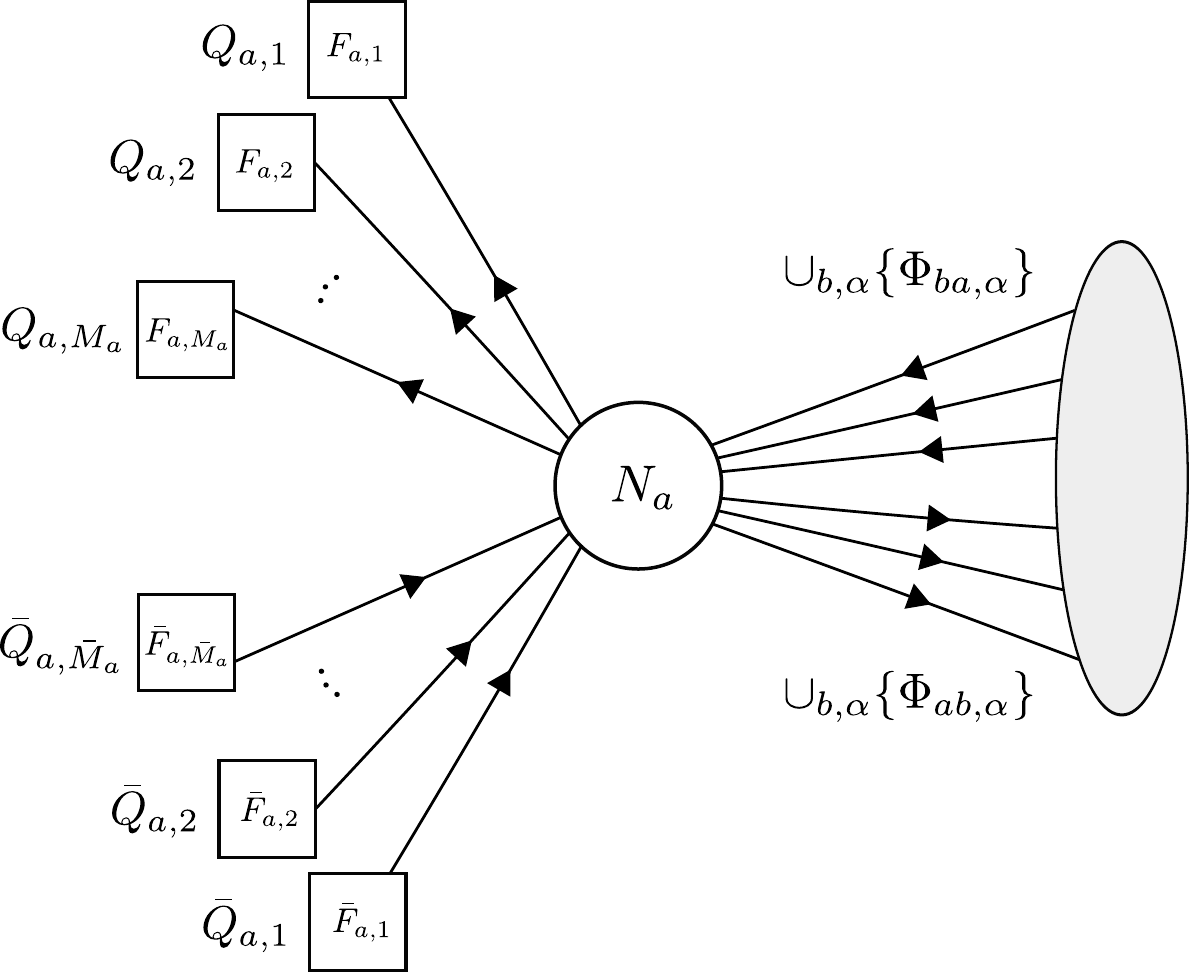}\\[1mm]
\caption{Representation of the arrows (fields) and square nodes (flavour symmetries) for a single gauge node labelled $a$. The shaded area represent the rest of the quiver.}
\end{center}
\end{figure}
In the generating function \(\Z\), the chemical potential of a generic bi-fundamental field \(\Phi_{ab,\alpha}\) will be denoted by \(x_{ab,\alpha}\), while to each quark \(Q_{a,\beta}\) and antiquark \(\bar Q_{a,\gamma}\) we will assign the chemical potential matrices \(\T_{a,\beta}\) and \(\bar \T_{a,\gamma}\) respectively, defined as
\begin{align}
\T_{a,\beta}=\text{diag}(t_{a,\beta, 1},t_{a,\beta, 2},...,t_{a,\beta, F_{a,\beta}})\,,\qquad
\bar \T_{a,\gamma}=\text{diag}(\bar t_{a,\gamma, 1},\bar t_{a,\gamma, 2},...,\bar t_{a,\gamma , \bar F_{a,\gamma}})\,.
\end{align}
The entries of these matrices encode all of the quark and antiquark chemical potentials: $t_{a,\beta, k}=e^{i\theta_{a,\beta, k}}$ is the chemical potential for a quark $Q_{a,\beta, k }$ charged under the $U(1)_k$ of the maximal torus $\prod_{j=1}^{F_{a,\beta}} U(1)_j\subset U(F_{a,\beta})$, while $\bar t_{a,\gamma, k}=e^{-i\theta_{a,\gamma, k}}$ is the chemical potential for an antiquark $\bar Q_{a,\gamma , k }$ charged under the $U(1)_k$ of $\prod_{j=1}^{\bar F_{a,\gamma}} U(1)_j\subset U(\bar F_{a,\gamma})$.\\
\begin{center}
\begin{table}[H]
\centering
\begin{tabular}{|l|c|c||c|c|} 
\hline
 & $U(N_a)$ & $U(N_b)$& $U(F_{a,\beta})$ & $U( \bar F_{a,\gamma})$ \\[2mm] \hline
$\Phi_{ab,\alpha}$ & $\bar \Box$ & $ \Box$ & $\textbf{1}$ & $\textbf{1}$ \\[2mm] \hline
$\Phi_{aa,\alpha}$ & $\text{Adj}$ & $\textbf{1}$ & $\textbf{1}$ & $\textbf{1}$ \\[2mm] \hline
$ Q_{a,\beta}$ & $\bar\Box$ &$\textbf{1}$& $\Box$ & $\textbf{1}$ \\[2mm]\hline 
$\bar Q_{a,\gamma}$& $ \Box$ & $\textbf{1}$ & $\textbf{1}$ & $\bar\Box$ \\[2mm] \hline
\end{tabular}
\caption{Gauge and flavour groups representations carried by \(\Phi_{ab,\alpha}\), \(Q_{a,\beta}\) and \(\bar Q_{a,\gamma}\). $\Box$, $\bar\Box$ and $\textbf{1}$ are respectively the fundamental, antifundamental and trivial representations of the correspondent group.}
\label{transf props}
\end{table}
\end{center}

\subsection{From gauge invariants to determinants and word counting }

For quiver gauge theories with bi-fundamental fields, the generating function 
$ \Z ( \{ x_{ab} \} ) $ for local holomorphic gauge invariant operators 
constructed from the chiral fields, is given by \cite{quivcalc}
\bea\label{Zresqc} 
\Z ( \{x_{ab,\alpha }\} ) = \prod_{i=1} F_0^{[n]} (\{ x_{ab} \rightarrow \sum_{ \alpha } x_{ab ; \alpha }^i  \}) \,.
\eea
 It is useful to introduce the 
{\it complete $n$-node quiver} which is a quiver that has $1$ edge for every specified start and end-point. 
An expression for $ F_0^{[n]} ( \{ x_{ab} \} ) $ was given as the inverse of a 
sum over permutations of subsets of the set of nodes of the $n$-node complete quiver. Equivalently this is an expression in terms of 
loops in the complete quiver
\begin{align}\label{F in loops in intro}
F_0^{[n]}(\{y_i\})=\left( 1+\sum_{\vv\subseteq V_n}\,\sum_{\sigma\in\text{Symm}(\vv)}\,\prod_{i\in\text{Cycles}(\sigma)}\,\left(-y_i\right)\right)^{-1}\,.
\end{align}
Here \(\vv\) is any subset of nodes of the quiver (except the empty set), and for the cycle \((abc\cdots d )\) \(y_{(abc\cdots d )}=x_{ab}x_{bc}\cdots x_{d a}\).
In this we observe, using standard matrix identities, that
\bea 
F_0^{[n]} = { 1 \over \det ( \mathbb 1_n - X_n ) } \,,
\eea 
where $X_n$ is an $n \times n $ matrix with entries $x_{ab}$. This formula is the subject of the Mac Mahon master theorem \cite{MacMahon}.

While the function $ \Z (\{ x_{ab , \alpha }\} ) $ counts gauge invariant operators, the gauge theory set-up does not immediately offer a combinatoric interpretation for $ F_0^{[n]} (\{ x_{ab}\} ) $. We give an interpretation of $ F_0^{[n]} $ in terms of word-counting problems associated with the complete $n$-node quiver. There are in fact two counting problems, one of them is a closed string counting problem. Consider a language where the words are made from letters which correspond to simple loops in the $n$-node quiver. These are loops that visit each node of the quiver no more than once. These letters equivalently correspond to cyclic permutations of any subset of integers $\{ 1, \cdots , n \}$. 
The words are constructed as strings, i.e. ordered sequences, of these letters with the additional equivalences introduced that letters corresponding to two simple loops $c$ and $c'$ commute if the loops do not share a node. We denote these letters by $ \hat y_c$. Then 
\bea 
\hat y_c \hat y_{c'} = \hat y_{c'} \hat y_c \,,
\eea 
if $c $ and $c'$ are loops that do not share a node. Any word contains a list of these letters $\hat y_{c_1} , \hat y_{c_2} \cdots \hat y_{c_k} $ with multiplicities $( m_1 , m_2 , \cdots , m_k )$. With these specified numbers, there is a multiplicity $ \mathcal {M} ( m_1 , \cdots , m_k ) $, of words since, in general, the order of the letters matters: if two loops $\hat y_c , \hat y_{c'}$ do share a node then $ \hat y_c \hat y_{c'} \ne \hat y_{c'} \hat y_c $. The expansion of $F_0^{[n]} $ in terms of the loop variables contains terms of the form $ y_{c_1}^{m_1} y_{c_2}^{m_2} \cdots y_{c_k}^{m_k} $ with coefficients, which are precisely 
the multiplicities of the words $ \mathcal { M } ( m_1 , \cdots , m_k )$. 

This is a remarkable new connection between a counting problem of words built from a partially commuting set of letters and the counting of gauge invariants. Since the letters correspond to simple loops, we call this the closed string word counting problem. Thus $ F_0^{ [n]} ( \{ x_{ab} \}  ) $ generates multiplicities of closed string words. In section \ref{subsec:WordCounting} we explain why this is true.
Along the way, we introduce another word counting formula based on letters corresponding to open string bits. 

\subsection{Generalization to flavoured quivers }

We extend the counting results to quivers that have bifundemental matter fields, as well as fundamental matter. 
We find again that the counting in the limit of large rank gauge groups is given as an infinite product. 
Each factor is obtained by making a simple substitution in a basic function $F^{[n]} ( \{ x_{ab} , t_a , \bar t_a \} )$, for the case quivers with $n$ gauge nodes. The function $F^{[n]} $ has an elegant expression in terms of matrices \(X_n\) and \(\Lambda_n\), whose matrix elements are
\begin{equation}\label{Lambda copy 2}
\left. X_n\right|_{ab}=
x_{ab}\,,\qquad\qquad
\left. \Lambda_n\right|_{ab}=t_a\bar t_b\,.
%
%
\end{equation}
Le us also define another $ n \times n $ matrix, 
\begin{align}
\chi_n\equiv (\mathbb 1_n-X_n)^{-1}\,.
\end{align}
In terms of these, \(F^{[n]}\) is the determinant
\begin{align}\label{hat final2 result}
 F^{[n]}(\{x_{ab}\},\{t_a\},\{\bar t_a\})=\det\left(\chi_n\,\exp\left[\chi_n\,\Lambda_n\right]\right)
 = \det ( \chi_n ) \exp \left ( { \rm tr } \left ( \chi_n \Lambda_n \right ) \right ) \,.
\end{align}
%
%
%
%
The generating function \(\Z\) can be obtained through the infinite product
\begin{align}\label{PExp like relation hat improved copy}
{\mathcal{Z}}(&\{x_{ab,\alpha}\},\{\T_{a,\beta}\},\{\bar \T_{a,\gamma}\})\nn
&=\prod_i F^{[n]}\left(\left\{x_{ab}\rightarrow \sum_{\alpha}x_{ab,\alpha}^i\right\},\left\{t_{a}\rightarrow\sum_{\beta}\frac{ \Tr(\T_{a,\beta}^{\,i})}{\sqrt i}\right\},\left\{\bar t_a\rightarrow\sum_{\gamma}\frac{\Tr(\bar \T_{a,\gamma}^{\,i})}{\sqrt i}\right\}\right)\,\,.
\end{align}
In the course of our derivation of $F^{[n]} $, we find the identity
\begin{align}\label{counting F in res}
F^{[n]}&=\det\left(\chi_n\,\exp\left[\chi_n\,\Lambda_n\right]\right)\nn
&=\sum_{\vec p\,}\,
\prod_{a=1}^n
\left( p_a + \sum_{b=1}^np_{ab}
\right)!
\left(\prod_{b=1}^n\frac{x_{ab}^{\,p_{ab}}}{p_{ab}!}\right)
\left(\frac{y_{a}^{\,p_{a}}}{p_{a}!}\right)
\left(\frac{\bar y_{a}^{\,\bar p_{a}}}{\bar p_{a}!}\right)
\,
\delta\left(p_{a}-\bar p_{ a} + 
\sum_{b=1}^n(p_{ab}-p_{ba})
\right)
\,,
\end{align}
with $\vec p=\bigcup_{a=1}^n\{\cup_{b=1}^n p_{ab},y_a^{p_a},\bar y_a^{\,\bar p_a}\}$. 
For the unflavoured case, this implies
\begin{align}\label{counting F_0 in res}
F^{[n]}_0&=\frac{1}{\det(\mathbb 1_n-X_n)}=\sum_{\vec p\,}\,
\prod_{a=1}^n
\left(\sum_{b=1}^np_{ab}
\right)!
\left(\prod_{b=1}^n\frac{x_{ab}^{\,p_{ab}}}{p_{ab}!}\right)
\delta\left(
\sum_{b=1}^n(p_{ab}-p_{ba})
\right)
\,,
\end{align}
where now $\vec p=\bigcup_{a,b=1}^n\{p_{ab}\}$. This formula is interpreted in section \ref{subsec:WordCounting} in terms of 
the counting of words built from partially commuting open string bits. The open string word counting 
has previously been studied in \cite{CarFoa} and its equivalence to the closed string word counting given.


\section{Group integral formula to partition sums }

In this section we will derive a contour integral formulation for the generating function \(\Z\). Our starting point is the group integral representation \cite{Sundborg:2000wp, Aharony:2003sx}
\begin{align}
\Z(\{x_{ab,\alpha}\},&\{t_{a,\beta,k}\},\{\bar t_{a,\gamma,k}\})=
\int \left(\prod_a d\,U_a\right)\nn
&\times\prod_a\,
\exp\left\{
\sum_{ i=1}^{ \infty } \frac{1}{i }
\left[
\sum_{b,\alpha}x_{ab,\alpha}^i \,\Tr (U_a^{\dagger i })\,\Tr (U_b^i )\right.\right.\nn
&\qquad\qquad\qquad\qquad
\left.\left.+\sum_\beta\,\sum_{k=1}^{F_{a,\beta}}t_{a,\beta , k}^i \,\Tr( U_a^{\dagger i })+
\sum_\gamma \,\sum_{k=1}^{\bar F_{a,\gamma}}\bar t_{a,\gamma, k}^i \,\Tr( U_a^i )
\right]
\right\}\,. 
\end{align}
 Here $x_{ab,\alpha}$ is the chemical potential for the $\Phi_{ab,\alpha}$ field, while $t_{a,\beta,k}=e^{i\theta_{a,\beta, k}}$ is the chemical potential for a quark $Q_{a,\beta, k }$ charged under the $U(1)_k$ of the maximal torus $\prod_{j=1}^{F_{a,\beta}} U(1)_j\subset U(F_{a,\beta})$. Analogously, $\bar t_{a,\gamma,k}=e^{-i\theta_{a,\gamma, k}}$ is the chemical potential for an antiquark $\bar Q_{a,\gamma, k }$ charged under the $U(1)_k$ of $\prod_{j=1}^{\bar F_{a,\gamma}} U(1)_j\subset U(\bar F_{a,\gamma})$. Expanding the generating function 
gives the \emph{counting function} 
$\N(\{n_{ab,\alpha}\},\{n_{a,\beta,k}\},\{\bar n_{a,\gamma,k}\}) $ for specified numbers $ n_{ab,\alpha}$ of 
bifundamentals $ \Phi_{ab, \alpha}$, $ n_{a,\beta,k} $ of quarks $ Q_{ a , \beta , k }$ and 
$ \bar n_{ a , \gamma , k } $ anti-quarks 
$\bar Q_{ a , \gamma , k } $:
\begin{align}\label{igt counting}
\Z(\{x_{ab,\alpha}\},\{t_{a,\beta,k}\},\{\bar t_{a,\gamma,k}\})&=
\sum_{\{n_{ab,\alpha}\}}\,\sum_{\{n_{a,\beta,k}\}}\,\sum_{\{\bar n_{a,\gamma,k}\}}\,
\N(\{n_{ab,\alpha}\},\{n_{a,\beta,k}\},\{\bar n_{a,\gamma,k}\})\nn
&\qquad\quad\times \left(
\prod_{a,b,\alpha}\,x_{ab,\alpha}^{n_{ab,\alpha}}\right)\,
\left(\prod_{a,\beta,k}\,t_{a,\beta,k}^{n_{a.\beta,k}}\right)\,
\left(\prod_{a,\gamma,k}\,\bar t_{a,\gamma,k}^{\,\,\bar n_{a,\gamma,k}}\right)\,.
\end{align}

The chemical potentials for the quark/antiquark matter content can be nicely encoded in the unitary matrices $\T_{a,\beta}=\text{diag}(t_{a,\beta,1},t_{a,\beta,2},...,t_{a,\beta,F_{a,\beta}})$ and $\bar \T_{a,\gamma}=\text{diag}(\bar t_{a,\gamma,1},\bar t_{a,\gamma,2},...,\bar t_{a,\gamma,\bar F_{a,\gamma}})$ respectively, so that
\begin{align}
\Z(\{x_{ab,\alpha}\},&\{\T_{a,\beta}\},\{\bar \T_{a,\gamma}\})=
\int \left(\prod_a d\,U_a\right)\,\prod_a\,
\exp\left\{
\sum_{ i =1}^{ \infty} \frac{1}{i }
\left[
\sum_{b,\alpha}x_{ab,\alpha}^i \,\Tr (U_a^{\dagger i })\,\Tr (U_b^i )\right.\right.\nn
&\qquad\qquad\qquad\qquad
\left.\left.+\sum_\beta\,\Tr( U_a^{\dagger i })\,\Tr(\T_{a,\beta}^i )+
\sum_\gamma \,\Tr( \bar \T_{a,\gamma}^{ i })\,\Tr( U_a^i )
\right]
\right\}\,. 
\end{align}
Using the shorthand notation $\int \left(\prod_a d\,U_a\right)\equiv \int$ and expanding the exponential function we get
\begin{align}
\Z(\{&x_{ab,\alpha}\},\{\T_{a,\beta}\},\{\bar \T_{a,\gamma}\})\nn
&=\int\prod_a
\left\{
\left(
\sum_{\{p_{ab,\alpha}^{(i)}\}_a}\,\prod_{b,\alpha}\,\frac{x_{ab,\alpha}^{\sum_i i p_{ab,\alpha}^{(i )}}}{\prod_i p_{ab,\alpha}^{(i )}!\,n^{p_{ab,\alpha}^{(i )}}}
\,\prod_i \,(\Tr\, U_a^{\dagger i })^{p_{ab,\alpha}^{(i )}}\,(\Tr\, U_b^i )^{p_{ab,\alpha}^{(i )}}
\right)\right.\\[3mm]
&\qquad\qquad\qquad\qquad
\times\left(
\sum_{\{p_{a,\beta}^{(i)}\}_a}\,\prod_{\beta}\,\frac{1}{\prod_i p_{a,\beta}^{(i)}!\,n^{p_{a,\beta}^{(i )}}}
\,\prod_i \,(\Tr\, U_a^{\dagger i })^{p_{a,\beta}^{(i )}}\,(\Tr\, \T_{a,\beta}^i )^{p_{a,\beta}^{(i )}}
\right)\nn
&\qquad\qquad\qquad\qquad\qquad\qquad
\times\left.\left(
\sum_{\{\bar p_{a,\gamma}^{(i )}\}_a}\,\prod_{\gamma}\,\frac{1}{\prod_i \bar p_{a,\gamma}^{(i )}!\,n^{\bar p_{a,\gamma}^{(i )}}}
\,\prod_i \,(\Tr\,\bar\T_{a,\gamma}^{ i })^{\bar p_{a,\gamma}^{(i )}}\,(\Tr\, U_a^i )^{\bar p_{a,\gamma}^{(i )}}
\right)\right\}\,,\nonumber
\end{align}
where $\sum_{\{p_{ab,\alpha}^{(i )}\}_a}\equiv \prod_{i ,b,\alpha}\,\sum_{p_{ab,\alpha}^{(i )}=0}^{\infty}$, $ \sum_{\{p_{a,\beta}^{(i)}\}_a} \equiv \prod_{ \beta , i } \sum_{p_{a , \beta , i } } $ and 
$ \sum_{\{\bar p_{a,\gamma}^{(i )}\}_a} \equiv \prod_{ \gamma , i } \sum_{ p_{ \gamma , i }} $. 
 Rearranging sums and collecting like terms, we obtain 
\begin{align}\label{gif tr}
\Z&(\{x_{ab,\alpha}\},\{\T_{a,\beta}\},\{\bar \T_{a,\gamma}\})\nn
&=\sum_{\{p_{ab,\alpha }^{( i )}\}}\,\sum_{\{p_{a,\beta}^{(i )}\}}\,
\sum_{\{\bar p_{a,\gamma}^{( i )}\}}\,
\int
\left(
\prod_{a,b,\alpha}\,\frac{x_{ab,\alpha}^{\sum_i i p_{ab,\alpha}^{(i )}}}{\prod_i p_{ab,\alpha}^{(i )}!\,n^{p_{ab,\alpha}^{(i )}}}\right)\,
\left(
\prod_{a,\beta}\,\frac{1}{\prod_i p_{a,\beta}^{(i )}!\,i^{p_{a,\beta}^{(i )}}}\right)\,
\left(
\prod_{a,\gamma}\,\frac{1}{\prod_i \bar p_{a,\gamma}^{(i )}!\,i ^{\bar p_{a,\gamma}^{(i )}}}\right)\nn
&\qquad
\times
\left\{
\prod_{a,i } \,(\Tr\,U_a^{\dagger i })^{\sum_{b,\alpha}p_{ab,\alpha}^{(i )}+\sum_\beta p_{a,\beta}^{(i )}}\,\,(\Tr\,U_a^i )^{\sum_{b,\alpha}p_{ba,\alpha}^{(i )}+\sum_\gamma \bar p_{a,\gamma}^{(i )}}
\right\}\\[3mm]
&\qquad\qquad\qquad
\times\prod_{a,i }\,\left[\prod_\beta (\Tr\,\T_{a,\beta}^i )^{ p_{a,\beta}^{(i )}}\right]\,\left[\prod_\gamma (\Tr\,\bar\T_{a,\gamma}^{ i })^{\bar p_{a,\gamma}^{(i )}}\right]\,.\nonumber
\end{align}
We now collect powers of $ x_{ab , \alpha } , \T_{ a , \beta }  , \bar \T_{ a , \gamma }$ 
denoted $ n_{ ab , \alpha } , n_{ a , \beta } , \bar n_{ a , \gamma }$, and introduce the quantities 
\begin{align}\label{vec p}
\vec p_{ ab, \alpha } = \cup_i \{ p_{ab, \alpha}^{(i)} \} ,\,\qquad\qquad 
\allowbreak\vec p_{a, \beta } = \cup_i \{  p_{a, \beta }^{(i)} \} ,\,\qquad\qquad \vec {\bar p}_{a , \gamma } = \cup_i \{  \bar p_{a , \gamma }^{\,(i)} \}\,.
\end{align}
These form partitions of $n_{ab , \alpha } ,\, n_{ a , \beta } ,\, \bar n_{ a , \gamma }$, which can be 
interpreted as cycle lengths of permutations $\sigma_{ab,\alpha}\in S_{n_{ab,\alpha}},\,\sigma_{a,\beta}\in S_{n_{a,\beta}}$ and $ \bar \sigma_{a,\gamma}\in S_{\bar n_{a,\gamma}}$ respectively. These cycle structures determine conjugacy classes denoted $ [\sigma_{ a b , \alpha } ] , [ \sigma_{ a , \beta } ] , [ \bar \sigma_{ a , \gamma } ] $. 
We have
\begin{align}\label{|p|}
\sum_{i=1}^\infty i p_{ab,\alpha}^{(i )}=n_{ab,\alpha}\,,\qquad\quad
 \vert  \vec p_{ab,\alpha}  \vert =\frac{n_{ab,\alpha}!}{\prod_i p_{ab,\alpha}^{(i )}!\,i^{p_{ab,\alpha}^{(i )}}}\,,
\end{align}
and similarly for \(\vec p_{a,\beta}\) and \(\vec{\bar p}_{a,\gamma}\). The second equation above gives the number of permutations with the specified cycle structure. 
We also use the identity
\begin{align}
\prod_i\,(\Tr\,U^i)^{ [\sigma]^{(i)}}=\sum_{R\vdash n\atop l(R)\leq N}\chi_R(\sigma)\chi_R(U)\,,\qquad \sigma\in S_n\,,\,\,\, U\in U(N)\,,
\end{align}
which follows from Schur-Weyl duality (see e.g. \cite{FulHar}): here $R$ is a partition of $n$ and $[\sigma]^{(i)}$ is the number of cycles of length $i$ in $\sigma$, which is a function of the conjugacy class $[\sigma ]$. The Young diagrams are constrained to have no more than $N$ rows, which is expressed as $ l(R) \le N$. 
This encodes the constraints following from finiteness of the ranks $N_a$. For $ n_a \le N_a$, these constraints can be dropped, which is the origin of simplifications at large $N_a$. 
Collecting powers of traces of $ U_a^{\dagger} $, this equation can be used to rewrite the traces in \eqref{gif tr} as
\begin{align}
\prod_i &\,(\Tr\,U_a^{\dagger i })^{\sum_{b,\alpha}p_{ab,\alpha}^{(i )}+\sum_\beta p_{a,\beta}^{(i )}}=
\sum_{R_a\vdash n_a\atop l(R_a)\leq N_a}\chi_{R_a}(\times_{b,\alpha} \sigma_{ab,\alpha}\times_\beta \sigma_{a,\beta})\chi_{R_a}(U_a^\dagger)\,,\nn
&\qquad\qquad\quad n_a=\sum_{b,\alpha}n_{ab,\alpha}+\sum_\beta n_{a,\beta}
\,,
\end{align}
and similarly for the other terms. 
The product of the permutations over $ b ,\, \alpha ,\, \beta $ describes an outer product of permutations acting on subsets of size $n_{a b , \alpha } ,\, n_{ a , \beta } $ of $n_a$. 
Using these definitions, we can write
\begin{align}\label{gif tr aft}
\Z&(\{x_{ab,\alpha}\},\{\T_{a,\beta}\},\{\bar \T_{a,\gamma}\})=
\sum_{\{n_{ab,\alpha}\}}\,\sum_{\{n_{a,\beta}\}\atop\{\bar n_{a,\gamma}\}}
\sum_{\{\vec p_{ab,\alpha}\}}\,\sum_{\{\vec p_{a,\beta}\}}\,
\sum_{\{\vec{\bar p}_{a,\gamma}\}}\\[3mm]
&\times\int
\left(
\prod_{a,b,\alpha}\,\frac{x_{ab,\alpha}^{n_{ab,\alpha}}}{n_{ab,\alpha}!}\,|\vec p_{ab,\alpha}|\right)\,
\left(
\prod_{a,\beta}\,\frac{1}{n_{a,\beta}!}\,|\vec p_{a,\beta}|\right)\,
\left(
\prod_{a,\gamma}\,\frac{1}{\bar n_{a,\gamma}!}\,|\vec {\bar p}_{a,\gamma}|\right)\nn
&
\times
\prod_a\,\left\{
\sum_{R_a\vdash n_a\atop l(R_a)\leq N_a}\sum_{S_a\vdash n_a\atop l(S_a)\leq N_a}\chi_{R_a}(\times_{b,\alpha}  \sigma_{ab,\alpha}\times_\beta  \sigma_{a,\beta})\,\chi_{S_a}(\times_{b,\alpha}  \sigma_{ba,\alpha}\times_\gamma  {\bar \sigma}_{a,\gamma})\,\chi_{R_a}(U_a^\dagger)\,
\chi_{S_a}(U_a)
\right\}\nn
&
\times\prod_{a}\,
\left\{\sum_{\left\{ r_{a,\beta}\vdash n_{a,\beta}\atop l(r_{a,\beta})\leq F_{a,\beta}\right\}_a}\,\prod_\beta \chi_{r_{a,\beta}}( \sigma_{a,\beta})\chi_{r_{a,\beta}}(\T_{a,\beta}) \right\}
\left\{\sum_{\left\{\bar r_{a,\gamma}\vdash \bar n_{a,\gamma}\atop l(\bar r_{a,\gamma})\leq \bar F_{a,\gamma}\right\}_a}\,\prod_\gamma \chi_{\bar r_{a,\gamma}}( {\bar \sigma}_{a,\gamma})\chi_{\bar r_{a,\gamma}}(\bar \T_{a,\gamma}) \right\}
\nonumber\,,
\end{align}
where \(\sigma_{ab,\alpha}\), \(\sigma_{a,\beta}\) and \(\bar\sigma_{a,\gamma}\) are representatives of the conjugacy classes specified by \(\vec p_{ab,\alpha}\), \(\vec p_{a,\beta}\) and \(\vec{\bar p}_{a,\gamma}\) respectively. We can now cast the sums over these vectors into sums over the permutations \(\sigma_{ab,\alpha}\in S_{n_{ab,\alpha}}\), \(\sigma_{a,\beta}\in S_{n_{a,\beta}}\) and \(\bar\sigma_{a,\gamma}\in S_{\bar n_{a,\gamma}}\).
We also use the symmetric group character expansion
\begin{align}
\chi_{R_a}&(\times_{b,\alpha}\sigma_{ab,\alpha}\times_\beta\sigma_{a,\beta})\nn
&=\sum_{\cup_{b,\alpha}\{r_{ab,\alpha}\vdash n_{ab,\alpha}\}\atop \cup_\beta\{r_{a,\beta}\vdash n_{a,\beta}\}}
g(\cup_{b,\alpha}r_{ab,\alpha}\cup_\beta r_{a,\beta};R_a)\left(\prod_{b,\alpha}\,\chi_{r_{ab,\alpha}}(\sigma_{ab,\alpha})\right)
\,\left(\prod_\beta\,\chi_{r_{,\beta}}(\sigma_{a,\beta})\right)\,,
\end{align}
and similarly for \(\chi_{S_a}(\times_{b,\alpha}  \sigma_{ba,\alpha}\times_\gamma  {\bar \sigma}_{a,\gamma})\). In the formula above, $g (\cup_{b,\alpha}r_{ab,\alpha}\cup_\beta r_{a,\beta};R_a) $ is a Littlewood-Richardson coefficient. This is the multiplicity of the representation $ \otimes_{ b , \alpha } r_{ a , b , \alpha } \otimes_{ \beta } r_{ a , \beta } $ of the subgroup $ \times_{ b , \alpha } 
S_{ n_{ a b , \alpha } } \times_{ \beta } S_{ n_{ a , \beta } } $ when the representation $R_a$ of $S_{n_a}$ is decomposed into irreducibles of the product subgroup. 
Finally, using use the $U(N)$ character orthogonality formula 
\bea 
\int d\,U\,\chi_{R}(U^\dagger)\chi_S(U)=\delta_{R,S}\,.
\eea 
we obtain
\begin{align}\label{gif sigma}
\Z&(\{x_{ab,\alpha}\},\{\T_{a,\beta}\},\{\bar \T_{a,\gamma}\})\nn
&=\sum_{\{n_{ab,\alpha}\}}\,\sum_{\{n_{a,\beta}\}\atop\{\bar n_{a,\gamma}\}}\,\sum_{\{\sigma_{ab,\alpha}\}}\,\sum_{\{\sigma_{a,\beta}\}}\,
\sum_{\{\bar \sigma_{a,\gamma}\}}
\left(
\prod_{a,b,\alpha}\,\frac{x_{ab,\alpha}^{n_{ab,\alpha}}}{n_{ab,\alpha}!}\,\right)\,
\left(
\prod_{a,\beta}\,\frac{1}{n_{a,\beta}!}\right)\,
\left(
\prod_{a,\gamma}\,\frac{1}{\bar n_{a,\gamma}!}\right)\nn
&\,
\times\sum_{\left\{R_a\vdash n_a\atop l(R_a)\leq N_a\right\}}\sum_{\{r_{ab,\alpha}\vdash n_{ab,\alpha}\}\atop\{s_{ab,\alpha}\vdash n_{ab,\alpha}\} }\,\sum_{\{s_{a,\beta}\vdash n_{a,\beta}\}\atop \{\bar s_{a,\gamma}\vdash \bar n_{a,\gamma}\}}\,
\left\{\prod_a
g(\cup_{b,\alpha}r_{ab,\alpha}\cup_\beta s_{a,\beta};R_a)\,g(\cup_{b,\alpha}s_{ba,\alpha}\cup_\gamma \bar s_{a,\gamma};R_a)\right\}\nn
&\qquad\times
\left(\prod_{a,b,\alpha}\,\chi_{r_{ab,\alpha}}(\sigma_{ab,\alpha})\chi_{s_{ba,\alpha}}(\sigma_{ba,\alpha})\right)
\,\left(\prod_{a,\beta}\,\chi_{s_{a,\beta}}(\sigma_{a,\beta})\right)
\,\left(\prod_{a,\gamma}\,\chi_{\bar s_{,\gamma}}(\bar \sigma_{a,\gamma})\right)
\nn
&\qquad\times
\sum_{\{r_{a,\beta}\vdash n_{a,\beta}\}\atop \{\bar r_{a,\gamma}\vdash \bar n_{a,\gamma}\}}
\left\{\prod_{a,\beta} \chi_{r_{a,\beta}}(\sigma_{a,\beta})\chi_{r_{a,\beta}}(\T_{a,\beta}) \right\}
\left\{\prod_{a,\gamma} \chi_{\bar r_{a,\gamma}}( {\bar \sigma}_{a,\gamma})\chi_{\bar r_{a,\gamma}}(\bar\T_{a,\gamma}) \right\}
\,. 
\end{align}
Note that we dropped the $l(r_{a,\beta})\leq F_{a,\beta}$ constraint on the sum over quark representations, since contributions coming from representations with $l(r_{a,\beta})> F_{a,\beta}$ are automatically zero due to the vanishing of $\chi_{r_{a,\beta}}(\T_{a,\beta})$ (similar comments hold for the sum over antiquark representations as well).

Finally, using the orthogonality of the symmetric group characters $\sum_{\sigma\in S_n} \chi_r(\sigma)\chi_s(\sigma)=n!\delta_{r,s}$, we get the formula
\begin{align}\label{igt}
\Z&(\{x_{ab,\alpha}\},\{\T_{a,\beta}\},\{\bar \T_{a,\gamma}\})=
\sum_{\{n_{ab,\alpha}\}}\,\sum_{\{n_{a,\beta}\}\atop\{\bar n_{a,\gamma}\}}\,\left(
\prod_{a,b,\alpha}\,x_{ab,\alpha}^{n_{ab,\alpha}}\right)\,
\sum_{\left\{R_a\vdash n_a\atop l(R_a)\leq N_a\right\}}\,\sum_{\{r_{ab,\alpha}\vdash n_{ab,\alpha}\}}\,\sum_{\{r_{a,\beta}\vdash n_{a,\beta}\}\atop \{\bar r_{a,\gamma}\vdash \bar n_{a,\gamma}\}}\\[3mm]
&\prod_a\,
g(\cup_{b,\alpha}r_{ab,\alpha}\cup_\beta r_{a,\beta};R_a)\,g(\cup_{b,\alpha}r_{ba,\alpha}\cup_\gamma \bar r_{a,\gamma};R_a)\
\left(\prod_{\beta}\chi_{ r_{a,\beta}}(\T_{a,\beta})  \right)
\left(\prod_{\gamma} \chi_{\bar r_{a,\gamma}}(\bar\T_{a,\gamma})  \right)
\,.\nonumber
\end{align}
Note that setting \(\T_{a,\beta}=t_{a,\beta}\,\mathbb 1_{a,\beta}\) (\(\bar\T_{a,\gamma}=\bar t_{a,\gamma}\,\mathbb 1_{a,\gamma}\)) gives an unrefined generating function, in which we no longer distinguish quark (antiquark) states charged under different $U(1)$ factors in the maximal torus of \(U(F_{a,\beta})\) (\(U(\bar F_{a,\gamma})\)). This unrefinement is immediately obtained from \eqref{igt} through the substitutions
\begin{align}
\chi_{ r_{a,\beta}}(\T_{a,\beta}) \rightarrow \text{dim}^{U(F_{a,\beta})}(r_{a,\beta})\,t_{a,\beta}^{n_{a,\beta}}\,,\qquad\quad
\chi_{ \bar r_{a,\gamma}}(\bar \T_{a,\gamma}) \rightarrow \text{dim}^{U(\bar F_{a,\gamma})}(\bar r_{a,\gamma})\,\bar t_{a,\gamma}^{\,\bar n_{a,\gamma}}
\,.
\end{align}
The $ \text{dim}^{ U(F)} ( r ) $ is the dimension of the representation $r$ of $U(F)$. 

For an $F$ dimensional unitary matrix \(\T\) with eigenvalues \((t_1,t_2,...,t_F)\) and a partition \(R\) of \(n\), we have
\begin{align}
\chi_R(\T)=\sum_{\sigma\in S_n}\frac{\chi_R(\sigma)}{n!}\prod_i (\Tr\,\T^i)^{[\sigma]^{(i)} }=\sum_{\{n_j\}}g(\cup_{j}[n_j];R)\,\prod_{j=1}^F\,t_j^{n_j}\,,
\end{align}
where $ n = \sum_j n_j $ and \([n_j]\) is the single-row totally symmetric representation of \(S_{n_j}\). These Littlewood-Richardson multiplicities for single-row representations and a general $R$ are called Kostka numbers \cite{FulHar}. Note also that the Littlewood-Richardson multiplicities satisfy \cite{FulHar,doubcos}
\begin{align}
\sum_s g(r_1,s;R)g(r_2,r_3;s)=g(r_1,r_2,r_3;R)\,.
\end{align}
Using these identities, we can write the counting function \(\N(\{n_{ab,\alpha}\},\{n_{a,\beta,k}\},\{\bar n_{a,\gamma,k}\})\) as
\begin{align}\label{counting f}
\N&(\{n_{ab,\alpha}\},\{n_{a,\beta,k}\},\{\bar n_{a,\gamma,k}\})\\[3mm]
&=\sum_{\left\{R_a\vdash n_a\atop l(R_a)\leq N_a\right\}}\,\sum_{\{r_{ab,\alpha}\vdash n_{ab,\alpha}\}}\,\prod_a\,
g(\cup_{b,\alpha}r_{ab,\alpha}\cup_{\beta,k} [n_{a,\beta,k}];R_a)\,g(\cup_{b,\alpha}r_{ba,\alpha}\cup_{\gamma,k} [\bar n_{a,\gamma,k}];R_a)\,,\nonumber
\end{align}
where \(n_a=\sum_{b,\alpha}n_{ab,\alpha}+\sum_{\beta,k} n_{a,\beta,k}\). 

We can give a pictorial interpretation of the counting function \eqref{counting f} as follows. 
\begin{itemize}
\item[\(i)\)] Choose the set of integers \(\cup_{a,b,\alpha}\{n_{ab,\alpha}\}\cup_{a,\beta,k}\{n_{a,\beta,k}\}\cup_{a,\gamma,k}\{\bar n_{a,\gamma,k}\}\)
These determine the numbers of elementary fields of various types in the composite operators under consideration.

\item[\(ii)\)] To all edges joining the gauge node \(a\) to the gauge node \(b\), associate a representation \(r_{ab,\alpha}\) of the symmetric group \(S_{n_{ab,\alpha}}\). 
\item[\(ii)\)] Divide each gauge node \(a\) into two components, \(a^+ \) and \(a^- \): the former collects all the edges 
 coming into the node $a$, while the latter collects all the edges leaving the node $a$. Connect 
\(a^+ \) to \(a^- \) by adding a directed edge carrying a representation \(R_a\) of \(S_{n_a}\), where \(n_a=\sum_{b,\alpha}n_{ab,\alpha}+\sum_{\beta,k} n_{a,\beta,k}\). The result is called \emph{split-node} quiver.
\item[\(iii)\)] To each \(a^- \) attach the Littlewood-Richardson coefficient \(g(\cup_{b,\alpha}r_{ab,\alpha}\cup_{\beta,k} [n_{a,\beta,k}];R_a)\); to each \(a^+ \) attach the Littlewood-Richardson coefficient \(g(\cup_{b,\alpha}r_{ba,\alpha}\cup_{\gamma,k} [\bar n_{a,\gamma,k}];R_a)\).
\item[\(iv)\)] Take the product of all the Littlewood-Richardson coefficients obtained in the previous step and sum over all possible representations \(\{R_a\}\) and \(\{r_{ab,\alpha}\}\), imposing finite \(N\) constraints \(l(R_a)\leq N_a\) at each gauge node \(a\).
\end{itemize}

\begin{figure}[H]
\begin{center}\includegraphics[scale=1.5]{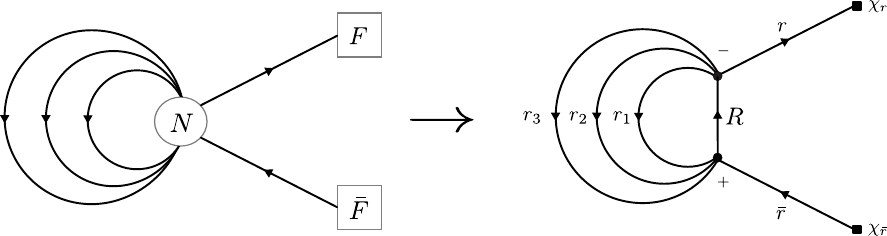}\\[1mm]
\caption{The $\N=1$ quiver and the corresponding split node diagram for a \(\N=2\) SQCD with an adjoint hypermultiplet.}\label{split node n=4 mat}
\end{center}
\end{figure}

As an example of the application of \eqref{igt}, consider an \(\N=2\) SQCD with an adjoint hypermultiplet. The \(\N=1\) quiver diagram for this gauge theory and its corresponding split node quiver are depicted in figure \ref{split node n=4 mat}.
The generating function for this model can then be readily obtained using \eqref{igt}:
\begin{align}
\Z(x_1,x_2,x_3,\T,\bar \T)&=
\sum_{n_1,n_2,n_3=0}^\infty\,\,\sum_{n,\bar n=0}^\infty\,
x_1^{n_1}\,x_2^{n_2}\,x_3^{n_3}\\[3mm]
&\qquad\times\,\sum_{R\vdash m\atop l(R)\leq N}\,\sum_{{{r_1\vdash n_1}\atop r_2\vdash n_2}\atop r_3\vdash n_3}\,\sum_{r\vdash n\atop \bar r\vdash \bar n}
g(r_1,r_2,r_3,r;R)\,g(r_1,r_2,r_3,\bar r;R)\,\chi_{ r}(\T)\, \chi_{\bar r}(\bar\T)\,,
\nonumber
\end{align}
with $m=n_1+n_2+n_3+n=n_1+n_2+n_3+\bar n$. On the other hand, using \eqref{counting f} we can write
the counting function 
\begin{align}
\N(&n_1,n_2,n_3,\{n_j\},\{\bar n_k\})\nn
&=\sum_{R\vdash m\atop l(R)\leq N}\,\sum_{{{r_1\vdash n_1}\atop r_2\vdash n_2}\atop r_3\vdash n_3}\,
g(r_1,r_2,r_3,[n_1],[n_2],\cdots ,[n_F];R)\,g(r_1,r_2,r_3,[\bar n_1],[\bar n_2],\cdots,[\bar n_{\bar F}];R)\,,
\end{align}
so that
\begin{align}
\Z\left(x_1,x_2,x_3,\{t_j\},\{\bar t_k\}\right)=\sum_{n_1,n_2,n_3}\,\sum_{\{n_j\}}\,\sum_{\{\bar n_k\}}&\,\,\N(n_1,n_2,n_3,\{n_j\},\{\bar n_k\})\nn
&\times\,x_1^{n_1}\,x_2^{n_2}\,x_3^{n_3}\,\left(\prod_{j=1}^F t_j^{n_j}\right)\left(\prod_{k=1}^{\bar F}\bar t_k^{\,\,\bar n_k}\right)\,.
\end{align}

Let us now consider the flavoured conifold gauge theory \cite{Ouyang:2003df, Levi:2005hh, Benini:2006hh, Bigazzi:2008zt}, whose quiver is depicted in figure \ref{flavoured conifold pic}:
\begin{figure}[H]
\begin{center}\includegraphics[scale=1.6]{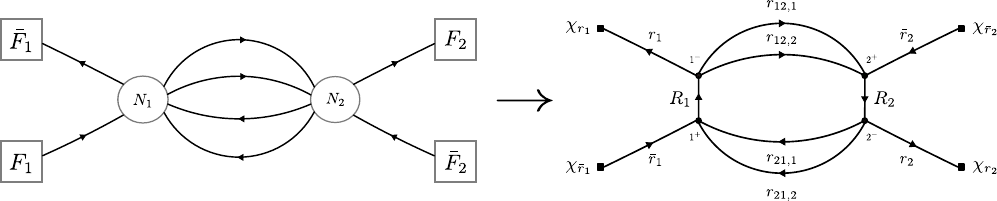}\\[1mm]
\caption{The flavoured conifold quiver and its split node quiver.}\label{flavoured conifold pic}
\end{center}
\end{figure}
Applying \eqref{igt}, we find that the generating function for the flavoured conifold is
\begin{align}
&\Z(x_{12,1},x_{12,2},x_{21,1},x_{21,2},\T_1,\T_2,\bar \T_1,\bar \T_2)\nn
&
=\sum_{n_{12,1},n_{12,2}=0}^\infty\,\,\sum_{n_{21,1},n_{21,2}=0}^\infty\,\,
x_{12,1}^{n_{12,1}}\,x_{12,2}^{n_{12,2}}\,x_{21,1}^{n_{21,1}}\,x_{21,2}^{n_{21,2}}\,
\sum_{R_1\vdash m_1\atop l(R_1)\leq N_1}\,\sum_{R_2\vdash m_2\atop l(R_2)\leq N_2}\,
\sum_{r_{12,1}\vdash n_{12,1}\atop r_{12,2}\vdash n_{12,2}}\,\sum_{r_{21,1}\vdash n_{21,1}\atop r_{21,2}\vdash n_{21,2}}\,
\sum_{r_1\vdash n_1\atop \bar r_1\vdash \bar n_1}\,\sum_{r_2\vdash n_2\atop \bar r_2\vdash \bar n_2}\nn
&\qquad\qquad\times
g(r_{12,1},r_{12,2},r_1;R_1)\,g(r_{21,1},r_{21,2},\bar r_1;R_1)\,\chi_{r_1}(\T_1)\,\chi_{\bar r_1}(\bar \T_1)\nn
&\qquad\qquad\qquad\times
g(r_{21,1},r_{21,2}, r_2;R_2)\,g(r_{12,1},r_{12,2},\bar r_2;R_2)\,\chi_{r_2}(\T_2)\,\chi_{\bar r_2}(\bar \T_2)\,,
\end{align}
where $m_1=n_{12,1}+n_{12,2}+n_1=n_{21,1}+n_{21,2}+\bar n_1$ and $m_2=n_{21,1}+n_{21,2}+n_2=n_{12,1}+n_{12,2}+\bar n_2$. As in the previous example, using \eqref{counting f} we get
\begin{align}
\N&(n_{12,1},\,n_{12,2},\,n_{21,1},\,n_{21,2},\{n_{1,j}\},\{n_{2,j}\},\{\bar n_{1,k}\},\{\bar n_{2,k}\})=
\sum_{R_1\vdash m_1\atop l(R_1)\leq N_1}\,\sum_{R_2\vdash m_2\atop l(R_2)\leq N_2}\,
\sum_{r_{12,1}\vdash n_{12,1}\atop r_{12,2}\vdash n_{12,2}}\,\sum_{r_{21,1}\vdash n_{21,1}\atop r_{21,2}\vdash n_{21,2}}
\nn
&\times\, g(r_{12,1},r_{12,2},[n_{1,1}],[n_{1,2}],\cdots,[n_{1,F_1}];R_1)\,g(r_{21,1},r_{21,2},[\bar n_{1,1}],[\bar n_{1,2}],\cdots,[\bar n_{1,\bar F_1}];R_1)\nn
&\quad\times
g(r_{21,1},r_{21,2},[n_{2,1}],[n_{2,2}],\cdots,[n_{2,F_2}];R_2)\,g(r_{12,1},r_{12,2},[\bar n_{2,1}],[\bar n_{2,2}],\cdots,[\bar n_{2,\bar F_2}];R_2)\,,
\end{align}
so that
\begin{align}
\Z(&x_{12,1},\,x_{12,2},\,x_{21,1},\,x_{21,2},\{t_{1,j}\},\{t_{2,j}\},\{\bar t_{1,k}\},\{\bar t_{2,k}\})\nn
&=\sum_{n_{12,1}\atop n_{12,2}}\,\,\sum_{n_{21,1}\atop n_{21,2}}\,\sum_{\{n_{1,j}\}}\,\sum_{\{n_{2,j}\}}\,\sum_{\{\bar n_{1,k}\}}\,\sum_{\{\bar n_{2,k}\}}\, \N(n_{12,1},\,n_{12,2},\,n_{21,1},\,n_{21,2},\{n_{1,j}\},\{n_{2,j}\},\{\bar n_{1,k}\},\{\bar n_{2,k}\})\nn
&\qquad\qquad\times\, x_{12,1}^{n_{12,1}}\,x_{12,2}^{n_{12,2}}\,x_{21,1}^{n_{21,1}}\,x_{21,2}^{n_{21,2}}\,
\left(\prod_{j=1}^{F_1} t_{1,j}^{n_{1,j}}\right)\,\left(\prod_{j=1}^{F_2} t_{2,j}^{n_{2,j}}\right)\,\left(\prod_{k=1}^{\bar F_1} \bar t_{1,k}^{\,\,\bar n_{1,k}}\right)\,\left(\prod_{k=1}^{\bar F_2} \bar t_{2,k}^{\,\,\bar n_{2,k}}\right)\,.
\end{align}

All of the previous formulae hold for any \(N\). In the next section we will drop the \(l(R_a)\leq N_a\) constraints, \(\forall a\), to focus on the large \(N\) case.

\subsection{The generating function $\mathcal Z$ and the building block $F^{[n]}$  }\label{The Generating Function}

Let us take the large \(N\) limit, for all the gauge groups of the theory. 
In appendix \ref{Derivation of the Gen Fun} we show that $\Z(\{x_{ab,\alpha}\},\{\T_{a,\beta}\},\{\bar \T_{a,\gamma}\})$ can be written as the multiple sum
\begin{align}\label{N p t}
\Z=&\sum_{\pmb p}\,\prod_i\,
\prod_{a}\,
\left(\prod_{b,\alpha}\frac{x_{ab,\alpha}^{n_{ab,\alpha}}}{p_{ab,\alpha}^{(i)}!}\right)\,\left(\prod_\beta\frac{ ( \Tr \, \T_{a,\beta}\,^i)^{\,p_{a,\beta}^{(i)}}}{p_{a,\beta}^{(i)}!}\right)\,\left(\prod_\gamma\frac{ ( \Tr \,  \bar\T_{a,\gamma}\,^i)^{\,\bar p_{a,\gamma}^{(i)}}}{\bar p_{a,\gamma}^{(i)}!}\right)
\nn
&\times
\cfrac{\left(\sum_{b,\alpha}p_{ab,\alpha}^{(i)}+\sum_{\beta} p_{ a,\beta}^{(i)}
\right)!}{i^{\sum_\beta p_{a,\beta}^{(i)}}}
 \,\,
\delta_{a}\left(
\sum_{b,\alpha}(p_{ab,\alpha}^{(i)}-p_{ba,\alpha}^{(i)})+\sum_{\beta}p_{a,\beta}^{(i)}-\sum_{\gamma}\bar p_{ a,\gamma}^{(i)}
\right)
\,,
\end{align}
where \(\pmb p=\cup_{ab,\alpha}\vec p_{ab,\alpha}\cup_{a,\beta}\vec p_{a,\beta}\cup_{a,\gamma}\vec{\bar p}_{a,\gamma}\), and the vectors \(\vec p_{ab,\alpha},\,\vec p_{a,\beta},\,\vec{\bar p}_{a,\gamma}\) are defined in \eqref{vec p}.

%
%
Crucially, we can now define the quantity
\begin{align}\label{first def of F in text}
F^{[n]}(\{x_{ab}\},\{t_a\},\{\bar t_a\}\})=&
\sum_{\vec p}
\prod_{a=1}^n
\left(\sum_{b=1}^np_{ba}+\bar p_{ a}
\right)!\,\,
\delta_{a}\left(
\sum_{b=1}^n(p_{ab}-p_{ba})+p_{a}-\bar p_{ a}
\right)\nn
&\qquad\qquad\times
\left(\prod_{b=1}^n\frac{x_{ab}^{ p_{ab}}}{p_{ab}!}\right)\,
\left(\frac{t_a^{p_{a}}}{p_{a}!}\right)\,\left(\frac{\bar t_a^{\,\bar p_{a}}}{\bar p_{a}!}\right)
\,,
\end{align}
with \(\vec p=\cup_{a,b}\{p_{ab}\}\cup_a\{p_{a},\,\bar p_{a}\}\), such that 
\begin{align}\label{PExp like relation}
{\mathcal{Z}}(&\{x_{ab,\alpha}\},\{\T_{a,\beta}\},\{\bar \T_{a,\gamma}\})\nn
&=\prod_i F^{[n]}\left(\left\{x_{ab}\rightarrow \sum_{\alpha}x_{ab,\alpha}^i\right\},\left\{t_a\rightarrow\sum_{\beta}\frac{ \Tr(\T_{a,\beta}^{\,i})}{i}\right\},\left\{\bar t_a\rightarrow\sum_{\gamma}\Tr(\bar \T_{a,\gamma}^{\,i})\right\}\right)\,\,.
\end{align}
From this equation we see that $ F^{[n]} $ is the building block of \(\Z\). Note that the  \(t\) coefficients in the RHS of \eqref{PExp like relation} are weighted by a \(i^{-1}\) coefficient, while the \(\bar t\) coefficients are not: in section \ref{full gen fun sec} we will derive a more symmetric version of this formula, where the weighting for chemical potentials of the quark and antiquark field is the same.

In appendix \ref{sec: contour int in app} we derive an expression for \(F^{[n]}\) in terms of contour integrals, namely
\begin{align}\label{F}
 F^{[n]}(\{x_{ab}\},\{t_a\},\{\bar t_a\})=
\left(\prod_{a=1}^n\oint_{\mathcal{C}_{a}}\frac{dz_{a}}{2\pi i}\right)
\prod_{a=1}^n\,
 I_a(\vec z;\vec x_{a},t_a,\bar t_a)\,,
\end{align}
in which
\begin{align}
 \vec{z}=(z_1,z_2,...,z_n)\,,\qquad \vec{x}_a=(x_{1a},x_{2a},...,x_{na})\,,
\end{align}
and
\begin{align}\label{I in  hat F}
I_a(\vec z;\vec x_{a},t_a,\bar t_a)=\cfrac{\exp\left(z_{a}\,t_{a}\right)}
{ z_{a}-\left(\bar t_{a}+\sum_{b}z_{b}\,x_{ba}\right)}\,.
\end{align}
We also obtained a pole prescription for the computation of these contour integrals: in the appendices \ref{computing sums} and \ref{r&c} we explain that only the $z_{a}$ pole coming from the $I _{a}$ term in the integrand has to be enclosed by $\C_{a}$.

As a last remark, note that all the variables $z_a,\,x_{ab},\,t_{a},\,\bar t_{a}$ in eq. \eqref{F} are charged under the $\prod_{a=1}^n U(1)_a\subset\prod_{a=1}^n U(N_a)$ subgroup of the theory as follows:
\begin{center}
\begin{table}[H]
\centering
\begin{tabular}{|c|c|c|} 
\hline
Variable & Charge& Subgroup of $\prod_a U(1)_a$ \\[2mm] \hline
$x_{ab} $&$(-1,1)$ &$ U(1)_a\times U(1)_b$\\ \hline
$t_{a} $&$-1$ &$ U(1)_a$\\ \hline
$\bar t_{a} $&$1$ &$ U(1)_a$\\ \hline
$z_a $&$1$ &$ U(1)_a$\\ \hline
\end{tabular}
\caption{$U(1)$ charges of the variables appearing in $F^{[n]}$.}
\label{charges in F}
\end{table}
\end{center}
The charge for the $x_{ab}$ coefficients comes from the fact that these variables are associated to fields leaving node $a$ and joining node $b$, thus transforming under $(\bar N_a,N_b)$ in the original theory. Similar comments holds for the charges of $t_{a}$ and $\bar t_{a}$, while the charge for $z_a$ has been chosen in such a way that the function $F^{[n]}$ is neutral under $\prod_a U(1)_a$, as it should be.

For completeness, let us write down the contour integral formulation for \(\Z\), which can be immediately obtained from \eqref{first def of F in text} by means of \eqref{PExp like relation}, and reads
\begin{align}\label{midstep4}
{\mathcal{Z}}=
&\prod_i\left(\prod_{a}
\oint_{\mathcal{C}_{a,i}}\frac{dz_{a,i}}{2\pi iz_{a,i}}\right)\prod_a\,
\cfrac{\exp\left(\cfrac{z_{a,i}\sum_\beta \Tr(\T_{a,\beta}\,^i)}{i}\right)}
{1-z_{a,i}^{-1}\left(\sum_{\gamma}\Tr(\bar \T_{a,\gamma}\,^i)+\sum_{b,\alpha}z_{b,i}\,x_{ba,\alpha}^i\right)}\,.
\end{align}
The simplification coming from using \(F^{[n]}\) in place of the latter is evident.
%
%
%
%
%


\section{The unflavoured case: contour integrals and paths on graphs}\label{The Matter-Free Case}

We now have to calculate the contour integral in $F^{[n]}$, that is, calculate residues. In an $n$-node quiver, each $z_a$ variable has $n$ poles, but not all of them have to be included in the contour $\C_a$. The constraints from the convergence of 
the sums in appendix \ref{sum over ab,alpha} instruct us on which poles to pick and which ones to discard. In appendix \ref{r&c} we show that they indeed give us a very simple and intuitive prescription: \emph{for all $a$, only the $z_a$ pole coming from the $I_a$ integrand has to be enclosed by $\C_a$}. 

We consider here the case in which we set $t_a=\bar t_a=0$ $\forall \,a$ in \eqref{F}, to get the quantity
\begin{align}\label{F0}
F^{[n]}(\{x_{ab}\},0,0)\equiv F_0^{[n]}(\{x_{ab}\})=
\left(\prod_{a}\oint_{\mathcal{C}_{a}}\frac{dz_{a}}{2\pi i}\right)
\prod_a\,
I_a(\vec z;\vec x_{a})\,,
\end{align}
where
\begin{align}\label{I in F0}
I_a(\vec z;\vec x_{a})=\cfrac{1}
{z_{a}-\sum_{b}z_{b}x_{b,a}}\,.
\end{align}
Recall that $ I_a(\vec z;\vec x_{a}) $ is a shorthand, which it will now be convenient to expand: 
\begin{align}
I_a(\vec z;\vec x_{a})=I_a(z_1,z_2,...,z_n;\vec x_{a})\,,
\end{align}
so that we can rewrite \eqref{F0} as
\begin{align}
\label{F0 notation expanded}
F_0^{[n]}(\{x_{ab}\})&=
\left(\prod_{a}\oint_{\mathcal{C}_{a}}\frac{dz_{a}}{2\pi i}\right)
\prod_a\,
I_a(z_1,z_2,...,z_n;\vec x_{a})
\,.
\end{align}
We want to compute contour integrals in eq. \eqref{F0 notation expanded}. Let us choose an ordering in which to compute such integrals: we choose the simplest one, that is we integrate over \(z_1,z_2,...,z_i,z_{i+1},...,z_n\) in this precise order. We will refer to this ordering as the `natural ordering'. With the pole prescription discussed in appendix \ref{r&c}, the $z_a$ integration picks up the $z_a$ pole in the $I_a$ integrand only. Then, after the first integral (the $z_1$ integral with our ordering choice) has been computed, eq. \eqref{F0 notation expanded} becomes
\begin{align}
F_0^{[n]}(\{x_{ab}\})&=
H_1(\vec x)
\left(\prod_{a>1}\oint_{\mathcal{C}_{a}}\frac{dz_{a}}{2\pi i}\right)
\prod_{a>1}\,
I_a(z_1^*,z_2,...,z_n;\vec x_{a})\,,
\end{align}
where we introduced the $H_1$ coefficient, outcome of the residue calculation, that depends only on the $x$ variables. After the integration has been done, $z_1$ is replaced by its pole equation 
\begin{align}
z_1^*=z_1^*(z_2,z_3,...,z_n;\vec x)
\end{align}
in all of the remaining integrands $I_a$ ($a>1$). The explicit form 
\bea 
z_1^* ( z_2 , z_3 , \cdots z_n ; \vec x ) =  { 1  \over ( 1 - x_{1,1}) } \, { \sum_{ b=2}^{n} z_b x_{b,1}  }
\eea
comes from solving $I_1^{-1} (\vec z;\vec x_1)=0$ for $z_1$. In the second step, we can solve $ I_2^{-1} (z_1 \rightarrow z_1^* ,  z_2 , z_3 \cdots , z_n ) = 0 $, which gives 
\bea 
z_2^* = { 1 \over {(1-x_{1,1})(1-x_{2,2})}-x_{1,2}x_{2,1} }\,  \sum_{ b  =3}^n \left ( x_{b,2} + { x_{b,1} x_{1,2} \over{1-x_{1,1}}  } \right ) \,.
\eea 
In the next step, we calculate $ I_3^{-1} ( z_1 \rightarrow z_1^* , z_2 \rightarrow z_2^* , z_3 , z_4 \cdots , z_n ) $ and we solve $I_3^{-1}=0$ to calculate $z_3^* ( z_4 , z_5 , \cdots z_n ) $. 

Generally, the explicit equation for each of the $z_j^*$ ($1\leq j\leq n$) comes from solving for $z_j$ the equation
\begin{align}\label{def pole j}
&I_j^{-1}(z_1^*,z_2^*,...,z_{j-1}^*,z_{j},z_{j+1},...,z_n;\vec x_j)=0
\end{align}
for each $j$. These pole equations are of the form 
\begin{align}\label{allpoles}
z_j^*(z_{j+1},z_{j+2},...,z_n;\vec{x})=\sum_{i>j}z_i\,a_{i,j}\,,
\end{align}
for some coefficients $a_{i,j} $, which are functions of $ \vec x $. 
It is useful however to introduce a different equation for the poles $z_j^*$. Note that $z_j^*$ is a function of the set $\{z_{j+1},z_{j+2},...,z_n\}$. If $r$ integrations have already been done, then the $z_j^* $ pole equations, with \(j\leq r\), can be expressed in terms of the remaining set of $z_a$, that is $\{ z_{r+1} , z_{r+2} \cdots , z_n \} $. The variables $z_k$ ($j\leq k\leq r$) appearing in (\ref{allpoles}) can be substituted with their respective pole equations $z_k^*$.
We can thus write 
\begin{align}\label{origin of recursion hat a}
z_{j}^*&\left(z_{j+1}^*,z_{j+2}^*,...,z_r^*,z_{r+1},...,z_n;\vec{x}\right)\nn
&\qquad\qquad\qquad=\sum_{i>r}z_i\,a_{i,j}+\sum_{\lambda=j+1}^r\,z_{\lambda}^*\left(z_{\lambda+1}^*,z_{\lambda+2}^*,...,z_r^*,z_{r+1},...,z_n;\vec{x}\right)\,a_{\lambda,j}
\,.
\end{align}
Repeated substitutions to eliminate the variables $ z_{ k }^* $ in favour of $ z_{ k' }^* $, for $k  <   k' \leq r  $, will lead to an expression of the form 
\begin{align}\label{allpoles same z}
z_\lambda^{*[r]}=z_\lambda^*(z_{r+1},...,z_n;\vec{x})=\sum_{i>r}z_i\, \hat a_{i,\lambda}^{[r]}\,,\qquad\lambda\leq r\,,
\end{align}
for some new $\hat a^{[r]}$ coefficients, functions of $ \vec x $, that we call \emph{pole coefficients}. Inserting this equation in \eqref{origin of recursion hat a} gives a recursive relation for $\hat a_{i,j}^{[r]}$:
\begin{subequations}\label{hatted a}
\begin{align}\label{formula recursion}
&\hat a_{i,j}^{[r]}=a_{i,j}+\sum_{\lambda=j+1}^r \hat a_{i,\lambda}^{[r]}\,a_{\lambda,j}\,,\qquad\quad i>r\,,\qquad j\leq r\leq n-1\,.
\end{align} 
There is no $\hat a^{[n]}$ coefficient, as can be seen from \eqref{allpoles same z}. We will in fact observe that \(z^{*}_n=0\).

Comparing (\ref{allpoles}) and (\ref{allpoles same z}) gives 
\begin{align}\label{initial formula recursion}
&\hat a_{i,r}^{[r]}=a_{i,r}\,,\qquad i>r \,,
\end{align}
\end{subequations}
and we will shortly derive 
\begin{align}\label{def of a}
a_{i,r}=\frac{\displaystyle x_{i,r}+\sum_{k=1}^{r-1}\hat a_{i,k}^{[r-1]}\,x_{k,r}}{\displaystyle 1-\left(x_{r,r}+\sum_{k=1}^{r-1}\hat a_{r,k}^{[r-1]}\,x_{k,r}\right)}\,,\qquad i>r\,.
\end{align}
Now, for fixed $r$, all of the $z_j^{*[r]}$ equations ($1\leq j\leq r$) in \eqref{allpoles same z} will be functions of the same set of $z_a$, that is $\{z_k,\,r<k\leq n\}$. With this notation, after $r$ integrations have been done, $F_0^{[n]}$ will read
\begin{align}\label{F mid}
F_0^{[n]}&=\prod_{j=1}^rH_j(\vec x)\,\left(\prod_{a>r}^n
\oint_{\mathcal{C}_{a}}\frac{dz_{a}}{2\pi i}\right) \,
\prod_{a>r}\,
I_a\left(z_1^{*[r]},z_2^{*[r]},...,z_{r}^{*[r]},z_{r+1},z_{r+2},...,z_n;\vec x_{a}\right)
\,,
\end{align}
where explicitly
\begin{align}
I_a\left(z_1^{*[r]},z_2^{*[r]},...,z_{r}^{*[r]},z_{r+1},z_{r+2},...,z_n;\vec x_{a}\right)=\cfrac{1}
{z_a-\left(\sum\limits_{b>r}z_b\,x_{b,a}+\sum\limits_{i=1,..,r}z_i^{*[r]}\,x_{i,a}\right)}\,.
\end{align}
Going back to eq. \eqref{F mid}, suppose we want now to calculate the $z_{r+1}$ integral. Consider then the equation
\begin{align}
I_{r+1}^{-1}&\left(z_1^{*[r]},z_2^{*[r]},...,z_{r}^{*[r]},z_{r+1},z_{r+2},...,z_n;\vec x_{r+1}\right)\nn
&\qquad\qquad\qquad\qquad\qquad=
z_{r+1}-\left(\sum\limits_{b>r}z_b\,x_{b,r+1}+\sum\limits_{i=1}^r z_i^{*[r]}\,x_{i,r+1}
\right)
=0\,,
\end{align}
and let us solve it for $z_{r+1}$. We have
\begin{align}
(1-x_{r+1,r+1})z_{r+1}&=\sum_{b>{r+1}}z_b\,x_{b,{r+1}}+\sum_{i=1}^r z_i^{*[r]}\,x_{i,{r+1}}\nonumber\\[2mm]
&=\sum_{b>{r+1}}z_b\,x_{b,{r+1}}+\sum_{i=1}^r \sum_{j>r}\,z_j\,\hat a_{j,i}^{[r]}\,x_{i,{r+1}}\nonumber\\[2mm]
&=\sum_{j>{r+1}}z_j\left(x_{j,{r+1}}+\sum_{i=1}^r \hat a_{j,i}^{[r]}\,x_{i,{r+1}}\right) 
+\sum_{i=1}^r z_{r+1}\,\hat a_{r+1,i}^{[r]}\,x_{i,{r+1}}\,.
\end{align}
Collecting terms we get
\begin{align}\label{kth pole}
&\left(1-\left(x_{{r+1},{r+1}}+\sum_{i=1}^r\hat a_{r+1,i}^{[r]}\,x_{i,{r+1}}\right)\right)z_{r+1}=\nn
&\qquad\qquad\qquad =\sum_{j>{r+1}}z_j \left(x_{j,{r+1}}+\sum_{i=1}^r \hat a_{j,i}^{[r]}\,x_{i,{r+1}}\right)\,,
\end{align}
so that we can finally write
\begin{align}
z_{r+1}^*=\sum_{j>{r+1}}z_j\,\frac{\displaystyle x_{j,{r+1}}+\sum_{i=1}^r \hat a_{j,i}^{[r]}\,x_{i,{r+1}} }
{\displaystyle 1-\left(x_{{r+1},{r+1}}+\sum_{i=1}^r \hat a_{r+1,i}^{[r]}\,x_{i,{r+1}}\right)}
=\sum_{j>{r+1}} z_j\,a_{j,r+1}\,.
\end{align}
Recalling the definition of the pole coefficients $ \hat a^{ [r]}_{i , \lambda  } $ from \eqref{allpoles same z}
and substituting $ r \rightarrow r-1$, this proves eq. \eqref{def of a}. It also shows that \(z_n^*=0\), as there is no \(z_{j}\) with \(j>n\) to sum over.

Inserting this result in \eqref{F mid} we get
\begin{align}\label{F mid r+1}
F_0^{[n]}&=\prod_{j=1}^rH_j(\vec x)\,\left(\prod_{a>r}^n
\oint_{\mathcal{C}_{a}}\frac{dz_{a}}{2\pi i}\right) \,
\nn
&\times
\frac{1}{\displaystyle \left(1-\left(x_{{r+1},{r+1}}+\sum_{i=1}^r \hat a_{r+1,i}^{[r]}\,x_{i,{r+1}}\right)\right)z_{r+1}-
\sum_{j\neq 1,..,r,{r+1}}z_j\left(x_{j,{r+1}}+\sum_{i=1}^r \hat a_{j,i}^{[r]}\,x_{i,{r+1}}\right)
}\nn
&\times\prod_{a>r+1}\,
I_a(z_1^{*[r]},z_2^{*[r]},...,z_{r}^{*[r]},z_{r+1},z_{r+2},...,z_n;\vec x_{a})\nn
&=\prod_{j=1}^rH_j(\vec x)\,\frac{1}{\displaystyle 1-\left(x_{{r+1},{r+1}}+\sum_{i=1}^r \hat a_{r+1,i}^{[r]}\,x_{i,{r+1}}\right)}\,
\oint_{\mathcal{C}_{r+1}}\frac{dz_{r+1}}{2\pi i} \,
\frac{1}{z_{r+1}-z_{r+1}^*
}\nn
&\times\left(\prod_{a>r+1}^n\oint_{\mathcal{C}_{a}}\frac{dz_{a}}{2\pi i}\right) \,
\,\prod_{a>r+1}\,
I_a(z_1^{*[r]},z_2^{*[r]},...,z_{r}^{*[r]},z_{r+1},z_{r+2},...,z_n;\vec x_{a})\nn
%
%
%
%
%
%
&\equiv\prod_{j=1}^{r+1}H_j(\vec x)\,
\left(\prod_{a>r+1}^n\oint_{\mathcal{C}_{a}}\frac{dz_{a}}{2\pi i}\right) \,
\,\prod_{a>r+1}\,
I_a(z_1^{*[r+1]},z_2^{*[r+1]},...,z_{r}^{*[r+1]},z_{r+1}^{*[r+1]},z_{r+2},...,z_n;\vec x_{a})\,,
\end{align}
where we called, in agreement with our initial definitions, 
\begin{align}\label{H r+1}
H_{r+1}(\vec x)=\left[ 1-\left(x_{{r+1},{r+1}}+\sum_{i=1}^r \hat a_{r+1,i}^{[r]}\,x_{i,{r+1}}\right)\right]^{-1}\,.
\end{align}
It is clear now that once all the integration have been done, $F_0^{[n]}$ will simply be the product
\begin{align}\label{final}
F_0^{[n]}=
\prod_{i=1}^nH_i(\vec x)
=\prod_{i=1}^{n}\left(1-x_{{{i}},{{i}}}-\sum_{q=1}^{i-1} \hat a_{{i},q}^{[i-1]}\,x_{q,{{i}}}\right)^{-1}\,.
\end{align}
In appendix \ref{Three Nodes Matter-Free Example residue way} we present an explicit example of the application of these formulae to a three node unflavoured quiver. From the last equation we can see how the pole coefficients $\hat a_{{i},q}^{[i-1]}$ play a central role in the computation of $F_0^{[n]}$. Our goal now is to rewrite them in a more compact and appealing form. For notational purposes it is useful now to define $G_{[n]}$ as the inverse of $F_0^{[n]}$; $G_{[n]}=\left({F_0^{[n]}}\right)^{-1}$.

Choosing any $1\leq r<n$, for all $n\geq p>r$ and $1\leq k\leq r$ we find an expression which can be interpreted in terms of paths on the complete $n$-node quiver:
\begin{align}
G_{[r]}\,\hat a_{p,k}^{[r]}&=G_{[r]\setminus k}\,x_{p,k}+\sum_{i=1\atop i\neq k}^rG_{[r]\setminus \{k,i\}}x_{p,i}x_{i,k}+\sum^r_{i,j=1\atop i\neq j\neq k}G_{[r]\setminus \{k,i,j\} }x_{p,i}x_{i,j}x_{j,k}\nn
&
+...+\sum^r_{i_1,i_2,..,i_t=1\atop i_1\neq i_2 \neq...\neq i_t\neq k}G_{[r]\setminus \{k,\cup_{h=1}^t i_h\} }x_{p,i_1}x_{i_1,i_2}x_{i_2,i_3}\cdots x_{i_{t-1},i_t}x_{i_t,k}+...\nn
&
...+\sum^r_{i_1,i_2,..,i_{r-1}=1\atop i_1\neq i_2 \neq...\neq i_{r-1}\neq k}x_{p,i_1}x_{i_1,i_2}x_{i_2,i_3}\cdots x_{i_{r-2},i_{r-1}}x_{i_{r-1},k}\,,
\end{align}
or, in a more compact form:
\begin{align}\label{guess}
G_{[r]}\,\hat a_{p,k}^{[r]}=\sum_{t=0}^{r-1}\left(\sum^r_{i_1,i_2,..,i_t=1\atop i_1\neq i_2 \neq...\neq i_t\neq k}G_{[r]\setminus \{k,\cup_{h=1}^t i_h\} }x_{p,i_1}x_{i_1,i_2}x_{i_2,i_3}\cdots x_{i_{t-1},i_t}x_{i_t,k}\right)\,.
\end{align}
with the convention that $G_{[0]}=1$. We prove this formula in appendix \ref{attempt}. 
For fixed $r< n$ we now describe the interpretation of each of the terms in the expansion of \eqref{guess} as a path on the complete $n$-node quiver. Each term is a product of two different pieces. The first one is the $G$ function of a quiver containing a certain subset $[r]\setminus \{k,\cup_{h=1}^ti_h\}$ of the first $[r]=\{1,2,...,r\}$ nodes. 
The second one is a string of $x_{ab}$ variables, which can be interpreted as an oriented open line on the quiver. It departs from a node $p$, which is \emph{not} included in the set \([r]\), passes through some $t$ intermediate nodes $i_h$ and arrives at node $k$, with $i_1,i_2,...,i_t,k \in [r]$.
\begin{figure}[H]
\begin{center}\includegraphics[scale=1.3]{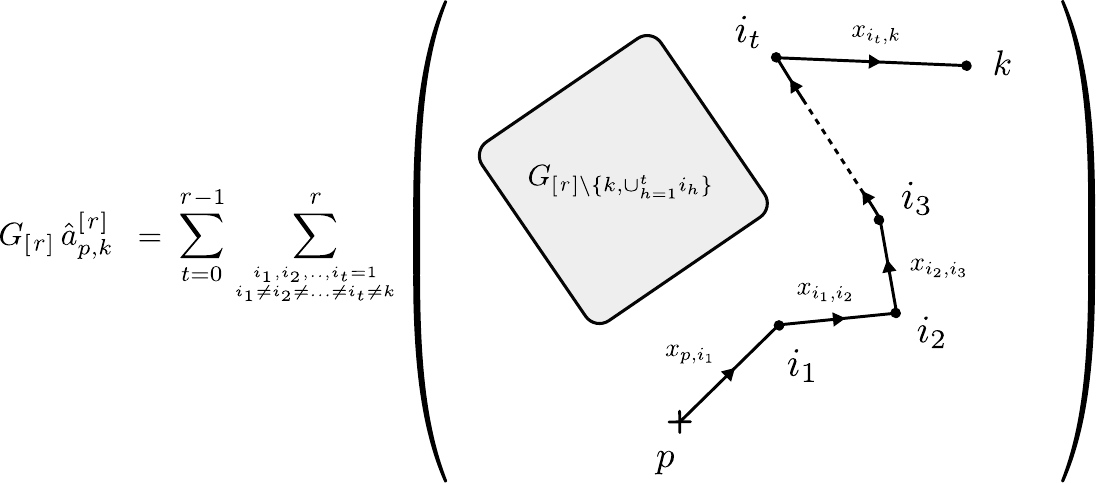}\\[1mm]
\caption{Pictorial interpretation of $G_{[r]}\,\hat a_{p,k}^{[r]}$. The starting point of the oriented open path, \(p\), belongs to the set \(\{r+1,r+2,...,n\}\).}
\label{poles coefficients pictorial interpretation}
\end{center}
\end{figure}
From here we also explicitly see that the pole coefficient $\hat a_{p,k}^{[r]}$ is charged under the $U(1)^n$ subgroup of the gauge group of the quiver. Since every $G_{[r]}$ has zero $U(1)^n$ charge, and the product of $x_{ab}$ coefficients $x_{p,i_1}x_{i_1,i_2}x_{i_2,i_3}\cdots x_{i_{t-1},i_t}x_{i_t,k}$ is charged under the $p$-th $U(1)$ and the $k$-th $U(1)$ as $(-1,1)$ respectively, the whole quantity $\hat a_{p,k}^{[r]}$ will carry a $(-1,1)$ charge under $ U(1)_p \times U(1)_k$, just like an $x_{p,k}$ variable would. These quantities are also helpful in writing down a recursive formula for $G_{[r]}$. Note that $G_{[r+1]}$ can be written as
\begin{align}
G_{[r+1]}&=G_{[r]}\left(1-x_{r+1,r+1}-\sum_{k=1}^{r}\hat a_{r+1,k}^{[r]}\,x_{k,r+1}\right)\nn
&=G_{[r]}\left(1-x_{r+1,r+1}\right)-\sum_{k=1}^{r}G_{[r]}\,\hat a_{r+1,k}^{[r]}\,x_{k,r+1}\,.\end{align}
The terms in the sum above are of the form \eqref{guess}, so that we can use it to bring $G_{[r+1]}$ into the form 
\begin{align}
G_{[r+1]}&
=G_{[r]}\left(1-x_{r+1,r+1}\right)\nn
&\qquad\qquad-\sum_{k=1}^{r}\sum_{t=0}^{r-1}\sum^r_{i_1,i_2,..,i_t=1\atop i_1\neq i_2 \neq...\neq i_t\neq k}G_{[r]\setminus \{k,\cup_{h=1}^t i_h\} }x_{r+1,i_1}x_{i_1,i_2}x_{i_2,i_3}\cdots x_{i_{t-1},i_t}x_{i_t,k}x_{k,r+1}\,,
\end{align}
and after relabelling some summation variables, we can write the this equation as
\begin{align}
G_{[r+1]}&=G_{[r]}-G_{[r]}x_{r+1,r+1}\nn
&\qquad\qquad-
\sum_{t=1}^{r}\sum^r_{i_1,i_2,..,i_t=1\atop  i_1\neq i_2 \neq...\neq i_t}G_{[r]\setminus \{\cup_{h=1}^t i_h\} }x_{r+1,i_1}x_{i_1,i_2}x_{i_2,i_3}\cdots x_{i_{t-1},i_t}x_{i_t,r+1}\,,
\end{align}
and since the second term on the RHS of this identity is just the $t=0$ term of the following sum, we finally have
\begin{align}\label{good G}
G_{[r+1]}&=G_{[r]}-\sum_{t=0}^{r}\sum^r_{i_1,i_2,..,i_t=1\atop  i_1\neq i_2 \neq...\neq i_t}G_{[r]\setminus \{\cup_{h=1}^t i_h\} }x_{r+1,i_1}x_{i_1,i_2}x_{i_2,i_3}\cdots x_{i_{t-1},i_t}x_{i_t,r+1}\,.
\end{align}
We can also give a similar formula for \emph{each} of the $H_l$ coefficients in the product \eqref{final}. We know that
\begin{align}\label{jnoj}
H_l(\vec x)=\left(1-x_{{{l}},{{l}}}-\sum_{q=1}^{l-1} \hat a_{{l},q}^{[l-1]}\,x_{q,{{l}}}\right)^{-1}\,,
\end{align}
and using $F^{[n]}_0=G_{[n]}^{-1}$ we can write
\begin{align}
H_l(\vec x)&=\left(1-F^{[l-1]}_0G_{[l-1]}x_{{l},{l}}-F^{[l-1]}_0\sum_{q=1}^{l-1} G_{[l-1]}\,\hat a_{l,q}^{[l-1]}\,x_{q,{l}}\right)^{-1}\nn
&=\left(1-F^{[l-1]}_0\left(G_{[l-1]}x_{{l},{l}}+\sum_{q=1}^{l-1} G_{[l-1]}\,\hat a_{l,q}^{[l-1]}\,x_{q,{l}}\right)\right)^{-1}\,.
\end{align}
We again get have terms like $G_{[l-1]}\,\hat a_{l,q}^{[l-1]}\,x_{q,l}$, which have the same structure of the ones encountered in the derivation of eq. \eqref{good G}. We can just redo the same steps done previously to bring the equation for the $H_l(\vec x)$ coefficient into the form
\begin{align}\label{res form}
H_l(\vec x)&=\left(1-F^{[l-1]}_0\,\sum_{t=0}^{l-1}\sum^{l-1}_{i_1,i_2,..,i_t=1\atop  i_1\neq i_2 \neq...\neq i_t}G_{[l-1]\setminus \{\cup_{h=1}^t i_h\} }x_{l,i_1}x_{i_1,i_2}x_{i_2,i_3}\cdots x_{i_{t-1},i_t}x_{i_t,l}\right)^{-1}\nn
&=
F_0^{[l-1]}\,\left(G_{[l-1]}-\sum_{t=0}^{l-1}\sum^{l-1}_{i_1,i_2,..,i_t=1\atop  i_1\neq i_2 \neq...\neq i_t}G_{[l-1]\setminus \{\cup_{h=1}^t i_h\} }x_{l,i_1}x_{i_1,i_2}x_{i_2,i_3}\cdots x_{i_{t-1},i_t}x_{i_t,l}\right)^{-1}\nn
&=
\frac{G_{[l-1]}}{G_{[l]}}\,,
\end{align}
where in the last step we used eq. \eqref{good G}. We can then rewrite eq. \eqref{final} as
\begin{align}
F^{[n]}_0=\prod_{i=1}^nH_i(\vec{x})=\prod_{i=1}^n \frac{G_{[i-1]}}{G_{[i]}}\,,
\end{align}
with $G_{[0]}=1$.

\subsection{$F_0^{[n]}$ and the sum over subsets}\label{F-and-recursions}
In this section we will prove the expression for $\left(F_0^{[n]}\right)^{-1}$ given in \cite{quivcalc}
\begin{align}\label{sj formula}
\left( F_0^{[n]} \right)^{-1} =1+\sum_{\vv \subseteq V_n}\,\sum_{\sigma\in\text{Sym}(\vv)} (-1)^{ C_{\sigma } } y_\sigma(\{x_{ab}\})\,,
\end{align}
where $\vv$ is any subset of the set of nodes $V_n=\{1,2,...,n\}$ of the quiver but the empty set, and $\text{Sym}(\vv)$ is the group of all the permutations of elements in $\vv$. $C_{\sigma } $ is the number of cycles in $ \sigma $. 
$y_\sigma(\{x_{ab}\})$ is a monomial built from the $x_{ab}$ coefficients as
\begin{align}\label{cycle f}
y_\sigma(\{x_{ab}\})&=\prod_i y_{\sigma^{(i)}}(\{x_{ab}\})\,,
\end{align}
where the product runs over the cycles $\sigma^{(i)}$ of the permutation $\sigma=\prod_i\sigma^{(i)}$, and
for a single cycle $ ( i_1 , i_2 , \cdots , i_k )$
\begin{align}\label{single cycle perm}
y_{(i_1,i_2,...i_k)}(\{x_{ab}\})=x_{i_1,i_2}x_{i_2,i_3 }\cdots x_{i_k ,i_1}\,.
\end{align}
For example, when $ \sigma = ( 1 2 ) ( 3) $, the permutation which swaps $1$ and $2$ and leaves $3$ fixed, then $y_{(12)(3)}(\{x_{ab}\})=x_{12}x_{21}x_{33}$. This equation has thus an interpretation in terms of \emph{loops} $\{y_c\}$ on a complete quiver, where each loop \(y_c\) corresponds to a cycle $c=(i_1 , \cdots , i_k )$ as in \eqref{single cycle perm}. Since these loops corresponds to cyclic permutations, they do not visit the same node more than once: for this reason we call them \emph{simple loops}, to distinguish them from more general closed paths.
In the following we will write the above formula as $ \tilde G_{ [n]} $ : 
\bea 
\tilde G_{ [n]} = 1+\sum_{\vv \subseteq V_n}\,\sum_{\sigma\in\text{Sym}(\vv)} (-1)^{ C_{\sigma } } y_\sigma(\{x_{ab}\})\,.
\eea 
To prove the identity \eqref{sj formula} we will show that the sequence $\tilde G_{[n]}$
obeys the same recursion relation \eqref{good G} satisfied by the $G_{[n]}$ coefficients obtained from the residue computations. We have 
\bea 
\tilde G_{[n+1] } =  1+ \sum_{ \vv \subset \{ 1 , \cdots , n+1 \} } \sum_{ \sigma \in Sym ( \vv ) } (-1)^{ C_{\sigma } } y_{ \sigma }  \,.
\eea
If the subset $\vv$ of $ \{ 1 , \cdots , n+1 \}$ does not include $n+1$, we have a sum which, together with 
the leading $1$, gives $G_{[n]} $. The remaining terms involve subsets which include the $\{n+1\}$ node. 
For such subsets, the permutation $ \sigma $ can either be of the product form $ \sigma' (n+1)$, where $ \sigma' $ is a permutation of $ \{ 1 , \cdots , n \} $ and $(n+1)$ is a single cycle of length one, or alternatively it is of the form $ \sigma' ( i_1 , i_2 , \cdots , i_k , n+1  ) $, with $ \sigma'$ a permutation of $ \{ 1 , \cdots , n  \} \setminus \{ i_1 , \cdots , i_k \} $ and $(i_1 , \cdots , i_k , n+1)$ a cycle of length $k+1$. The first type of term gives 
\bea 
- \tilde G_{ [n]} y_{n+1} =  - \tilde G_{ [n]} x_{n+1 , n+1}  \,.
\eea
The second type of term gives 
\bea 
 - \sum_{ k=1}^{ n } \sum_{ i_1 \ne i_2 \ne \cdots \ne  i_k =1}^n   \tilde G_{ [n] \setminus \{ i_1 , \cdots , i_k \} } y_{ i_1 , \cdots , i_k , n+1} \,.
\eea
Collecting the terms we find 
\bea 
\tilde G_{[n+1]} = \tilde G_{[n]} ( 1 - x_{n+1 ,  n+1} )  - \sum_{ k=1}^n 
\sum_{ i_1 \ne i_2 \ne \cdots i_k =1}^{n}   \tilde G_{ [n] \setminus \{ i_1 , \cdots  , i_k \} } x_{i_1 i_2 } x_{i_2 i_3 } \cdots x_{i_{k} , n+1} x_{n+1 , i_1 } \,.
\eea
 This proves that the guessed formula $ \tilde G_{[n]} $ satisfies the same recursion 
relation as $G_{[n]} $. It is evident that $ G_{[1]} = \tilde G_{[1]} = 1 $. This proves that 
$ \tilde G_{[n]} = G_{ [n]} $, \(\forall\, n\).

\subsection{$F_0^{[n]}$ and determinants}\label{subsec:F0Dets} 

Equation \eqref{sj formula} 
can be used to recast $ F^{[n]}_0(\{x_{ab}\})$ as a determinant expression given by
\begin{align} 
 F^{[n]}_0(\{x_{ab}\}) = \frac{1}{\det\left(\mathbb 1_n- X_n\right)}\,,
\end{align}
where $\mathbb{1}_n$ is the $n$ dimensional identity matrix and $ X_n$ is a $n\times n$ matrix defined by
\begin{equation}\label{Lambda copy}
\left. X_n\right|_{ij}=
x_{ij}\,,\qquad
1\leq(i,j)\leq n\,.
\end{equation}
The following identity for the expansion of \(\det(\mathbb 1_n-X_n)\) in terms of sub-determinants of \(X_n\), 
or equivalently characters of \(X_n\) associated with single-column Young diagrams, is useful: 
\begin{align} 
 \det\left(\mathbb 1_n- X_n\right)& = \sum_{ k=0}^{n } ( -1)^k \chi_{ [1^k] } ( X_n ) \cr 
& =
 \sum_{ k=0}^n  { ( -1)^k } \sum_{ i_1,i_2,..., i_k = 1 }^{ n } \, \sum_{ \sigma \in S_k } 
 \,\frac{(-1)^{ \sigma }}{k!} x_{i_1 i_{\sigma (1)} } x_{i_2 i_{\sigma(2)} } \cdots 
 x_{ i_k i_{\sigma(k )} }  \,.
\end{align}
This expansion is organized according to the number of $1$'s picked up from the matrix $  ( \mathbb 1_n- X_n)  $ in calculating its determinant. When we pick $n-k$ of the $1$'s, we have the sum of the sub-determinants constructed from blocks of size $k$ from the matrix $X_n$. When we pick $n-k$ of these $1$ valued entries, we have the sum of the sub-determinants constructed from blocks of size $k$ from the matrix $X_n$. 
The sign $ (-1)^{ \sigma } $ is the parity of the permutation. Because of the antisymmetrisation \(\sum_\sigma(-1)^\sigma\), the sum over $ i_1 ,i_2,... , i_k$ can be restricted to run over the set $ i_1 \neq i_2\neq... \neq i_k$, so that it can be rewritten as a sum over subsets $\vv_k$ of $k$ different integers from $ \{ 1 , \cdots , n \} $. 
For each choice of subset there is factor of $k! $ for ways of assigning $ i_1, \cdots , i_k$ to the elements of the subset. 
Hence 
\begin{align}
\det\left(\mathbb 1_n- X_n\right) = \sum_{ k=0}^n ( -1)^{k} \sum_{ \vv_k } \sum_{ \sigma \in Sym \left (  \vv_k \right ) } (-1)^{ k - C_{ \sigma } } y_{ \sigma } \,.
\end{align}
$ Sym ( \vv_k  )$ is the symmetric group of permutations of elements in $ \vv_k$. 
Here we have used the fact that the parity of a permutation $ \sigma $ can be written in terms of the number of cycles as $ (-1)^{ \sigma } = (-1)^{ k - C_{ \sigma } } $ and we also used the definition of $y_{\sigma}$. The expression \eqref{sj formula} now follows. 

\subsection{Word counting and the building block $F_0^{[n]} $}\label{subsec:WordCounting} 

The generating function $ \Z (\{ x_{ab ; \alpha }\} ) $ for gauge invariant operators for unflavoured quiver theories has been given as an infinite product built from a building block $ F_{0}^{[n]} (\{x_{ab}\} ) $. This has been expressed in terms of a determinant of the matrix $( \mId_n -  X_n)  $, where $(X_n)|_{ab}=x_{ab}$. 

After expanding $ F_{0}^{[n]} (\{x_{ab}\} ) $ in a power series in the variables 
$x_{ab}$, it is natural to ask if the coefficients in this series have a combinatoric interpretation as counting something. The answer does not immediately follow from 
the combinatoric interpretation of $ \Z ( \{x_{ab ; \alpha }\} ) $ in terms of gauge invariants, nevertheless, the coefficients in the expansion of $ F_{0}^{[n]} (\{x_{ab}\} ) $ are themselves positive. This follows from the Cauchy-Littlewood formula for the expansion of the inverse determinant:
\bea 
{ 1 \over \det ( \mId_n - X_n ) } = \sum_{ k=0}^{ \infty} { 1 \over k! } \sum_{ i_1 \cdots i_k =1 }^{n  } \sum_{ \sigma  \in S_k } x_{ i_1 , i_{ \sigma(1)} } x_{i_2 , i_{ \sigma(2)}} 
\cdots x_{ i_k , i_{\sigma (k) } } \,.
\eea
This strongly suggests that there should be a combinatoric interpretation in terms of properties of graphs. We will find that there are in fact two combinatoric interpretations: both in terms of word counting related to the quiver with one directed edge for every specified start and end-point. We will call the latter the {\it complete $n$-node quiver}. 
We will refer to these two as the {\it charge conserving open string word (COSW) counting} problem and the {\it closed string word (CSW) counting} problem. It turns out that the equivalence between these two word counting problems is a known mathematical result! This gives a new connection between word counting problems and gauge theory.

To motivate the CSW interpretation, let us take the simple case of $n=2$, for which we have
\bea 
F_{0}^{[2]} ( x_{11} , x_{12} , x_{21} , x_{22} ) = { 1 \over ( 1 - x_{11} - x_{22} - x_{12} x_{21} + x_{11} x_{22} )} \,.
\eea 
The denominator depends on variables 
\bea 
  y_{11} = x_{11} \,,\qquad\quad
  y_{12} = x_{12}x_{21} \,,\qquad\quad 
  y_{22} = x_{22} \,.
\eea
These variables are associated with closed loops in a graph with two nodes, and one edge 
for every pair of specified starting and end points. Let us first set $y_{12} = 0 $: we have 
\bea 
{ 1 \over ( 1 - y_{11} - y_{22} + y_{11} y_{22}  )} = { 1 \over ( 1- y_{11} ) } { 1 \over ( 1 - y_{22} ) }  \,.
\eea
Expanding in powers of $y_{11} , y_{22} $, we see 
\bea 
{ 1 \over ( 1 - y_{11} - y_{22} + y_{11} y_{22}  )} =\sum_{m_1 =0}^{ \infty } y_{11}^{m_1} \sum_{ m_2 =0}^{\infty } y_{22}^{ m_2 } \,.
\eea
We describe the CSW interpretation in this simple case. Take the letters $ \hat y_{11}  , \hat y_{22}$ and consider arbitrary strings of these, with the condition that 
\bea\label{comy11y22}  
\hat y_{11} \hat y_{22} = \hat y_{22} \hat y_{11} \,.
\eea
A general word is characterized by the number $m_{11} $ and $m_{22} $ of $ \hat y_{11} , \hat y_{22}$. 
With these numbers specified, the commutation relation can be used to write any such word as 
\bea 
( \hat y_{11})^{m_{11} }  ( \hat y_{22} )^{m_{22} } \,.
\eea
There is thus, precisely one word with content $ (m_{11} , m_{22} )$. Thus the coefficient of $y_{11}^{ m_{11} } y_{22}^{ m_{22} } $ is equal to the number of words in a language made from letters $ \hat y_{11} , \hat y_{22}$. The words are sequences of these letters, with the commutation relation (\ref{comy11y22}). Now set $ y_{11} =0 $
\begin{align}
 F_{ 0 }^{[2]} ( y_{11} = 0 , y_{12} , y_{22} ) &= { 1 \over ( 1 - y_{12} - y_{22} ) } = \sum_{ m = 0 }^{ \infty } ( y_{12} + y_{22}  )^{ m }  = \sum_{  m =0 }^{ \infty } \sum_{ m_{12} = 0 }^m 
{ m! \over m_{12} ! m_{22} ! }  y_{12}^{ m_{12} } y_{22}^{ m_{22} } \nn 
& = \sum_{ m_{22} = 0 }^{ \infty } \sum_{ m_{12} = 0 }^{ \infty } {  ( m_{12} + m_{22} ) ! \over m_{12} ! m_{22}! }  y_{12}^{ m_{12} } y_{22}^{ m_{22} }\,.
\end{align}
In this case, we can consider letters $ \hat y_{12} , \hat y_{22} $, without imposing the commutation condition. Then a general word with specified numbers $m_{12} , m_{22} $ is the number of sequences we can write with $m_{12} , m_{22} $ copies of $\hat y_{12} ,\hat  y_{22} $. Each word corresponds to one way of placing the $m_{12}$ objects of one kind and $m_{22}$ objects of another kind in $ m_{12} + m_{22}$ positions. This shows that the number of words is $ {  ( m_{12} + m_{22} ) ! \over m_{12} ! m_{22}! } $ in agreement with the coefficient above.


These simple examples illustrate a general interpretation of all the coefficients 
in the expansion of $ F_0^{[n]}$, in terms of the cycle variables $ y_{c}$. Consider the complete $n$-node quiver. To each simple closed loop $c$ on the graph, associate a variable $ \hat y_c $. 
If we label the nodes of the graph $ \{ 1 , \cdots , n \} $, every cyclic permutation of a subset of the nodes corresponds to a simple loop on the graph. These simple loops visit each node no more than once. To define the CSWs, we associate a letter to $\hat y_c$ to every simple loop. We impose the relation 
\bea\label{comrelycycp}  
\hat y_c \hat y_{c'} = \hat y_{c'}  \hat y_{c} 
\eea
for every pair of simple loops \(c,\,c'\) that have no node in common. The letters which do share a node
are treated as non-commuting, while the letters that do not share a node are treated as commutative. Then we consider strings containing $m_c$ copies of the letter $ \hat y_c$. A simple guess, based on the above examples, is that the coefficient of $ \prod_c y_c^{ m_c} $ in the expansion of $ F_0^{[n]} $ is exactly equal to the number of distinct words build from the letters $ \hat y_c$ with specified numbers $m_c$ for each letter. This word counting interpretation is called closed string word counting since the loops can be thought as closed strings made from open strings which are the edges extending between nodes. The validity of this interpretation will be explained by using its equivalence to an open string word counting. 

Appendix \ref{sec:FandCSW} gives more examples of direct checks of this connection between closed string word counting and the building block function $F_{0}^{[n]} $.

From the derivation of the generating function of gauge invariants we know that 
\begin{align}\label{counting F_0}
F^{[n]}_0(\{x_{ab}\})&=\frac{1}{\det(\mathbb 1_n-X_n)}=\sum_{\vec p\,}\,
\prod_{a=1}^n
\left(\sum_{b=1}^np_{ab}
\right)!
\left(\prod_{b=1}^n\frac{x_{ab}^{\,p_{ab}}}{p_{ab}!}\right)
\delta\left(
\sum_{b=1}^n(p_{ab}-p_{ba})
\right)
\,.
\end{align}
This gives another way to see that the coefficients in the expansion are positive, and in fact integers. 
Consider the coefficient of $\prod_{a,b}x_{ab}^{p_{ab}}$, which is 
\begin{align}\label{term in exp}
\prod_a\,
\frac{\left(\sum_{b=1}^np_{ab}
\right)!}{\prod_{b=1}^n p_{ab}!}\,
\delta\left(
\sum_{b=1}^n(p_{ab} - p_{ba})
\right)\,.
\end{align}

This leads directly to the open string word counting. Consider letters $ \hat x_{ab}$ 
corresponding to each directed edge, going from $ a $ to $b$ in the complete $n$-node quiver.
We will call these { \it open string bits}. Then consider words which are sequences of these letters. These words will be called open string words. We impose the commutation condition 
\bea\label{osw-crels}  
\hat x_{ab} \hat x_{a' b' } = \hat x_{a' b' } \hat x_{a b }
\eea
for $ a \ne a'$. So sequences which differ by such a swap are counted as the same word. 
Thus, string bits which have different starting points do not commute. Two different string bits with the same starting 
point do not commute. For each starting point $a$ the factor 
\begin{align}\label{term in exp copy}
\frac{\left(\sum_{b=1}^np_{ab}
\right)!}{\prod_{b=1}^n p_{ab}!}\,
\end{align}
counts the number of sequences containing $p_{ab}$ copies of $ \hat x_{ab}$. 
Defining 
\bea 
p_a = \sum_{ b } p_{ab} = \sum_{ b } p_{ba} \,,
\eea
an open string word will take the form 
\begin{align}\label{canORD}  
& w_o =  \hat x_{ 1 a_1 } \hat x_{1 a_2 } \cdots \hat x_{ 1 a_{p_1} } ~~~
  \hat x_{ 2 a_{ p_1 +1}  } \hat x_{2 a_{p_1+2 }  } \cdots \hat x_{ 2 a_{p_1 +p_2 } }
~~ \cdots ~~ \hat x_{ n a_{p_1 + \cdots + p_{n-1} +1 }}\cdots \hat x_{ n a_{ p_1 + \cdots + p_n } }\,.
\end{align}
The open string bits with different starting points commute, so we have used that commutativity to place all the ones starting at $1$ to the far left, the ones starting from $2$ next, and so on. The integers $a_1 , \cdots , a_{ \sum_i p_i } $ will contain $p_1 $ copies of $1$, $p_2$ copies of $2$ etc. This condition says that the sequence of open string bits that appear in the expansion of $F_0^{[n]} $ contains as many bits with starting point $i$ as with end points as $i$. We will refer to this as charge conserving open string words. So we have shown that the $F_0^{[n]} $ counts charge-conserving open string words. Remarkably, Cartier and Foata proved that charge-conserving open string words are in 1-1 correspondence with closed string words ! This is theorem 3.5 in Cartier-Foata \cite{CarFoa}.

We refer the reader to \cite{CarFoa} for the formal proof. Here we explain, with examples, the meaning of this equivalence between the counting of charge-conserving open string words (COSW) and closed string words (CSW). Given an a CSW, it is easy to write down a corresponding COSW. Take for example 
\bea 
\hat y_{11} \,\hat y_{12} \,\hat y_{11} \,\hat y_{22} \,\hat y_{123}  = \hat y_{11}\, \hat y_{12} \,\hat y_{22}\, \hat y_{11}\, \hat y_{123} \,.
\eea
Write these closed-string letters in terms of open string bits: 
\bea 
&&\hat y_{11} = \hat x_{11}\,,\quad\,\, \hat y_{22} = \hat x_{22}\,,\quad\,\,
\hat y_{12} = \hat x_{12} \hat x_{21}\,,\quad\,\, 
 \hat y_{123} = \hat x_{12} \hat x_{23} \hat x_{31} \,.
\eea
The word of interest becomes 
\bea 
\hat x_{11}  \hat x_{12} \hat x_{21} \hat x_{11} \hat x_{22} \hat x_{12} \hat x_{23} \hat x_{31} = 
\hat x_{11} \hat x_{12}\hat x_{11} \hat x_{12} ~~ \hat x_{21}\hat x_{22}  \hat x_{23} ~~ \hat x_{31} \,.
\eea
We have used the commutativity to arrange as in \eqref{canORD}. A CSW determines in this way a unique COSW. 

The reverse is also true. A COSW determines a unique CSW. The general proof is non-trivial \cite{CarFoa}. We just illustrate with some examples here. Consider some COSW with specified numbers of starting (and end-) points of particular types, say three starting and ending at $1$, two at $2$ and three at $3$. These words are of the form 
\bea 
\hat x_{ 1 , \tau (1 ) } \hat x_{1 , \tau (1) } \hat x_{ 1 , \tau (1 ) }  ~~~ 
\hat x_{2 , \tau ( 2 ) }  \hat x_{2 , \tau ( 2 ) } ~~~ \hat x_{3 , \tau ( 3 ) }  \hat x_{3 , \tau ( 3 ) } \hat x_{3  , \tau ( 3 ) } \,.
\eea
Here $ \tau $ is a permutation in $S_{8}$, which should be thought of as moving the integers $ \{ 1 , 2, 3 \} $ from their initial positions $ (1,1,1,2,2,3,3,3) $ to a new position. When $ \tau $ is the identity we have the COSW 
\bea 
\hat x_{ 11} \hat x_{11} \hat x_{11} ~~ \hat x_{22} \hat x_{22} ~~~ \hat x_{33} \hat x_{33} \hat x_{33} 
=  \hat y_{ 11} \hat y_{11} \hat y_{11} ~~ \hat y_{22} \hat y_{22} ~~~ \hat y_{33} \hat y_{33} \hat y_{33} \,.
\eea
Suppose now $ \tau = (1,2,3,4,5,6,7,8) $, a cyclic permutation. The COSW is 
\bea 
\hat x_{ 1 3 } \hat x_{11} \hat x_{11} ~~ \hat x_{21} \hat x_{22} ~~~ \hat x_{32} \hat x_{33} \hat x_{33} \,.
\eea 
If we map this to closed string words, this will involve two copies of $ \hat y_{11} $, two copies of $ \hat y_{33} $, and $ \hat y_{132} = \hat x_{13} \hat x_{32} \hat x_{21} $. The unique CSW is 
\bea 
\hat y_{1 3 2} ~~ \hat y_{11} \hat y_{11} ~~ \hat y_{22} ~~ \hat y_{33} \hat y_{33} \,.
\eea
In arriving at this, we did a re-arrangement which moves the $\hat x_{32} $ across the $ \hat x_{22}$. This is allowed, since the open string bits commute when they have different strating point. i.e. different first index. The reader is encouraged to play with different choices of $ \tau $. It is easy to see that permutations $ \tau $ in $S_8$ are a somewhat redundant way to parametrize the COSW. In fact it is a coset of $S_8$
by $ S_3 \times S_2 \times S_3$ that parametrizes the COSW. For any choice of $ \tau $, 
there is always a CSW, i.e a list of $\hat y_{ c} $ for different cycles, arranged in a specific order (modulo the commutation relations \eqref{comrelycycp}), which agrees with the COSW after re-arrangements allowed by the commutation \eqref{osw-crels}. This is guaranteed by theorem 3.5 of \cite{CarFoa}.

We have focused on the combinatoric interpretation of $ F_{0}^{[n]} ( \{x_{ab}\} ) $, in terms 
of the complete quiver graph. This basic building block generates the counting of gauge 
invariants at large $N$ for any quiver, after taking an infinite product with the substitutions in \eqref{Zresqc}. 
If we are interested in a quiver where there is no edge going from $a$ to $b$, these substitutions involve setting $ x_{ab} \rightarrow 0$ for that pair of nodes. It is instructive to consider the quantity 
\bea 
\mathcal{ F}_0^{[n]}  ( \{x_{ab ; \alpha } \}) = F_0^{[n]}  ( \{x_{ab} \rightarrow \sum_\alpha x_{ab, \alpha  }\} )  \,,
 \eea 
which is not an infinite product, but knows about the connectivity of any chosen quiver graph, with general multiplicities (possibly zero) between any specified start and end-node. 
This quantity has an interpretation in terms of word counting of open string words, as it follows immediately from \eqref {counting F_0}:
\begin{align}\label{counting mathcal F_0}
\mathcal F^{[n]}_0(\{x_{ab,\alpha}\})&
=\sum_{\vec p\,}\,
\prod_{a=1}^n
\left(\sum_{b=1}^n\sum_{\alpha=1}^{M_{ab}} p_{ab,\alpha}
\right)!
\left(\prod_{b=1}^n\,\prod_{\alpha=1}^{M_{ab}} \frac{x_{ab,\alpha}^{\,p_{ab,\alpha}}}{p_{ab,\alpha}!}\right)
\delta\left(
\sum_{b=1}^n\sum_{\alpha=1}^{M_{ab}}(p_{ab,\alpha}-p_{ba,\alpha})
\right)
\,.
\end{align}
We again have the basic rule that different open string letters corresponding to string bits with the same starting point do not commute. Again by invoking the Cartier-Foata theorem we see that, for any quiver, it is possible to map the open word counting problem to a closed word counting problem, in which string letters corresponding to simple loops which share a node do not commute.

The building block $F_0^{[n]} ( \{x_{ab}\} ) $ gives the counting of gauge invariants 
at large $N$, by means of a simple combinatoric operation involving an infinite product and elementary substitutions. One of our motivations for developing a combinatorial interpretation for $F_0^{[n]} ( \{x_{ab}\} ) $, is that it highlights an interesting analogy with a deformation of the 
counting problems considered here. We have focused on the counting of all holomorphic invariants made from chiral fields in an $\N=1$ theory. In many of the $\N=1$ theories of interest in AdS/CFT, the general holomorphic invariants form the chiral ring in the limit of zero superpotential, but beyond that, one wants to impose super-potential relations. In these cases, the counting of chiral gauge invariant operators leads to the $N$-fold symmetric product of the ring of functions on non-compact Calabi-Yau spaces \cite{Benvenuti:2006qr}.
In the large $N$ limit, the plethystic exponential gives the counting in terms of the counting at $N=1$. The $N=1$ counting is a simple building block of the large $N$ counting. It has a physical interpretation as the ring of functions on the CY and the plethystic exponential has an interpreation in terms of the bosonic statistics of many identical branes. 

The procedure of taking an infinite product and making substitutions, that we have developed 
for the $ N \rightarrow \infty$ counting at zero superpotential, can be viewed as an analog of
the plethystic exponential. In this analogy the function $ F_0^{ [n]} (\{ x_{ab}\} ) $ corresponds to the $U(1)$ counting, which is the same as counting holomorphic functions on a CY. 
The counting problems we have solved also correspond to some large $N$ geometries: namely 
the spaces of multiple matrices, subject to gauge invariance constraints.
There is no symmetric product structure in this geometry, but there is nevertheless a simple analog of the plethystic exponential. There is no physical interpretation of $ F_0^{[n]} ( \{x_{ab}\} ) $ as a gauge theory partition function, but there is nevertheless an interpretation in terms of string word counting partially commuting string letters. A deeper understanding and interpretation of these analogies will undoubtedly be fascinating.


\section{The flavoured case: from contour integrals to a determinant expression}\label{full gen fun sec}

We now turn to the full picture, that is we allow for quarks and antiquarks.
Take then eq. \eqref{F}:
\begin{align}\label{Fcopy}
F^{[n]}(\{x_{ab}\},\{ t_{a}\},\{\bar t_{a}\})=
\left(\prod_{a}\oint_{\mathcal{C}_{a}}\frac{dz_{a}}{2\pi i}\right)
\prod_a\,
I_a(x_{ab}, t_{a},\bar t_{a})\,,
\end{align}
where
\begin{align}\label{denwmat}
I_a(\vec z;\vec x_{a}, t_{a},\bar t_{a})=\cfrac{\exp\left(z_{a}t_{a}\right)}
{z_{a}-\left(\bar t_{a}+\sum\limits_{b=1}^nz_{b}\,x_{b,a}\right)}\,.
\end{align}
Again we have to compute residues. First of all note that the numerator of \eqref{denwmat} is regular in $z_a$, so that the only poles may come from its denominator. We can simplify the next steps by using a trick: let us rename $\bar t_a\equiv x_{0,a} $ and multiply it by a dummy variable, $z_0$. Pictorially, this would consist of taking all the open (fundamental matter) edges in the quiver and joining them to a fictitious node, that we call $`0$ node'. For consistency, let us also rename \(t_a \equiv x_{a,0}\)\,. Using this notation we can rewrite eq. \eqref{denwmat} as
\begin{align}\label{denwmat0}
I_a(\vec z;\vec x_{a},t_{a},\bar t_{a})=\cfrac{\exp\left(z_{a}\, x_{a,0}\right)}
{z_{a}-\left(z_0\,x_{0,a}+\sum\limits_{b=1}^nz_{b}\,x_{b,a}\right)}
=\cfrac{\exp\left(z_{a} \,x_{a,0}\right)}
{z_{a}-\left(\sum\limits_{b=0}^nz_{b}\,x_{b,a}\right)}\,.
\end{align}
where it is understood that $z_0$ will be set to $1$ after the $z_i$ ($1\leq i \leq n$) integrals have been done. 
This means that the intermediate expressions arising from successive integrations will take the same form as in the unflavoured case of 
Section \ref{The Matter-Free Case}. In particular the pole prescription still holds unaltered. 

With this formalism, eq. \eqref{allpoles same z} becomes
\begin{align}\label{allpoles same z w/ m}
z_j^{*[r]}=z_j^*(z_{r+1},...,z_n,z_0;\vec{x})=\sum_{i>r\atop \cup\{i=0\}}z_i\, \hat a_{i,j}^{[r]}\,,
\end{align}
and correspondingly eq. \eqref{kth pole} gets modified as
\begin{align}\label{kth pole w/m}
&\left(1-\left(x_{{r+1},{r+1}}+\sum_{i=1}^r\hat a_{r+1,i}^{[r]}\,x_{i,{r+1}}\right)\right)z_{r+1}=\nonumber\\[2mm]
&\qquad\qquad\qquad =\sum_{j>{r+1}\atop \cup\{j=0\}}z_j \left(x_{j,{r+1}}+\sum_{i=1}^r \hat a_{j,i}^{[r]}\,x_{i,{r+1}}\right)\,.
\end{align}
We can then proceed in the exact same fashion as in section \ref{The Matter-Free Case}. The only manifestly different piece in the integrand are the numerators of \eqref{denwmat0}. To highlight the similarity to the unflavoured case, we write 
\begin{align}
\tilde I_a(\vec z;\vec x_{a})=\frac{1}{\displaystyle z_{a}-\left(x_{0,a}+\sum\limits_{b=1}^nz_{b}\,x_{b,a}\right)}\equiv
\frac{1}{\displaystyle z_{a}-\sum\limits_{b=0}^nz_{b}\,x_{b,a}}\,.
\end{align}
such that
\begin{align}
I_a(\vec z;\vec x_{a},t_{a},\bar t_{a})=\cfrac{\exp(z_a t_a)}
{z_{a}-\left(\bar t_{a}+\sum\limits_{b=1}^nz_{b}\,x_{b,a}\right)}\equiv
\exp(z_a \,x_{a,0}) \,\tilde I_a(\vec z;\vec x_{a})\,.
\end{align}
For the flavoured case the equation corresponding to \eqref{F mid} would then be
\begin{align}\label{F mid w/m}
F^{[n]}&=\prod_{j=1}^rH_j(\vec x)\,\left(\prod_{a>r}^n
\oint_{\mathcal{C}_{a}}\frac{dz_{a}}{2\pi i}\right) \,
\left(\prod_{k=1}^r\exp\left(z_k^{*[r]}\, x_{k,0}\right)\right)\nn
&\qquad\qquad\qquad\times
\prod_{a>r}\,
\tilde I_a(z_1^{*[r]},z_2^{*[r]},...,z_{r}^{*[r]},z_{r+1},z_{r+2},...,z_n;\vec x_{a})\,
\exp(z_a x_{a,0})\,,
\end{align}
where in exact analogy with \eqref{H r+1}
\begin{align}
H_{j}(\vec x)=\left( 1-\left(x_{{j},{j}}+\sum_{i=1}^{j-1} \hat a_{j,i}^{[j-1]}\,x_{i,{j}}\right)\right)^{-1}\,.
\end{align}
Again, we see that the only addition in comparison to the unflavoured case is the product over the exponential functions. 
After the $n$ integrations have been done, using the definition in eq. \eqref{allpoles same z w/ m}, we have
\begin{align}\label{allpoles same z w/ m r=n}
z_k^{*[n]}=z_k^*(z_0;\vec{x}_k)=\sum_{i>n\atop \cup\{i=0\}}z_i\, \hat a_{i,k}^{[n]}\equiv
z_0\, \hat a_{0,k}^{[n]}\,.
\end{align} 
At this point we set $z_0=1$. Eq. \eqref{allpoles same z w/ m r=n} becomes
\begin{align}
z_k^{*[n]}=\hat a_{0,k}^{[n]}\,,
\end{align}
so that
\begin{align}\label{E_k formula 0free}
\prod_{k=1}^n\exp\left(z_k^{*[n]}\, x_{k,0}\right)=\prod_{k=1}^r\exp{\left(\hat a_{0,k}^{[n]}\, x_{k,0}\right)}\,.
\end{align}
We can then say that $F$ is the product
\begin{align}\label{mat}
F^{[n]}=
\prod_{j=1}^n H_j(\vec x)\,\exp\left(z_j^{*[n]}\, x_{j,0}\right)
=\prod_{j=1}^{n}\left(\frac
{\displaystyle
\exp{\left(\hat a_{0,j}^{[n]}\, x_{j,0}\right)}}
{\displaystyle 
1-x_{{j},{j}}-\sum_{i=1}^{j-1} \hat a_{j,i}^{[j-1]}\,x_{i,{j}}
}\right)\,,
\end{align}
where \(x_{p,0}=t_p\) and \(x_{0,p}=\bar t_{p}\). As expected, by setting all the fundamental matter field chemical potentials to zero we return to the unflavoured case.

In Appendix \ref{using guess in matter} we show that the numerator of this formula 
has the form 
\begin{align}
\exp \left ( \sum_{ j =1}^n \hat a_{0,j}^{[n]} t_j \right ) = 
\exp\left(\sum_{p,q=1}^n\,t_p\bar t_q\, \frac{(-1)^{p+q}\,M_{p,q}}{\det(\mathbb 1_n- X_n)}\right)\,,
\end{align}
where $M_{p,q}$ is the \((p,q)\) minor\footnote{We recall that the \((p,q)\) minor $M_{p,q}$ of a square matrix \(A\) is defined as the determinant of the matrix obtained from removing the \(p\)-th row and \(q\)-th column from \(A\).} of the matrix \((\mathbb{1}_n-X_n)\).
We can then write
\begin{align}\label{final2 mod minors}
F^{[n]}=\frac{1}{\det(\mathbb{1}_n-X_n)}\,\exp\left(\sum_{p,q=1}^n\, t_p\bar t_q\,\,\frac{(-1)^{p+q}\,M_{p,q}}{\det(\mathbb 1_n- X_n)} \right)\,.
\end{align}
A second expression for the same quantity was also given in Appendix \ref{using guess in matter}, and it reads
\begin{align}\label{final2}
 F^{[n]}=F^{[n]}_0\,\exp\left( t_p\bar t_q\, \partial^{p,q}\,\log F_0^{[n]}\vphantom{\sum}\right)\,,
\end{align}
where we used Einstein summation on $p,q$, and $\partial^{p,q}=\frac{\partial}{\partial x_{pq}}$.

Note that, as in the unflavoured case, we can write \(F^{[n]}\) as a determinant of a suitable matrix, which encodes all the information of the quiver under study. Since
\begin{align}
\frac{(-1)^{p+q}\,M_{p,q}}{\det(\mathbb 1_n- X_n)}=\left.\left( \mathbb 1_n- X_n\right)^{-1}\right|_{q,p}\,,
\end{align}
if we introduce the \(n\times n\) matrices \(\chi_n\) and \(\Lambda_n\), defined by
\begin{align}
\left.\chi_n\right|_{p,q}=\left.(\mathbb 1_n-X_n)^{-1}\right|_{p,q}\,,\qquad\left. \Lambda_n\right|_{p,q}= t_p\bar t_q\,,
\end{align}
then we can write
\begin{align}\label{hat final2 result in flavoured section}
 F^{[n]}(\{x_{ab}\},\{t_a\},\{\bar t_a\})=\det\chi_n\,
\exp\left[\Tr\left(\chi_n\,\Lambda_n\right)\right]\,.
\end{align}
Finally, the last equation can be put in the determinant form
\begin{align}\label{det mat fin}
F^{[n]}(\{x_{ab}\},\{t_a\},\{\bar t_a\})=\det\left(\chi_n\,\exp\left[\chi_n\,\Lambda_n\right]\right)\,.
\end{align}
The generating function $\mathcal Z$ is obtained from $F$ using eq. \eqref{PExp like relation}. However, from e.g. eq. \eqref{det mat fin} we see that $ t_a,\bar t_{b}$ always appear pairwise, so that we can rewrite \eqref{PExp like relation} in the more symmetric form already anticipated in eq. \eqref{PExp like relation hat improved copy}, that is
\begin{align}\label{PExp like relation hat improved}
{\mathcal{Z}}(&\{x_{ab,\alpha}\},\{\T_{a,\beta}\},\{\bar \T_{a,\gamma}\})\nn
&=\prod_i F^{[n]}\left(\left\{x_{ab}\rightarrow \sum_{\alpha}x_{ab,\alpha}^i\right\},\left\{t_{a}\rightarrow\sum_{\beta}\frac{ \Tr(\T_{a,\beta}^{\,i})}{\sqrt i}\right\},\left\{\bar t_a\rightarrow\sum_{\gamma}\frac{\Tr(\bar \T_{a,\gamma}^{\,i})}{\sqrt i}\right\}\right)\,\,.
\end{align}
This is the final expression for our large \(N\) generating function.


\section{Some examples}
We will now present some simple applications of our counting formulae, for the large \(N\) limit.

\subsection{One node quiver}
\begin{figure}[H]
\begin{center}\includegraphics[scale=1.1]{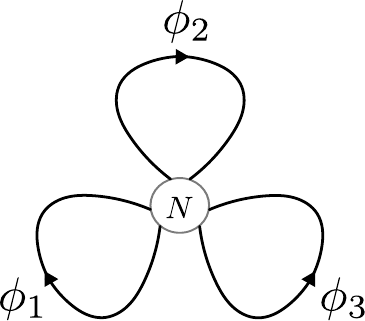}\caption{$d=4$, $\N=4$ SYM quiver}\label{N=4quiver} \end{center}
\end{figure}

Rewriting the chemical potentials of the fields as $x_{11}\rightarrow x,\,t_1\rightarrow t,\,\bar t_1\rightarrow \bar t$, we have
\begin{align}
\chi_1=\frac{1}{1-x}\,,\qquad\quad \Lambda_1=t\bar t\,,
\end{align}
so that
\begin{align}
F^{[1]}=\det(\chi_1\exp[\chi_1\Lambda_1])=\det(\chi_1)\,\exp\left[\Tr(\chi_1\Lambda_1)\right]&=
\cfrac{e^{\cfrac{t\bar t}{1-x}}}{1-x}\,.
\end{align}
The large $N$ generating function is then
\begin{align}
\mathcal Z(\{x_{\alpha}\},\{\T_{\beta}\},\{\bar \T_{\gamma}\})=\prod_i
\cfrac{
\exp\left(\cfrac{ {\sum_{\beta,\gamma} \Tr(\T_\beta^i)\,\Tr(\bar\T_\gamma^i)}}{{i(1-\sum_{\alpha}x_{\alpha}^i)}}\right)
}
{1-\sum_{\alpha}x_{\alpha}^i}\,.
\end{align}
For the $d=4$, $\N=4$ SYM theory with quiver shown in Figure \ref{N=4quiver}
\noindent the $\Z$ function is
\begin{align}
\mathcal Z_{SYM}(x_1,x_2,x_3)=\prod_i\frac{1}{1-x_1^i-x_2^i-x_3^i}\,.
\end{align}
For the SQCD model, described by the quiver
\begin{figure}[H]
\begin{center}\includegraphics[scale=1.1]{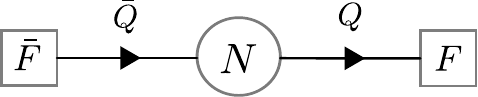}\caption{$d=4$ $\N=1$ SQCD quiver}\end{center}
\end{figure}
\noindent 
the generating function is instead
\begin{align}
\mathcal Z_{SQCD}(\T,\bar \T)=\prod_i\exp\left\{\frac{1}{i}\,\Tr(\T^i)\,\Tr(\bar T^i)\right\}=\prod_{j=1}^F\prod_{k=1}^{\bar F}\left(1-t_j\bar t_k\right)^{-1}\, , 
\end{align}
where we used
\begin{align}
 \T=\text{diag}(t_1,t_2,...,t_F)\,,\qquad\quad
 \bar\T=\text{diag}(\bar t_1,\bar t_2,...,\bar t_{\bar F})\,.
\end{align}
Note that if in the last example we do not distinguish the $U(1)\subset U(F)$ charges of the quarks and the $U(1)\subset U(\bar F)$ charges of the antiquarks, that is we set $\T=t\,\mathbb{1}_F$ and $\bar T=\bar t\,\mathbb{1}_{\bar F}$, we get
\begin{align}
\mathcal Z_{SQCD}(t\,\mathbb{1}_F,\bar t\,\mathbb{1}_{\bar F})=(1-t\bar t\,)^{-F\bar F}\,,
\end{align}
which was already derived in \cite{Gray:2008yu}, using different counting methods.

\begin{figure}[H]
\begin{center}\includegraphics[scale=1.1]{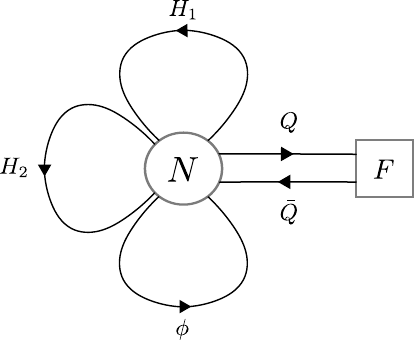}\caption{\(\N=2\) SQCD with and adjoint hypermultiplet.}\label{N=2 theory quiver}
\end{center}
\end{figure}

An interesting gauge theory can be obtained by adding  fundamental matter to \(\N=4\) SYM \cite{KarKat02,EF1012}. This operation breaks half of the supersymmetries leaving an \(\N=2\) theory, which in turn we can describe with the \(\N=1\) quiver \cite{Cremonesi:2014xha} in figure \ref{N=2 theory quiver}
The \(\N=2\) theory has a vector multiplet $\V$ (1 complex scalar $\phi$) and an hypermultiplet $\mathcal H$ (two complex scalars $H_1,\,H_2$) both in the adjoint of $U(N)$. A second hypermultiplet $\mathcal Q$ is in the bifundamental $U(N)\times U(F)$, where $U(F)$ is a global (non-dynamical) flavour symmetry (two complex scalars $Q,\,\bar Q$, transforming in opposite way under the symmetry group). The large $N$ generating function for this quiver, that we denote by $\mathcal Z_{\N=2}(x_1,x_2,x_3,\T,\bar \T)$, is given by
\begin{align}
\mathcal Z_{\N=2}(x_1,x_2,x_3,\T,\bar \T)=\prod_i \cfrac{\exp\left[\cfrac{\Tr(\T^i)\,\Tr(\bar\T^i)}{i\left(1-x_1^i-x_2^i-x_3^i\right)}\right]}{1-x_1^i-x_2^i-x_3^i}\,.
\end{align}
The first terms in the expansion of the unrefined $\mathcal Z_{\N=2}(x_1,x_2,x_3,t\,\mathbb 1_F,\bar t\,\mathbb 1_F)$ read
\begin{align}\label{1node exp}
\mathcal Z_{\N=2}(x_1,x_2,x_3,t\,\mathbb 1_F,\bar t\,&\mathbb 1_F)=1+x_1+x_2+x_3+F^2 t \bar t+2x_1x_2+2x_1x_3+2x_2x_3+2 F^2 t \bar tx_1\nn
&+2 F^2 t \bar tx_2+2 F^2 t \bar tx_3+6\,x_1x_2x_3+\frac{F^2}{2}\left(1+F^2\right) t^2 \bar t^2\\[3mm]
&+6F^2 t\bar t x_1x_2+6F^2 t\bar t x_1x_3+6F^2 t\bar t x_2x_3+\frac{F^2}{2}\left(1+3F^2\right) t^2 \bar t^2x_1+...\nonumber
\end{align}
Let us now check explicitly the validity of our generating function for some of these coefficients, in the large \(N\) limit. Let us start off by considering just one quark/antiquark pair and one adjoint scalar, say $H_1$. The Gauge Invariant Operators (GIOs) we can build out of these fields are
\begin{align}
(\phi)(\bar QQ)^k_l\,,\qquad \,(\bar Q \,\phi \,Q)^k_l\,,
\end{align}
where upper and lower indices belong to the fundamental and antifundamental of $U(F)$ and $U(\bar F)$ respectively, and round brackets denote $U(N)$ indices contraction. The total number of GIOs for this given configuration is $2F^2$. We see that this value is the same one of the coefficient $t\bar t x_1$, so that we have a first test of the validity of \eqref{1node exp}. Consider now the situation in which we only have two pairs of quarks/antiquarks. The only GIOs we can form are of the form
\begin{align}
(\bar QQ)^{k_1}_{l_1}\,(\bar QQ)^{k_2}_{l_2}\,,
\end{align}
using the same convention of the example above for the flavour and gauge indices. This is just a product of two matrix elements of the same \(F\) dimensional matrix \((\bar Q Q)\). The total number of inequivalent GIOs is then $\frac{1}{2}F^2\left(1+F^2\right) $: once again this is the same coefficient of the term $(t\bar t)^2$ in \eqref{1node exp}. As a last example, suppose added to the last configuration a single field $\phi$. The GIOs we can form would then be
\begin{align}
(\phi)\,(\bar QQ)^{k_1}_{l_1}\,(\bar QQ)^{k_2}_{l_2}\,,\qquad (\bar Q\,\phi\,Q)^{k_1}_{l_1}\,(\bar QQ)^{k_2}_{l_2}\,.
\end{align}
The one on the left consists brings a total of $\frac{F^2}{2}\left(1+F^2\right) $ GIOs, while the one on the right adds another $F^2$ GIOs to the final quantity, which then reads
\begin{align}
\frac{F^2}{2}\left(1+F^2\right) +F^2=\frac{F^2}{2}\left(1+3F^2\right)\,,
\end{align}
In agreement with the coefficient of $(t \bar t)^2x_1$ in \eqref{1node exp}.

\subsection{Two node quiver}
We now present some two-node quiver examples.
From the definitions in \eqref{Lambda copy 2} we can immediately write
\begin{equation}
\chi_2=\left(\mathbb{1}_2- X_2\right)^{-1}=\frac{1}{\det\left(\mathbb{1}_2- X_2\right)}
\left(
\begin{array}{ll}
1-x_{22} & x_{12}\\
x_{21}&1-x_{11}
\end{array}
\right)\,,
\end{equation}
and
\begin{equation}
\Lambda_2=
\left(
\begin{array}{ll}
t_1\bar t_1 & t_1\bar t_2\\
t_2\bar t_1&t_2\bar t_2
\end{array}
\right)\,;
\end{equation}
so that, from \eqref{hat final2 result}:
\begin{align}
F^{[2]}&=\det(\chi_2\exp[\chi_2\Lambda_2])=\det(\chi_2)\exp[\Tr(\chi_2\Lambda_2)]\nn
&=\cfrac{
\exp\left(
\cfrac{
t_1\bar t_1(1-x_{22})+t_1\bar t_2x_{21}+t_2\bar t_1x_{12}+t_2\bar t_2(1-x_{11})
}{1-x_{11}-x_{22}-x_{12}x_{21}+x_{11}x_{22}}
\right)
}
{1-x_{11}-x_{22}-x_{12}x_{21}+x_{11}x_{22}}\,.
\end{align}
Finally, recalling \eqref{PExp like relation hat improved copy}, we can get the large $N$ generating function from $F^{[2]}$ by mapping
\begin{subequations}
\begin{align}
&x_{11}\rightarrow \sum_{\alpha=1}^{M_{11}}x_{11,\alpha}^i\,,\qquad
x_{12}\rightarrow \sum_{\alpha=1}^{M_{12}}x_{12,\alpha}^i\,,\qquad
x_{21}\rightarrow \sum_{\alpha=1}^{M_{21}}x_{21,\alpha}^i\,,\qquad
x_{22}\rightarrow \sum_{\alpha=1}^{M_{22}}x_{22,\alpha}^i\,,\\[3mm]
&t_{k}\rightarrow i^{-1/2}\sum_{\beta=1}^{M_{k}}\Tr(\T_{k,\beta}^i)\,,\,\,\,\,k=1,2\,,\qquad
\bar t_{k}\rightarrow i^{-1/2}\sum_{\gamma=1}^{\bar M_{k}}\Tr(\bar\T_{k,\gamma}^i)\,,\,\,\,\,k=1,2\,,
\end{align}
\end{subequations}
and by taking the product over $i$ from 1 to $\infty$. 

The most famous two-node quiver is Klebanov and Witten's conifold gauge theory, consisting of the gauge group $U(N)\times U(N)$ and four bifundamental fields: two of them, $A_1$ and $A_2$, in the representation $(\bar\Box,\,\Box)$ and the remaining two, $B_1$ and $B_2$, in the representation $(\Box,\,\bar\Box)$ of the gauge group. Here we consider the deformation of such a model obtained by allowing flavour symmetries, which is sometimes called `flavoured conifold' \cite{Ouyang:2003df, Levi:2005hh, Benini:2006hh, Bigazzi:2008zt}
\begin{figure}[H]
\begin{center}\includegraphics[scale=1.5]{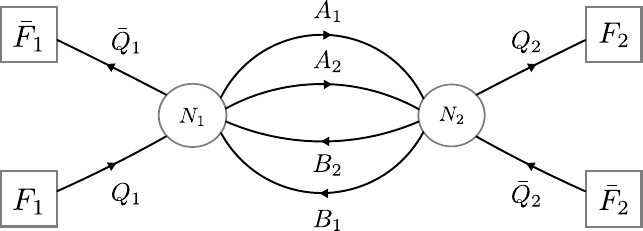}\caption{The flavoured conifold gauge theory.}
\end{center}
\end{figure}

We now choose a different notation for the chemical potentials of the fields, to accord to more standard conventions:
\begin{equation}
\begin{array}{lllll}
&x_{12,1}\rightarrow a_1\,,\qquad\quad &x_{12,2}\rightarrow a_2\,,\qquad \quad
&x_{21,1}\rightarrow b_1\,,\qquad\quad &x_{21,2}\rightarrow b_2\,,\nn
&\T_{1,1}\rightarrow q_1\,,\qquad\quad&\T_{2,1}\rightarrow q_2\,,\qquad\quad
&\bar \T_{1,1}\rightarrow \bar q_1\,,\qquad\quad&\bar \T_{2,1}\rightarrow \bar q_2\,.
\end{array}
\end{equation}
The first terms in the power expansion of \(\mathcal Z_{\text{Flavoured}\atop\text{ Conifold}}(a_1,a_2,b_1,b_2,q_1,q_2,\bar q_1,\bar q_2)\) in the large \(N\) limit then read
\begin{align}
\mathcal Z_{\text{Flavoured}\atop\text{ Conifold}}=&1+
a_1b_1+
2 a_1^2 b_1^2+2 a_1^2 b_1 b_2+2 a_2 a_1 b_1^2+a_1 b_2+2 a_2^2 b_1^2+a_2 b_1+a_2 b_2+\Tr_{F_1}(q_1)\Tr_{\bar F_1}(\bar q_1)\nn
&+2a_1 b_1 \Tr_{F_1}(q_1)\Tr_{\bar F_1}(\bar q_1)+2a_2 b_1 \Tr_{F_1}(q_1)\Tr_{\bar F_1}(\bar q_1)+4 a_1^2 b_1^2 \Tr_{F_1}(q_1)\Tr_{\bar F_1}(\bar q_1)\nn
&+6 a_1 a_2 b_1^2 \Tr_{F_1}(q_1)\Tr_{\bar F_1}(\bar q_1)+a_1\Tr_{F_1}(q_1)\Tr_{\bar F_2}(\bar q_2)+12a_1a_2b_1b_2\Tr_{F_1}(q_1)\Tr_{\bar F_1}(\bar q_1)+...\nn
\end{align}

\subsection{Three node quiver: $dP_0$}
The del Pezzo \(dP_0\) gauge theory (obtained from \(D3\) branes on \(\mathbb C_3/\mathbb Z_3\) orbifold singularities \cite{Douglas:1997de}) contains nine bifundamental fields charged under the \(U(N_1)\times U(N_2)\times U(N_3)\) gauge group as represented in the following quiver, in which we also added flavour symmetry:
\begin{figure}[H]
\begin{center}\includegraphics[scale=1.05]{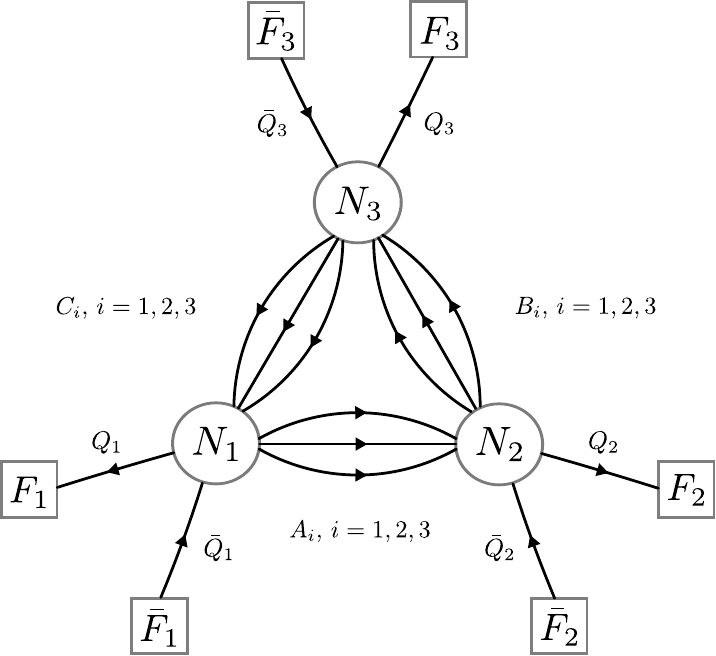}\caption{Flavoured \(dP_0\) gauge theory.}
\label{dP0 fig}
\end{center}
\end{figure}
We refer to this theory as the flavoured \(dP_0\) theory. Using the convention for the chemical potentials of the fields
\begin{equation}
\begin{array}{llllll}
&x_{12,1}\rightarrow a_1\,,\qquad\quad &x_{12,2}\rightarrow a_2\,,\qquad \quad
&x_{12,3}\rightarrow a_3\,,\qquad\quad &\T_{1,1}\rightarrow q_1\,,\qquad\quad&\bar \T_{1,1}\rightarrow \bar q_1\,,\nn
&x_{23,1}\rightarrow b_1\,,\qquad\quad &x_{23,2}\rightarrow b_2\,,\qquad \quad
&x_{23,3}\rightarrow b_3\,,\qquad\quad &\T_{2,1}\rightarrow q_2\,,\qquad\quad&\bar \T_{2,1}\rightarrow \bar q_2\,,\nn
&x_{31,1}\rightarrow c_1\,,\qquad \quad &x_{31,2}\rightarrow c_2\,,\qquad \quad
&x_{31,3}\rightarrow c_3\,,\qquad\quad &\T_{3,1}\rightarrow q_3\,,\qquad\quad&\bar \T_{3,1}\rightarrow \bar q_3\,,\nn
\end{array}
\end{equation}
we can write the generating function for the flavoured \(dP_0\) theory in the large \(N\) limit as:
\begin{align}
&\mathcal Z_{dP_0\text{ Flav.} }=\prod_i\,\frac{1}{1-\sum\limits_{j,k,l=1}^3(a_j\,b_k\,c_l)^i}\nn
&\times\exp
\left(
\frac{
\Tr_{F_1}(q_1^i) \left(\Tr_{\bar F_1}(\bar q_1^i) +\Tr_{\bar F_3}(\bar q_3^i) \sum\limits_{p=1}^3 a_p^i+\Tr_{\bar F_2}(\bar q_2^i) \sum\limits_{p,q=1}^3 c_p^i a_q^i\right)
}{i\left(1-\sum\limits_{j,k,l=1}^3(a_j\,b_k\,c_l)^i\right)}\right.\nn
&\qquad\qquad\left.+\frac{
\Tr_{F_2}(q_2^i) \left(\Tr_{\bar F_2}(\bar q_2^i) +\Tr_{\bar F_1} (\bar q_1^i) \sum\limits_{p=1}^3 b_p^i+\Tr_{\bar F_3}(\bar q_3^i) \sum\limits_{p,q=1}^3 b_p^i a_q^i\right)
}{i\left(1-\sum\limits_{j,k,l=1}^3(a_j\,b_k\,c_l)^i\right)}\right.\nn
&\qquad\qquad\qquad\left.
+\frac{
\Tr_{F_3}(q_3^i) \left(\Tr_{\bar F_3}(\bar q_3^i) +\Tr_{\bar F_2}(\bar q_2^i)  \sum\limits_{p=1}^3c_p^i+\Tr_{\bar F_1}(\bar q_1^i) \sum\limits_{p,q=1}^3 b_p^i c_q^i\right)
}{i\left(1-\sum\limits_{j,k,l=1}^3(a_j\,b_k\,c_l)^i\right)}
\right)\,.
\end{align}


\section{Summary and outlook}

In this paper, we have revisited the counting of local holomorphic operators in general quiver gauge theories 
with bi-fundamental fields, which was started in \cite{quivcalc} (see also related recent work \cite{Koch:2014yga}), focusing on the infinite product formula obtained for the limit of large $N$. This was extended to flavoured quivers, which include fundamental matter. The flavoured quivers have gauge nodes and flavour nodes. The gauge nodes are associated with
unitary group factors in the gauge symmetry. The flavour nodes are associated with unitary global symmery factors. 
 We used Schur-Weyl duality relating the representation theory of unitary groups to permutation groups
in order to convert integrals over the gauge unitary groups for the counting into permutation sums. The sums involved multiple 
permutations with constraints. These constraints were expressed by introducing contour integrals. This lead to 
an analogous infinite product formula for these flavoured quivers \eqref{PExp like relation hat improved copy}. 
 For any quiver with $n$ gauge nodes, all the factors in the infinite product are obtained by substitutions in one function $ F^{[n]} ( \{x_{ab}\} ,\{ t_a\} ,\{ \bar t_a\} )$, with $a,b$ ranging over the $n$ nodes. 
The building block $ F^{[n]} ( \{x_{ab}\} ,\{ t_a\} ,\{ \bar t_a\} )$ was found to be closely related to 
$F_0^{[n]} ( \{x_{ab}\} )$. The determinant and cofactors of the matrix $ ( \mathbb 1_n - X_n ) $ played a prominent role in these formulae. We also obtained results for the counting of local operators at finite $N$ in terms of Young diagrams and Littlewood-Richardson coefficients.

The permutation group-theoretic counting of operators at finite $N$ is the first step in the construction of an orthogonal basis of operators for an inner product related to two-point functions between holomorphic and anti-holomorphic operators in free field theory \cite{quivcalc}. The next step is the computation of three and higher point functions in the free theory. The orthogonal bases were found to be given by simple rules involving Young diagrams and related group theoretic multiplicities attached to the quiver diagram itself. Three-point functions were given by cutting and gluing of the diagrams. The counting and the construction of the orthogonal bases were given in terms of two-dimensional topological field theories (TFT2) of Dijkgraaf-Witten type, based on permutations and equipped with appropriate defects. Given the close similarity we have found between the counting formulae for flavoured and unflavoured quivers, we expect that the results on orthogonal bases, correlators and TFT2 \cite{quivcalc} will also have simple generalizations from unflavoured quivers to flavoured quivers. An interesting avenue would be to explore 
the links between this permutation group TFT2 approach to correlators with methods based on integrable models (see e.g \cite{Foda,BKMO1211}).

The flavoured counting at large $N$ is determined by $F^{[n]} $, which is closely related to $F_0^{[n] } $, which in turn we have related to word counting problems associated to the complete $n$-node quiver. One formulation of the word counting problem was in terms of words made from letters corresponding to simple closed loops on the quiver. The letters do not commute if they share a node, otherwise they commute. In another formulation, the letters correspond to edges of the quiver. Distinct letters do not commute if they share a starting point. These open string bits form words, a subset of which obey a charge conservation condition. A non-trivial combinatoric equivalence between the open string and closed string counting problems is given by the Cartier-Foata theorem. In this paper, we have come across 
these string-word-counting problems in connection with the counting formula for gauge invariants. It is natural to ask if such words, and their monoidal structure, are relevant beyond the counting of gauge theory invariants. One often finds that mathematical structures relevant to counting a class of objects are also relevant in the understanding of the interactions of such objects (see e.g. \cite{doubcos,gigravosc} for a concrete application of this idea). Do the string-words found here play a role in interactions, namely in the computation of correlators of gauge-invariant operators in the free field limit and at weak coupling? Since the work of Cartier-Foata has subsequently been related to statistical physics models \cite{heaps,heaps2}, this underlying mathematical structure could reveal 
new connections between four dimensional gauge theory and statistical physics. 

In the context of AdS/CFT, comparisons between the counting of a class of local gauge-invariant operators 
and the spectrum of brane fluctuations was initiated in \cite{KarKat02,EF1012,AFR1312}. 
These papers considered the simplest quiver gauge theory, namely $N=4$ SYM, and 
the additional fundamental matter corresponds to the addition of $7$-branes in the dual $AdS_5 \times S^5 $ background. The results presented here should be useful for generalizations of 
these results, such as increasing the number of $7$-branes, and more substantially, going beyond the $N=4$ SYM as starting point to more general quiver theories. The finite $N$ aspects 
of counting, where operators are labelled by Young diagrams, should be related to giant gravitons. This will require the investigation of $3$-brane giant gravitons in $AdS_5 \times S^5$, in the presence of the probe $7$-branes. Some discussion of such configurations is initiated in the conclusions of \cite{Kruczenski:2003be}. Such detailed comparisons for the general class of flavoured quiver theories we considered here would undoubtedly deepen our understanding of AdS/CFT.

In this paper we have counted general holomorphic gauge invariant operators in quiver theories with fundamental matter. These also form the space of chiral operators in the theory in the absence of a superpotential. When we turn on a superpotential, equivalence classes related by setting to zero the derivatives of the super-potential, form the chiral ring \cite{CDSW,lutay}. This jump in the spectrum of chiral primaries has been discussed in the context of AdS/CFT in \cite{Kinney:2005ej}. An important future direction is to understand this jump in quantitative detail. We have found that the quiver diagram defining a theory contains powerful information on the counting of operators in the theory, and the weighted adjacency matrix played a key role in giving a general form for the generating functions at large $N$. It would be interesting to look for analogous general formulae, involving the weighted adjacency matrix, along with the superpotential data, for the case of chiral rings at non-zero superpotential. In a similar vein we may ask if indices in superconformal theories, for general quivers, can be expressed in terms of the weighted adjacency matrix. It will be interesting to investigate this theme in existing examples of index computations for quivers (e.g. \cite{Nakayama:2005mf,rastelli,Kim:2009wb}). Beyond counting questions, the transition to non-zero superpotential poses the question of the exact form of BPS operators. In cases where the 1-loop dilatation operator is known, such as $\N=4$ SYM, we can find the BPS operators by solving for the null eigenstates among the holomorphic operators. Partial results at large $N$ as well as finite $N$, building on the knowledge of free field bases of operators, are available in \cite{BHR1,BHR2,doubcos,gigravosc,Brown:2010pb, Pasukonis:2010rv,Kimura:2010tx,Kimura:2012hp}. A similar treatment should be possible for orbifolds of $\N=4$. 

The counting of chiral operators with and without superpotential is of interest in studying the Hilbert series of moduli spaces arising from super-symmetric gauge-theories \cite{Gray:2008yu,Hanany:2008sb,Jokela:2011vg}. These moduli spaces often have an interpretation in terms of branes. Quiver gauge theories, with and without fundamental matter, have been studied in this context. The formulae obtained here, for finite $N$ as well as large $N$, will be expected to have applications in the study of these moduli spaces. Another potential application of the present counting techniques is in the thermodynamics of AdS/CFT or toy models thereof, e.g. \cite{Harmark:2014mpa}.

\vskip2cm

\begin{centerline}
{ \bf Acknowledgements} 
\end{centerline} 

\vskip.4cm 

We thank Robert de Mello Koch, Yang-Hui He, Vishnu Jejjala, James McGrane, Gabriele Travaglini and Brian Wecht for useful discussions. SR is supported by STFC consolidated grant ST/L000415/1 ``String Theory, Gauge Theory \& Duality." PM is supported by a Queen Mary University of London studentship.

\vskip2cm


\addtocontents{toc}{\vspace{2cm}}
\addcontentsline{toc}{section}{Appendices}

\appendix
\section{Generating function}\label{computing sums}

\subsection{Derivation of the generating function}\label{Derivation of the Gen Fun}

In this appendix we will derive eq. \eqref{N p t}. Our starting point will be eq. \eqref{igt}:
\begin{align}\label{igt copy}
\Z&(\{x_{ab,\alpha}\},\{\T_{a,\beta}\},\{\bar \T_{a,\gamma}\})=
\sum_{\{n_{ab,\alpha}\}}\,\sum_{\{n_{a,\beta}\}\atop\{\bar n_{a,\gamma}\}}\,
\left(
\prod_{a,b,\alpha}\,x_{ab,\alpha}^{n_{ab,\alpha}}\right)\,
%
%
%
%
%
%
\sum_{\left\{R_a\vdash n_a\atop l(R_a)\leq N_a\right\}}\,\sum_{\{r_{ab,\alpha}\vdash n_{ab,\alpha}\}}\,\sum_{\{r_{a,\beta}\vdash n_{a,\beta}\}\atop \{\bar r_{a,\gamma}\vdash \bar n_{a,\gamma}\}}\\[3mm]
&\prod_a\,
g(\cup_{b,\alpha}r_{ab,\alpha}\cup_\beta r_{a,\beta};R_a)\,g(\cup_{b,\alpha}r_{ba,\alpha}\cup_\gamma \bar r_{a,\gamma};R_a)\
\left(\prod_{\beta}\chi_{ r_{a,\beta}}(\T_{a,\beta})  \right)
\left(\prod_{\gamma} \chi_{\bar r_{a,\gamma}}(\bar\T_{a,\gamma})  \right)
\,,\nonumber
\end{align}
in which we will take the large $N$ limit, in such a way that we will be allowed to drop the constraints on the sums over \(R_a\). The derivation will involve well known symmetric group identities. In particular, we will use the equation
\begin{subequations}
\begin{align}\label{dimum}
\chi_R(U)=\sum_{\sigma\in S_n}\frac{\chi_R(\sigma)}{n!}\prod_i\,(\Tr\, U^i)^{[\sigma]^{(i)}}\,,\qquad U\in U(F)
\end{align}
where $R$ is a partition of $n$, $[\sigma]^{(i)}$ is the number of cycles of length $i$ in the conjugacy class $[\sigma]$ of the permutation $\sigma\in S_n$ and \(\Tr(U)\) is the trace taken in the fundamental representation of \(U(F)\). We will also use the formulae
\begin{align}\label{proddel}
\quad\sum_{R\vdash n}\chi_R(\sigma)\chi_R(\tau)=\sum_{\gamma\in S_n}\delta(\gamma\sigma\gamma^{-1}\tau^{-1})\,,
\end{align}
and
\begin{align}\label{chr}
\quad g(\cup_a r_a;R)=\left ( \prod_a \sum_{\sigma_a} \right )  \chi_R(\times_a \sigma_a)
{  \prod_a \frac{\chi_{r_a}(\sigma_a)} 
{n_a!}} \,.
\end{align}
\end{subequations}
Here $r_a$ are partitions of $n_a$ and \(\delta(\sigma)\) is the symmetric group delta function, which equals one iff $\sigma$ is the identity permutation. With these relations we can rewrite $\Z$ in \eqref{igt copy} as
\begin{align}\label{first step in gf}
\Z
=&\sum_{\vec n}\,\sum_{\vec\sigma,\,\vec\sigma '}\,\prod_{a}\,
\left(
\prod_{b,\alpha}\sum_{ r_{ab,\alpha}}\frac{x_{ab,\alpha}^{\sum_i i[\sigma_{ab,\alpha}]^{(i)}}}{(n_{ab,\alpha}!)^2}\,
\chi_{r_{ab,\alpha}}(\sigma_{ab,\alpha})\chi_{r_{ab,\alpha}}(\sigma'_{ab,\alpha})
\right)\nn
&\quad\times\left(
\prod_\beta \sum_{ r_{a,\beta}}
\frac{\prod_i(\Tr\,\T_{a,\beta}^{\,i})^{[\sigma_{a,\beta}]^{(i)}}}{(n_{a,\beta}!)^2}
\chi_{r_{a,\beta}}(\sigma_{a,\beta})
\chi_{r_{a,\beta}}(\sigma'_{a,\beta})
\right)\nn
&\qquad\qquad\times\left(
\prod_\gamma \sum_{r_{a,\gamma}}
\frac{\prod_i(\Tr\, \bar\T_{a,\gamma}^{\,i})^{[\bar\sigma_{a,\gamma}]^{(i)}}}{(\bar n_{a,\gamma}!)^2}
\chi_{\bar r_{a,\gamma}}(\bar\sigma_{a,\gamma})\chi_{\bar r_{a,\gamma}}(\bar\sigma'_{a,\gamma})
\right)\nn
&\qquad\qquad\qquad\times\sum_{R_a}\,
\chi_{R_{a}}(\times_{b,\alpha}\sigma_{ab,\alpha}\times_{\beta}\sigma_{a,\beta}')
\chi_{R_{a}}(\times_{b,\alpha}\sigma'_{ba,\alpha}\times_{\gamma}\bar\sigma_{a,\gamma}')\,,
\end{align}
where we defined
\begin{align}\label{vector sigma}
&\vec\sigma=\cup_{a,b,\alpha}\{\sigma_{ab,\alpha}\}\cup_{a,\beta}\{\sigma_{a,\beta}\}\cup_{a,\gamma}\{\bar\sigma_{a,\gamma}\}\,,\nn
&\sigma_{ab,\alpha}\in S_{n_{ab,\alpha}}\,,\qquad
\sigma_{a,\beta}\in S_{n_{a,\beta}}\,,\qquad
\bar\sigma_{a,\gamma}\in S_{\bar n_{a,\gamma}}\,,
\end{align}
and similarly
\begin{align}
\vec n=\cup_{a,b,\alpha}\{n_{ab,\alpha}\}\cup_{a,\beta}\{n_{a,\beta}\}\cup_{a,\gamma}\{\bar n_{a,\gamma}\}
\,.
\end{align}
Summing over the representations $R_a,r_{ab,\alpha},r_{a,\beta},\bar r_{ a,\gamma}$ then gives, using \eqref{proddel}
\begin{align}
\Z=\sum_{\vec n}&\,\sum_{\vec\sigma,\,\vec\sigma '}\sum_{\vec\rho}\,\prod_{a}
\left(
\prod_{b,\alpha}
\frac{x_{ab,\alpha}^{\sum_i i[\sigma_{ab,\alpha}]^{(i)}}}{(n_{ab,\alpha}!)^2}
\,
\delta\left(\rho_{ab,\alpha}\,\sigma_{ab,\alpha}\,\rho_{ab,\alpha}^{-1}\,{\sigma^{\prime \,-1}_{ab,\alpha}}\right)
\right)\nn
&\quad\times
\left(
\prod_\beta
\frac{\prod_i(\Tr\,\T_{a,\beta}^{\,i})^{[\sigma_{a,\beta}]^{(i)}}}{(n_{a,\beta}!)^2}\,
\delta\left(\rho_{a,\beta}\,\sigma_{a,\beta}\,\rho_{a,\beta}^{-1}\,{\sigma^{\prime\,-1}_{a,\beta}}\right)
\right)\nn
&\qquad\times\left(
\prod_\gamma
\frac{\prod_i(\Tr\,\bar \T_{a,\gamma}^{\,i})^{[\bar\sigma_{a,\gamma}]^{(i)}}}{(\bar n_{a,\gamma}!)^2}
\delta\left(\bar\rho_{ a,\gamma}\,\bar\sigma_{ a,\gamma}\,\bar\rho_{ a,\gamma}^{-1}\,{{{\bar\sigma}_{ a,\gamma}}}^{\prime\,-1}\right)
\right)\nn
&\qquad\quad\times
\sum_{\Gamma_a}
\delta\left(\Gamma_{a}(\times_{b,\alpha}\sigma_{ab,\alpha}\times_{\beta}\sigma_{a,\beta}')\Gamma_{a}^{-1}{(\times_{b,\alpha}\sigma'_{ba,\alpha}\times_{\gamma}\bar\sigma_{ a,\gamma}')}^{-1}\right)\,,
\end{align}
with
\begin{align}
&\vec\rho=\cup_{a,b,\alpha}\{\rho_{ab,\alpha}\}\cup_{a,\beta}\{\rho_{a,\beta}\}\cup_{a,\gamma}\{\bar\rho_{a,\gamma}\}\,,\nn
&\rho_{ab,\alpha}\in S_{n_{ab,\alpha}}\,,\qquad
\rho_{a,\beta}\in S_{n_{a,\beta}}\,,\qquad
\bar\rho_{a,\gamma}\in S_{\bar n_{a,\gamma}}\,.
\end{align}
If we now sum over the $\vec\sigma'$ permutations we get, redefining the dummy $\Gamma_a$ permutations as \(\Gamma_a\rightarrow \left(\times_{b,\alpha}\rho_{ba,\alpha}\times_\gamma \bar\rho_{a,\gamma}\right)\Gamma_a\left(\times_{b,\alpha}1\times_\beta \rho_{a,\beta}\right)^{-1} \) :
\begin{align}
\Z=\sum_{\vec n}\,\sum_{\vec\sigma,\,\vec\rho}\,\prod_{a}
&\left(
\prod_{b,\alpha}
\frac{x_{ab,\alpha}^{\sum_i i[\sigma_{ab,\alpha}]^{(i)}}}{(n_{ab,\alpha}!)^2}
\right)
\left(
\prod_\beta
\frac{\prod_i(\Tr\,\T_{a,\beta}^{\,i})^{[\sigma_{a,\beta}]^{(i)}}}{(n_{a,\beta}!)^2}
\right)
\left(
\prod_\gamma
\frac{\prod_i(\Tr\,\bar \T_{a,\gamma}^{\,i})^{[\bar\sigma_{a,\gamma}]^{(i)}}}{(\bar n_{a,\gamma}!)^2}
\right)\nn
&\qquad\times\sum_{\Gamma_a}\,
\delta\left(\Gamma_{a}(\times_{b,\alpha}\sigma_{ab,\alpha}\times_{\beta}\sigma_{a,\beta})\Gamma_{a}^{-1}{(\times_{b,\alpha}\sigma_{ba,\alpha}\times_{\gamma}\bar\sigma_{a,\gamma})}^{-1}\right)\,.
\end{align}
Finally, by summing over the now trivial $\vec\rho$ permutations we obtain
\begin{align}\label{cc}
\Z=\sum_{\vec n}\,\sum_{\vec\sigma}\prod_{a}&
\left(\prod_{b,\alpha}\frac{x_{ab,\alpha}^{\sum_i i[\sigma_{ab,\alpha}]^{(i)}}}{n_{ab,\alpha}!}\right)
\left(\prod_\beta \frac{1}{n_{ a,\beta}!}\right)
\left(\prod_\gamma \frac{1}{\bar n_{ a,\gamma}!}\right)\nn
&\qquad\qquad\times\,H_a(\{\sigma_{ab,\alpha}\},\{\sigma_{a,\beta}\},\{\bar\sigma_{a,\gamma}\};\{\T_{a,\beta}\},\{\bar \T_{a,\gamma}\})\,,
\end{align}
where we defined
\begin{align}
H_a(\{\sigma_{ab,\alpha}\},\{\sigma_{a,\beta}\},&\{\bar\sigma_{a,\gamma}\};\{\T_{a,\beta}\},\{\bar \T_{a,\gamma}\})=
\left(\prod_{\beta,i} (\Tr\,\T_{a,\beta}^{\,i})^{[\sigma_{a,\beta}]^{(i)}}\right)\left(\prod_{\gamma,i} (\Tr\,\bar\T_{a,\gamma}^{\,i})^{[\bar \sigma_{a,\gamma}]^{(i)}}\right)\nn
&\qquad\quad\times
\sum_{{\Gamma_a}}\delta\left(\Gamma_{a}(\times_{b,\alpha }\sigma_{ab,\alpha }\times_{\beta }\sigma_{a,\beta })\Gamma_{a}^{-1}{(\times_{b,\alpha }\sigma_{ba,\alpha }\times_{\gamma }\bar \sigma_{a,\gamma })}^{-1}\right)\,.
\end{align}
Eq. \eqref{cc} is a function of the conjugacy class of the permutations \(\sigma\), rather than of the permutations themselves. Exploiting this fact we can rewrite it as follows. Let us introduce the vectors of integers \(\vec p_{ab,\alpha}=\cup_{i}\{p_{ab,\alpha}^{(i)}\}\), \(\vec p_{a,\beta}=\cup_{i}\{p_{a,\beta}^{(i)}\}\) and \(\vec {\bar p}_{a,\gamma}=\cup_{i}\{\bar p_{a,\gamma}^{\,(i)}\}\). Here \(p_{ab,\alpha}^{(i)}\) is the number of cycles of length \(i\) in the permutation \(\sigma_{ab,\alpha}\), while \(p_{a,\beta}^{(i)}\) and \(\bar p_{a,\gamma}^{(i)}\) are the number of cycles of length \(i\) in the permutations \(\sigma_{a,\beta}\) and \(\bar\sigma_{a,\gamma}\) respectively. In accordance with eq. \eqref{vector sigma} we have
\begin{align}\label{|p|2}
\sum_{i=1}^\infty i p_{ab,\alpha}^{(i )}=n_{ab,\alpha}\,,\qquad\quad
 \vert  \vec p_{ab,\alpha}  \vert =\frac{n_{ab,\alpha}!}{\prod_i p_{ab,\alpha}^{(i )}!\,i^{p_{ab,\alpha}^{(i )}}}\,,
\end{align}
and similarly for \(\vec p_{a,\beta}\) and \(\vec{\bar p}_{a,\gamma}\). For notational purposes, it will be convenient to introduce the compact shorthand \(\pmb p=\cup_{ab,\alpha}\vec p_{ab,\alpha}\cup_{a,\beta}\vec p_{a,\beta}\cup_{a,\gamma}\vec{\bar p}_{a,\gamma}\). With this notation we can rewrite \eqref{cc} as
\begin{align}\label{cc w H 2}
\mathcal{Z}=\sum_{\vec n}\,
\sum_{\pmb p}\,
\prod_{a}
\left(
\prod_{b,\alpha}
\frac{|\vec p_{ab,\alpha}|}{n_{ab,\alpha}!}\,x_{ab,\alpha}^{\sum_i ip_{ab,\alpha}^{(i)}}
\right)&
\left(
\prod_{\beta}
\frac{|\vec p_{a,\beta}|)}{n_{a,\beta}!}
\right)
\left(
\prod_{\gamma}
\frac{|\vec{\bar p}_{a,\gamma}|}{\bar n_{a,\gamma}!}
\right)\nn
&\times H_a(\{\vec p_{ab,\alpha}\},\{\vec p_{a,\beta}\},\{\vec{\bar p}_{ a,\gamma}\};\{\T_{a,\beta}\},\{\bar \T_{a,\gamma}\})\,,
\end{align}
where now \( H_a(\{\vec p_{ab,\alpha}\},\{\vec p_{a,\beta}\},\{\vec{\bar p}_{ a,\gamma}\};\{\T_{a,\beta}\},\{\bar \T_{a,\gamma}\})\) reads, after summing over the \(\Gamma_a\) permutations
\begin{align}\label{H in perms}
&H_a(\{\vec p_{ab,\alpha}\},\{\vec p_{a,\beta}\},\{\vec{\bar p}_{ a,\gamma}\};\{\T_{a,\beta}\},\{\bar \T_{a,\gamma}\})=\prod_i\left(\prod_\beta (\Tr\,\T_{a,\beta}^{\,i})^{p_{a,\beta}^{(i)}}\right)\left(\prod_\gamma (\Tr\,\bar\T_{a,\gamma}^{\,i})^{[ {\bar p}_{a,\gamma}^{(i)}}\right)\nn
&\times
\delta_{a}\left(
\sum_{b,\alpha}(p_{ab,\alpha}^{(i)}-p_{ba,\alpha}^{(i)})+\sum_{\beta}p_{a,\beta}^{(i)}-\sum_{\gamma}\bar p_{ a,\gamma}^{(i)}
\right)\,\,
i\,\,^{\sum\limits_{b,\alpha}p_{ba,\alpha}^{(i)}+\sum\limits_{\gamma}\bar p_{ a,\gamma}^{(i)}}\left(\sum_{b,\alpha}p_{ba,\alpha}^{(i)}+\sum_{\gamma}\bar p_{ a,\gamma}^{(i)}
\right)!\,.
\end{align}
Using \eqref{H in perms} and \eqref{|p|2} in \eqref{cc w H 2} gives then
\begin{align}\label{kd a}
\Z&=
\sum_{\pmb p}\prod_i
\prod_{a}
\cfrac{\left(\sum_{b,\alpha}p_{ba,\alpha}^{(i)}+\sum_{\gamma}\bar p_{ a,\gamma}^{(i)}
\right)!}{i^{\sum_\beta p_{a,\beta}^{(i)}}}
\left(\prod_{b,\alpha}\frac{x_{ab,\alpha}^{ ip_{ab,\alpha}^{(i)}}}{p_{ab,\alpha}^{(i)}!}\right)
\left(\prod_\beta\frac{(\Tr\,\T_{a,\beta}^{\,i})^{p_{a,\beta}^{(i)}}}{p_{a,\beta}^{(i)}!}\right)\nn
&\times
\left(\prod_\gamma\frac{(\Tr\,\bar \T_{a,\gamma}^{\,i})^{\bar p_{a,\gamma}^{(i)}}}{\bar p_{a,\gamma}^{(i)}!}\right)
\,
\delta_{a}\left(
\sum_{b,\alpha}(p_{ab,\alpha}^{(i)}-p_{ba,\alpha}^{(i)})+\sum_{\beta}p_{a,\beta}^{(i)}-\sum_{\gamma}\bar p_{ a,\gamma}^{(i)}
\right)
\,,
\end{align}
which is eq. \eqref{N p t}.

Note that if we define the function \(F^{[n]}(\{x_{ab}\},\{t_a\},\{\bar t_a\})\) as
\begin{align}\label{first def of F}
F^{[n]}(\{x_{ab}\},\{t_a\},\{\bar t_a\}\})=&
\sum_{\vec p}
\prod_{a=1}^n
\left(\bar p_{ a}+\sum_{b=1}^np_{ba}
\right)!\,\,
\delta_{a}\left(
p_{a}-\bar p_{ a}+\sum_{b=1}^n(p_{ab}-p_{ba})
\right)\nn
&\qquad\qquad\times
\left(\prod_{b=1}^n\frac{x_{ab}^{ p_{ab}}}{p_{ab}!}\right)\,
\left(\frac{t_a^{p_{a}}}{p_{a}!}\right)\,\left(\frac{\bar t_a^{\,\bar p_{a}}}{\bar p_{a}!}\right)
\,,
\end{align}
where now \(\vec p \equiv \cup_{a,b}\{p_{ab}\}\cup_a\{p_{a},\,\bar p_{a}\}\), we can immediately obtain the generating function \(\Z\) \eqref{kd a} through the relation
\begin{align}\label{PExp like relation in app}
{\mathcal{Z}}(&\{x_{ab,\alpha}\},\{\T_{a,\beta}\},\{\bar \T_{a,\gamma}\})\nn
&=\prod_i F^{[n]}\left(\left\{x_{ab}\rightarrow \sum_{\alpha}x_{ab,\alpha}^i\right\},\left\{t_a\rightarrow\sum_{\beta}\frac{ \Tr(\T_{a,\beta}^{\,i})}{i}\right\},\left\{\bar t_a\rightarrow\sum_{\gamma}\Tr(\bar \T_{a,\gamma}^{\,i})\right\}\right)\,\,.
\end{align}
In fact, the RHS of \eqref{PExp like relation in app} reads
\begin{align}\label{rhs Z series}
\prod_i&
\sum_{\vec p}
\prod_{a}
\left(\bar p_{ a}+\sum_{c}p_{ca}
\right)!\,\,
\delta_{a}\left(
p_{a}-\bar p_{ a}+\sum_{c}(p_{ac}-p_{ca})
\right)\nn
&\qquad\qquad\times
\left(\prod_{b}\frac{\left(\sum_{\alpha}x_{ab,\alpha}^i\right)^{ p_{ab}}}{p_{ab}!}\right)\,
\left(\frac{\left(\sum_{\beta}\frac{ \Tr(\T_{a,\beta}^{\,i})}{i}\right)^{p_{a}}}{p_{a}!}\right)\,\left(\frac{\left(\sum_{\gamma}\Tr(\bar \T_{a,\gamma}^{\,i})\right)^{\,\bar p_{a}}}{\bar p_{a}!}\right)
\,,
\end{align}
and through the identity
\begin{align}\label{relation for sum copy}
\left(\sum_{a=1}^n z_a\right)^k=\sum_{\vec p}\delta\left(k-\sum_{a=1}^np_{a}\right)k!\,\prod_{a=1}^n\frac{z_a^{p_a}}{p_a!}\,,\qquad \vec p=(p_1,p_2,...,p_n)
\end{align}
we can write \eqref{rhs Z series} as
\begin{align}
&\prod_i
\sum_{\vec p}
\prod_{a}
\left(\bar p_{ a}+\sum_{b}p_{ba}
\right)!\,\,
\delta_{a}\left(
p_{a}-\bar p_{ a}+\sum_{b}(p_{ab}-p_{ba})
\right)\nn
&\qquad\times
\left(\sum_{\vec p_{ab}^{\,(i)}}\,\delta\left(p_{ab}-\sum_{\alpha }p_{ab,\alpha }^{\,(i)}\right)\,\prod_{b,\alpha}\frac{x_{ab,\alpha}^{i p_{ab,\alpha}^{\,(i)}}}{p_{ab,\alpha}^{\,(i)}!}\right)\\[3mm]
&\times
\left(\sum_{\vec \rho_a^{\,(i)}}\,\delta\left(p_a-\sum_{\beta  }p_{a,\beta }^{\,(i)}\right)\,\prod_{\beta}\,\frac{ (\Tr\,\T_{a,\beta}^{\,i})^{p_{a,\beta}^{\,(i)}}}{i^{p_{a,\beta}^{\,(i)}}\,p_{a,\beta}^{\,(i)}!}\right)\,
\left(\sum_{\vec {\bar \rho}_a^{\,(i)}}\,\delta\left(\bar p_a-\sum_{\gamma  }\bar p_{a,\gamma }^{\,(i)}\right)\,\prod_{\gamma}\,\frac{ (\Tr\,\bar \T_{a,\gamma}^{\,i})^{\bar p_{a,\gamma}^{\,(i)}}}{\bar p_{a,\gamma}^{\,(i)}!}\right)
\,,\nonumber
\end{align}
where \(\vec p_{a}^{\,(i)}=\cup_{b,\alpha}\{p_{ab,\alpha}^{\,(i)}\}\), \(\vec \rho_a^{\,(i)}=\cup_\beta\{p_{a,\beta}^{\,(i)}\}\) and \( \vec{\bar \rho}_a^{\,(i)}=\cup_\gamma\{\bar p_{a,\gamma}^{\,(i)}\}\). Summing over \(\vec p\) gives, exploiting the second, third and fourth Kronecker deltas in the expression above
\begin{align}
&\prod_i
\prod_{a}
\sum_{\vec p_{ab}^{\,(i)}}\,\sum_{\vec \rho_a^{\,(i)}}\,\sum_{\vec {\bar \rho}_a^{\,(i)}}\,
\left(\sum_{b,\alpha }p_{ba,\alpha }^{\,(i)}+\sum_{\gamma }\bar p_{ a,\gamma}^{\,(i)}
\right)!\,\,
\delta_{a}\left(
\sum_{b,\alpha }(p_{ab,\alpha }^{\,(i)}-p_{ba,\alpha }^{\,(i)})+\sum_{\beta } p_{a,\beta }^{\,(i)}-\sum_{\gamma }\bar p_{ a,\gamma }^{\,(i)}
\right)\nn
&\qquad\times
\left(\prod_{b,\alpha}\frac{x_{ab,\alpha}^{i p_{ab,\alpha}^{\,(i)}}}{p_{ab,\alpha}^{\,(i)}!}\right)
\left(\prod_{\beta}\,\frac{ (\Tr\,\T_{a,\beta}^{\,i})^{p_{a,\beta}^{\,(i)}}}{i^{p_{a,\beta}^{\,(i)}}\,p_{a,\beta}^{\,(i)}!}\right)\,
\left(\prod_{\gamma}\,\frac{ (\Tr\,\bar \T_{a,\gamma}^{\,i})^{\bar p_{a,\gamma}^{\,(i)}}}{\bar p_{a,\gamma}^{\,(i)}!}\right)
\nn
&=
\sum_{\pmb p}\prod_i
\prod_{a}
\cfrac{\left(\sum_{b,\alpha }p_{ba,\alpha }^{\,(i)}+\sum_{\gamma }\bar p_{ a,\gamma }^{\,(i)}
\right)!}{i^{\sum_\beta p_{a,\beta}^{\,(i)}}}
\left(\prod_{b,\alpha}\frac{x_{ab,\alpha}^{ ip_{ab,\alpha}^{\,(i)}}}{p_{ab,\alpha}^{\,(i)}!}\right)
\left(\prod_\beta\frac{(\Tr\,\T_{a,\beta}^{\,i})^{p_{a,\beta}^{\,(i)}}}{p_{a,\beta}^{\,(i)}!}\right)\nn
&\qquad\times
\left(\prod_\gamma\frac{(\Tr\,\bar \T_{a,\gamma}^{\,i})^{\bar p_{a,\gamma}^{\,(i)}}}{\bar p_{a,\gamma}^{\,(i)}!}\right)
\,
\delta_{a}\left(
\sum_{b,\alpha }(p_{ab,\alpha }^{\,(i)}-p_{ba,\alpha }^{\,(i)})+\sum_{\beta }p_{a,\beta }^{\,(i)}-\sum_{\gamma }\bar p_{ a,\gamma }^{\,(i)}
\right)=\Z
\,,
\end{align}
where in the second equality we used \(\pmb p \equiv \cup_{ab,\alpha}  \vec p_{ab,\alpha}  
\cup_{a,\beta}\vec p_{a,\beta}\cup_{a,\gamma}\vec{\bar p}_{a,\gamma}\) and the third one follows from \eqref{kd a}. We can now appreciate how every property of \(\Z\) is determined by the \(F^{[n]}\) function, which will play the role of fundamental building block of the generating function. In the following we will then focus mainly on the latter, which will improve the clarity of the exposition: the generating function \(Z\) can be obtained at any time through the relation \eqref{PExp like relation in app}.


\subsection{A contour integral formulation for $F^{[n]}$}\label{sec: contour int in app}
All of the Kronecker deltas $\delta_a$ in eq. \eqref{first def of F} ensure that, at each node $a$ in the quiver, there are as many fields flowing in as there are flowing out, ensuring the balance of the incoming and outgoing edge variables \(p_{ab},\,p_a,\,\bar p_a\). Using the contour integral resolution of the Kronecker delta 
\begin{align}\label{delta integral resolution}
\delta_a=\oint_{\C}\frac{dz}{2\pi iz}z^a\,,
\end{align}
where $\C$ is a closed path that encloses the origin, we can write a contour integral formulation for $F^{[n]}$, and thus for $\mathcal Z$. Let us then use \eqref{delta integral resolution} in \eqref{first def of F}, to get
\begin{align}\label{delta integral resolution usage}
F^{[n]}=
&\sum_{\vec p}
\prod_{a}
\left(\bar p_{ a}+\sum_{c}p_{ca}
\right)!
\left(\prod_{b}\frac{x_{ab}^{ p_{ab}}}{p_{ab}!}\right)
\left(\frac{t_a^{p_{a}}}{p_{a}!}\right)
\left(\frac{\bar t_a^{\,\bar p_{a}}}{\bar p_{a}!}\right)
\,
\oint_{\mathcal{C}_{a}}\frac{dz_{a}}{2\pi iz_{a}}z_{a}^{ p_{a}-\bar p_{ a}+\sum_{b}\left(p_{ab}-p_{ba}\right)}\,,
\end{align}
or, conveniently rearranging the integrands above,
\begin{align}\label{midstep}
F^{[n]}=
\sum_{\vec p}
&\left(\prod_{a}
\oint_{\mathcal{C}_{a}}\frac{dz_{a}}{2\pi iz_{a}}\right)\prod_a\,
\left(\prod_{b}\cfrac{\left(z_{b}x_{ba}z_{a}^{-1}\right)^{p_{ba}}}{p_{ba}!}\right)
\left(\frac{\left(z_{a}t_a\right)^{p_{a}}}{p_{a}!}\right)\nn
&\qquad\times
\left(\bar p_{ a}+\sum_{c}p_{ca}
\right)!
\left( \cfrac{\left(z_{a}^{-1}\,\bar t_a\right)^{\bar p_{a}}}{\bar p_{a}!}\right)\,.
\end{align}
Summing over the $p_{a}$ s gives the exponentials
\begin{align}\label{sum norm}
\sum_{p_{a}} 
\cfrac{\left(z_{a}t_a\right)^{p_{a}}}{p_{a}!}
=\exp\left(z_{a}t_a\right)\,,
\end{align}
while it is a little bit trickier to sum over the $\bar p_{a}$ s. Using the identity
\begin{align}
\left(\bar p_{ a}+\sum_{c}p_{ca}
\right)!=
\left(\sum_{c}p_{ca}\right)!\left(1+\sum_{c}p_{ca}\right)_{\left(\bar p_{ a}\right)}\,,
\end{align}
where $(a)_n=a(a+1)\cdots (a+n-1)$ is the Pochhammer symbol, we can rewrite \eqref{midstep} as
\begin{align}\label{midstep2}
F^{[n]}&=
\left(\prod_{a }\sum_{\vec p_a,\, \bar p_a}
\oint_{\mathcal{C}_{a}}\frac{dz_{a}}{2\pi iz_{a}}\right)\prod_a\,
\left(\sum_{c}p_{ca}\right)!
\left(\prod_{b}\cfrac{\left(z_{b}x_{ba}z_{a}^{-1}\right)^{p_{ba}}}{p_{ba}!}\right)
\nn
&\qquad\qquad\times
\exp\left(z_{a}t_a\right)
\left(1+\sum_{c}p_{ca}\right)_{\left(\bar p_{ a}\right)}\left(\cfrac{\left(z_{a}^{-1}\,\bar t_a\right)^{\bar p_{a}}}{\bar p_{a}!}
\right)
\,,
\end{align}
where we also used \eqref{sum norm}. In the following section \ref{sum over anti y} we show that
\begin{align}\label{sum bar}
\sum_{\bar p_a} &\left(1+\sum_{c}p_{ca}\right)_{\left(\bar p_{ a}\right)}
\left(\cfrac{\left(z_{a}^{-1}\,\bar t_a\right)^{\bar p_{a}}}{\bar p_{a}!}\right)
=\left(
\frac{1}{1-z_{a}^{-1}\,\bar t_a}
\right)^{1+\sum\limits_{c}p_{ca}}\,.
\end{align}
We impose absolute convergence of the sums on the LHS, which ensures (by Fubini's theorem) that we can swap the sum and integral symbols in \eqref{midstep2}. Using \eqref{sum bar} in \eqref{midstep2}, we can write $F^{[n]}$ as
\begin{align}\label{midstep3}
F^{[n]}=
&\left(\prod_{a}\sum_{\vec p_a}
\oint_{\mathcal{C}_{a}}\frac{dz_{a}}{2\pi iz_{a}}\right)\prod_a\,
\cfrac{\exp\left(z_{a}t_a\right)}{1-z_{a,i}^{-1}\,\bar t_a}
\,
\left(\sum_{c}p_{ca}\right)!\prod_{b}\cfrac{1}{p_{ba}!}\left(\frac{z_{b}x_{ba}z_{a}^{-1}}{1-z_{a}^{-1}\,\bar t_a}\right)^{p_{ba}}\,.
\end{align}
Now we just have to compute the $p_{ab}$ sums. In section \ref{sum over ab,alpha} we show that 
\begin{align}\label{sum bif text}
\left(\prod_b\sum_{p_{ba}}\right)&
\left(\sum_{b }p_{ba}\right)!\prod_{b}\cfrac{1}{p_{ba}!}\left(\frac{z_{b}x_{ba}z_{a}^{-1}}{1-z_{a}^{-1}\,\bar t_a}\right)^{p_{ba}}
=\frac{1-z_{a}^{-1}\,\bar t_a}{1-z_{a}^{-1}\left(\bar t_a+\sum_{b}z_{b}\,x_{ba}\right)}\,,
\end{align}
where again we impose the absolute convergence of all the sums on the LHS, for the same reason just discussed. Eq. \eqref{midstep3} has now become
\begin{align}\label{midstep4 app}
F^{[n]}(\{x_{ab}\},\{t_a\},\{\bar t_a\})=
&\left(\prod_{a}
\oint_{\mathcal{C}_{a}}\frac{dz_{a}}{2\pi iz_{a}}\right)\prod_a\,
\cfrac{\exp\left(z_{a}t_a\right)}
{1-z_{a}^{-1}\left(\bar t_a+\sum_{b}z_{b}\,x_{ba}\right)}\,.
\end{align}
We can rewrite the latter equation more compactly as
\begin{align}
\label{G fin app}
F^{[n]}(\{x_{ab}\},\{t_a\},\{\bar t_a\})=
\left(\prod_{a}\oint_{\mathcal{C}_{a}}\frac{dz_{a}}{2\pi i}\right)
\prod_a\,
I_{a}(\vec z;\vec x_{a},t_{a},\bar t_a)\,,
\end{align}
where \(\vec{z}=(z_1,z_2,...,z_n)\), $n$ being the number of nodes of the quiver, \(\vec{x}_a=\cup_b\{x_{ba}\}\) and
\begin{align}
I_{a}(\vec z;\vec x_{a},t_{a},\bar t_a)=
\cfrac{\exp\left(z_{a}t_a\right)}
{z_{a}-\left(\bar t_a+\sum_{b}z_{b}\,x_{ba}\right)}\,.
\end{align}
Eq. \eqref{F} is thus obtained.


\subsubsection{Summing over $\bar p_{a}$}
\label{sum over anti y}
We want to prove eq \eqref{sum bar}
\begin{align}\label{sum bar copy}
\sum_{\bar p_a} &\left(1+\sum_{c}p_{ca}\right)_{\left(\bar p_{ a}\right)}
\left(\cfrac{\left(z_{a}^{-1}\,\bar t_a\right)^{\bar p_{a}}}{\bar p_{a}!}\right)
=\left(
\frac{1}{1-z_{a}^{-1}\,\bar t_a}
\right)^{1+\sum\limits_{c}p_{ca}}\,,
\end{align}
for any node \(a\) of the quiver. We also have to take care about the convergence of all the sums on the LHS of this equation. These $z_a$ variables will eventually be integrated over  closed curves $\C_a $ in the complex plane, which we will use to compute the contour integrals in \eqref{midstep2} through residues theorem. As discussed in the previous section, we require absolute convergence of the sums on the LHS,
\begin{align}\label{fub0}
\sum_{\bar p_a} &\left(1+\sum_{c}p_{ca}\right)_{\left(\bar p_{ a}\right)}
\left|\cfrac{\left(z_{a}^{-1}\,\bar t_a\right)^{\bar p_{a}}}{\bar p_{a}!}\right|
<\infty
\,.
\end{align}
Throughout this section we will therefore restrict to the $z_a$ that satisfy this constraint. With the mappings
\begin{align}\label{not for sum over bar}
&x\rightarrow1+\sum_{c}p_{ca}\\[2mm]
&z\rightarrow z_{a}^{-1}\,\bar t_a%
\end{align}
the equality \eqref{sum bar copy} reads
\begin{align}\label{sum over bar y simple}
\sum_{p}\,(x)_{\left(p
\right)}\frac{z^p}{p!}=\frac{1}{\left(1- z\right)^x}\,.
\end{align}
This is a known identity, and can be derived with the chain of equalities
\begin{align}\label{mtc formula}
\frac{1}{(1-z)^x}=\sum_{p=0}^\infty\left(p+x-1\atop p\right)z^p=
\sum_{p=0}^\infty\frac{(p+x-1)!}{(x-1)!\,p!}\, z^p=
\sum_{p=0}^\infty\frac{(x)_{(p)}}{p!}\,z^p\,.
\end{align}
The first step above holds only when \(|z|<1\). Our proposition is thus proven.

\subsubsection{Summing over $p_{ab}$}
\label{sum over ab,alpha}
We want now to prove \eqref{sum bif text}, for each node \(a\) of the quiver:
\begin{align}\label{sum over bif copy}
\left(\prod_b\sum_{p_{ba}}\right)&
\left(\sum_{b }p_{b a}\right)!\prod_{b}\cfrac{1}{p_{ba}!}\left(\frac{z_{b}x_{ba}z_{a}^{-1}}{1-z_{a}^{-1}\,\bar t_a}\right)^{p_{ba}}
=\frac{1-z_{a}^{-1}\,\bar t_a}{1-z_{a}^{-1}\left(\bar t_a+\sum_{b}z_{b}\,x_{ba}\right)}\,.
\end{align}
As in the previous section, we work in a region $ {\mathcal U}_\text{{a.c}}  \subset \mathbb{C}^n $ (parametrized by  $z_1 , \cdots , z_n $) where the sums converge absolutely:
\begin{align}\label{fub0-bif}
\left(\prod_b\sum_{p_{ba}}\right)&
\left(\sum_{b }p_{b a}\right)!\prod_{b}\cfrac{1}{p_{ba}!}\left|\frac{z_{b}x_{ba}z_{a}^{-1}}{1-z_{a}^{-1}\,\bar t_a}\right|^{p_{ba}}
<\infty
\,.
\end{align}

Let us then prove the simpler identity
\begin{align}\label{sum over bif copy simple 2}
\sum_{\vec p}
\left(\sum_{b }p_{b }\right)!
\prod_{b}
\frac{1}{p_{b}!}\, \left(\frac{z_{b}}{1-y}\right)^{p_{b}}=
\frac{1-y}{1-\left(y+\sum\limits _{b}z_{b}\right)}
\,,
\end{align}
with \(\vec p=\cup_{b}\{p_{b}\}\), which turns into \eqref{sum over bif copy} through the mappings
\begin{align}\label{not for sum bifs 2}
 z_{b}\rightarrow z_{b}x_{ba}\,z_{a}^{-1}\,,\qquad
 y\rightarrow z_{a}^{-1}\,\bar t_a\,,\qquad
 p_{b}\rightarrow p_{ba}\,.
\end{align}
Similarly, the condition for absolute convergence \eqref{fub0-bif} becomes
\begin{align}\label{fubs}
\sum_{\vec p}
\left(\sum_{b }p_{b }\right)!
\prod_{b}
\frac{1}{p_{b}!}\, \left|\frac{z_{b}}{1-y}\right|^{p_{b}}<\infty
\,.
\end{align}

We will prove \eqref{sum over bif copy simple 2} twice, starting from its right hand side, by choosing two different ways of factorising the ratio 
\begin{align}
\frac{1-y}{1-\left(y+\sum\limits _{b}z_{b}\right)}=\frac{1-y}{(1-z_{a})-\left(y+\sum\limits _{b\neq a}z_{b}\right)}\,.
\end{align}
In the first one we will factor out the term \((1-y)/(1-z_a)\) and in the second one the term \((1-y)/(1-(y+\sum _{b\neq c}z_{b}))\). We will then expand in power series the remaining part of each expression, to obtain two different power expansions. The upshot is that we will obtain two different sets of constraints for the convergence of the power series. Both sets of constraints will hold in the region of absolute convergence \eqref{fub0-bif}, and they will determine the pole prescription for the contour integrals in \eqref{F}.

\subsubsection*{First factorisation}

We start from the RHS of eq. \eqref{sum over bif copy simple 2}. We are going to factor out the term \((1-y)/(1-z_a)\) and expand in a power series the remaining part of the expression. Let us then write
\begin{align}\label{2nd method steps}
\frac{1-y}{1-\left(y+\sum\limits _{b}z_{b}\right)}=
\frac{1-y}{1- z_{a}}\cfrac{1}{1-\left(
\cfrac{y+\sum_{b\neq a}z_{b}}{1- z_{a}}
\right)}\,,
\end{align}
and let us expand the second factor on the RHS above to get
\begin{align}\label{2mtd step 2}
\frac{1-y}{1-\left(y+\sum\limits _{b}z_{b}\right)}=
\frac{1-y}{1- z_{a}}
\,\sum_{n=0}^\infty \left(
\cfrac{y+\sum_{b\neq a}z_{b}}{1- z_{a}}
\right)^n\,,
\end{align}
with the constraint
\begin{align}\label{2mtd cstr raw}
\left|
\cfrac{y+\sum_{b\neq a}z_{b}}{1- z_{a}}
\right|<1\,.
\end{align}
We now rewrite eq. \eqref{2mtd step 2} as
\begin{align}\label{2mtd two pieces}
\frac{1-y}{1-\left(y+\sum\limits _{b}z_{b}\right)}=
(1-y)\,\sum_{n=0}^\infty \left(y+\sum_{b\neq a}z_{b}\right)^n
\cfrac{1}{\left(1- z_{a}\right)^{n+1}}\,,
\end{align}
in order to expand the two terms $\left(y+\sum_{b\neq a}z_{b}\right)^n$ and $\left(1- z_{a}\right)^{-(n+1)}$ separately. For the first one we get
\begin{align}\label{1st piece in 2 mtd}
\left(y+\sum_{b\neq a}z_{b}\right)^n=
\left(\prod_{b\neq a}\,\sum_{p_{b}=0}^\infty\right)\,\sum_{p_y=0}^\infty\,\frac{y^{p_y}}{p_y!}\,
n!\,
\delta\left(
n-p_y-\sum_{b \neq a}p_{b }
\right)
\prod_{b\neq a}\,
\frac{ z_{b}^{p_{b}}}{p_{b}!}\,,
\end{align}
while for the second one, using eq. \eqref{mtc formula}, we obtain
\begin{align}\label{2nd piece in 2 mtd}
\cfrac{1}{\left(1- z_{a}\right)^{n+1}}&=
\sum_{p_{a}=0}^\infty
\frac{\left(p_{a}+n\right)!}{n!}\,\frac{z_{a}^{p_{a}}}{p_{a}!}\,.
\end{align}
The last equality is valid for $|z_{a}|<1$. Inserting eqs. \eqref{1st piece in 2 mtd} and \eqref{2nd piece in 2 mtd} into eq. \eqref{2mtd two pieces}, and rearranging the order of the sums\footnote{Since we are only considering $\{z_a\}$ variables that satisfy absolute convergence condition \eqref{fubs}, this is a legitimate operation.} to let the sum over \(n\) act first we get
\begin{align}\label{2mtd mid 2}
\frac{1-y}{1-\left(y+\sum\limits _{b}z_{b}\right)}&=
(1-y)\,\sum_{n=0}^\infty 	 
\sum_{\vec p}\sum_{p_y=0}^\infty\frac{y^{p_y}}{p_y!}\,\delta\left(
n-p_y-\sum_{b \neq a}p_{b }
\right)
\left(p_{a}+n\right)!
\prod_{b}\frac{z_{b}^{p_{b}}}{{p_{b}!}}\nn
&=
(1-y)\,
\sum_{\vec p}\sum_{p_y=0}^\infty\frac{y^{p_y}}{p_y!}\,
\left(p_y+\sum_{b }p_{b }\right)!
\prod_{b}\frac{z_{b}^{p_{b}}}{{p_{b}!}}\,.
\end{align}
Now, since
\begin{align}\label{2mtd sum mid}
\sum_{n=0}^\infty \frac{(a+n)!}{n!}y^n=a!\,\frac{1}{(1-y)^{1+a}}\,,\qquad |y|<1
\end{align}
we can sum over $p_y$ in the last line of eq. \eqref{2mtd mid 2} to obtain
\begin{align}\label{2mtd final}
\frac{1-y}{1-\left(y+\sum\limits _{b}z_{b}\right)}&=
(1-y)\,
\sum_{\vec p}\,
\frac{\left(\sum_{b }p_{b }\right)!}
{(1-y)^{1+\sum_{b }p_{b }}}
\prod_{b}\frac{z_{b}^{p_{b}}}{{p_{b}!}}=
\sum_{\vec p}\,
\frac{\left(\sum_{b }p_{b }\right)!}
{\prod_{b }(1-y)^{p_{b }}}
\prod_{b}\frac{z_{b}^{p_{b}}}{{p_{b}!}}\nn
&=
\sum_{\vec p}\,
\left(\sum_{b }p_{b }\right)!
\prod_{b}
\frac{1}{{p_{b}!}}
\left(\frac{z_{b}}{1-y}\right)^{p_{b}}\,,
\end{align}
together with the constraint 
\begin{align}\label{2mtd cstr y}
|y|<1\,.
\end{align}
Eq. \eqref{2mtd final} is exactly eq. \eqref{sum over bif copy simple 2}, which becomes our initial proposition \eqref{sum over bif copy} through the substitutions \eqref{not for sum bifs 2}. In the steps presented above, we got three constraints:
\begin{align}\label{2mtd constr simp}
\left\{\left|
\cfrac{y+\sum_{b\neq a}z_{b}}{1- z_{a}}
\right|<1\,\right\}
\,,\quad\,
\left\{\left|
 z_{a}
\right|<1\right\}
\,,\quad\,
\left\{|y|<1\right\}\,.
\end{align}
The first one becomes, through the substitutions \eqref{not for sum bifs 2},
\begin{align}\label{2mtd constr almost}
&\left\{\left|
\cfrac{z_{a}^{-1}\,\bar t_a
+\sum\limits_{b\neq a}z_{b}x_{ba}\,z_{a}^{-1}}{1-x_{aa}}
\right|<1\,\right\}
\,,
\end{align}
which we can also write as the set $\K_a^+$
\begin{align}\label{2mtd constr}
\K_a^+=&\left\{z_{a}\in\mathbb{C}\,\text{s.t.}\,|z_{a}|>\left|
\cfrac{\bar t_a
+\sum\limits_{b\neq a}z_{b}x_{ba}}{1-x_{aa}}
\right|\,\right\}
\,.
\end{align}
We stress that the set of \(\cup_a\{z_a\}\) that satisfy the latter constraint \emph{includes}
the region of absolute convergence  $ {\mathcal U}_\text{{a.c}}  \subset \mathbb{C}^n $. 
In this region $ {\mathcal U}_\text{{a.c} } $,  the exchanges of orders of  summation 
we performed earlier are valid by the Fubini's theorem.  

\subsubsection*{Second factorisation}
We will now show \eqref{sum over bif copy} in a different way, again starting from the RHS of eq. \eqref{sum over bif copy simple 2}. This time we factor out the term \((1-y)/(1-(y+\sum _{b\neq c}z_{b}))\), to expand in a power series the remaining part of the expression. Let us then begin by writing
\begin{align}
\frac{1-y}{1-\left(y+\sum\limits _{b}z_{b}\right)}&=
\frac{1-y}{\left(1-\left(y+\sum\limits _{b\neq c}z_{b}\right)\right)- z_{c}}\nn
&=
\frac{1-y}{1-\left(y+\sum\limits _{b\neq c}z_{b}\right)}\,
\cfrac{1}{1-\cfrac{ z_{c}}{1-\left(y+\sum\limits _{b\neq c}z_{b}\right)}}\,.
\end{align}
Now we expand the second term in the line above in power series, to get
\begin{align}
\frac{1-y}{1-\left(y+\sum\limits _{b}z_{b}\right)}&=
\frac{1-y}{1-\left(y+\sum\limits _{b\neq c}z_{b}\right)}\,
\sum_{n=0}^\infty
\left(
\cfrac{ z_{c}}{1-\left(y+\sum\limits _{b\neq c}z_{b}\right)}
\right)^n\,,
\end{align}
along with the constraint
\begin{align}
\left|
\cfrac{ z_{c}}{1-\left(y+\sum\limits _{b\neq c}z_{b}\right)}
\right|<1\,.
\end{align}
All these steps are similar to the ones in eqs. \eqref{sum over bif copy simple 2}-\eqref{2mtd cstr raw}. Proceeding in the same fashion we first write
\begin{align}\label{3mtd mid}
\frac{1-y}{1-\left(y+\sum\limits _{b}z_{b}\right)}&=
(1-y)\,
\sum_{n=0}^\infty
 z_{c}^n
\frac{1}
{\left(1-\left(y+\sum\limits _{b\neq c}z_{b}\right)\right)^{n+1}}
\,.
\end{align}
Then we expand the rational part of the RHS in power series, rearranging the order of the sums in such a way that the sum over \(k\) acts first, to get
\begin{align}\label{2pc 3mt}
&\frac{1}
{\left(1-\left(y+\sum\limits _{b\neq c}z_{b}\right)\right)^{n+1}}=
\sum_{k=0}^\infty\frac{(k+n)!}{k!\,n!}\,\left(y+\sum_{b\neq c}z_{b}\right)^k\nn
&\qquad\qquad\qquad=
\sum_{k=0}^\infty\frac{(k+n)!}{k!\,n!}\,
\left(\prod_{b\neq c}\sum_{p_{b}=0}^\infty\right)\,\sum_{p_y=0}^\infty\,
\frac{y^{p_y}}{p_y!}\,
\delta\left(
k-p_y-\sum_{b \neq c }p_{b }
\right)\,k!
\prod_{b\neq c}
\frac{z_{b}^{p_{b}}}{p_{b}!}\nn
&\qquad\qquad\qquad=
\left(\prod_{b\neq c}\sum_{p_{b}=0}^\infty\right)\,\sum_{p_y=0}^\infty\,\frac{y^{p_y}}{p_{y}!}\,
\frac{\left(n+p_y+\sum_{b \neq c }p_{b }\right)!}{n!}
\prod_{b\neq c}\frac{z_{b}^{p_{b}}}{p_{b}!}\,, 
\end{align}
together with the constraint (coming from the first equality)
\begin{align}
\left|y+\sum\limits _{b\neq c}z_{b}\right|<1\,.
\end{align}
Using eq. \eqref{2pc 3mt} in \eqref{3mtd mid} we get
\begin{align}\label{3mtd mid 2}
\frac{1-y}{1-\left(y+\sum\limits _{b}z_{b}\right)}&=
(1-y)\,
\sum_{n=0}^\infty\,
\sum_{\vec p}\,\sum_{p_y=0}^\infty\,\frac{y^{p_y}}{p_{y}!}\,\delta\!\left(
p_{c}-n
\right)\!\!
\left(n+p_y+\sum_{b \neq c }p_{b }\right)!
\prod_{b}\frac{z_{b}^{p_{b}}}{p_{b}!}\nn
&=
(1-y)\,
\sum_{\vec p}\,\sum_{p_y=0}^\infty\,\frac{y^{p_y}}{p_{y}!}\,
\left(p_y+\sum_{b }p_{b }\right)!
\prod_{b}\frac{z_{b}^{p_{b}}}{p_{b}!}\,,
\end{align}
where we again rearranged the order of the sums to let the sum over \(n\) act first. Now this equation is identical to eq. \eqref{2mtd mid 2}, and we know that if we impose the constraint
\begin{align}
|y|<1
\end{align}
\eqref{3mtd mid 2} is enough to prove \eqref{sum over bif copy}. Our initial proposition is again proven. 

In the derivation we got, among others, the constraint
\begin{align}\label{k1 fec}
\left\{\left|\cfrac{ z_{c}}{1-\left(y+\sum\limits _{b\neq c}z_{b}\right)} \right|<1\right\}\,,
\end{align}
which with the substitutions \eqref{not for sum bifs 2} becomes
\begin{align}\label{3mtd constr almost}
&\left\{\left|\cfrac{ z_{c}x_{ca}\,z_{a}^{-1}}{1-\left(z_{a}^{-1}\,\bar t_a+\sum\limits _{b\neq c}z_{b}x_{ba}\,z_{a}^{-1}\right)} \right|<1\right\}\,.
\end{align}
The same quantity can also be described in terms of the set $k^-_{c,a}$, defined as
\begin{align}\label{3mtd constr}
k^-_{c,a}=&\left\{z_{c}\in\mathbb{C}\,\text{s.t.}\,\left|z_{c} \right|<
\left|\cfrac{z_{a}-\left(\bar t_a+\sum\limits _{b\neq c}z_{b}x_{ba}\right)}{ x_{ca}}\right|
\right\}\,.
\end{align}
This constraint has to be interpreted in the same manner as the one in \eqref{2mtd constr}: $k^-_{c,a}$ \emph{includes} the region $ \mathcal{U}_{\text{a.c} } $ that makes \eqref{sum over bif copy simple 2} absolutely convergent. 

Fixing node $a$, the derivation above holds for any $c\neq a$. This means that we can obtain constraints like the one in \eqref{k1 fec} for all the nodes $c\neq a$ of the quiver, that we can impose \emph{all at the same time}. We can then define the quantity
\begin{align}\label{exclude}
\mathcal K^-_a=\bigcap_{c\neq a}\,\,k^-_{c,a}=
\bigcap_{c\neq a}\,\,
\left\{z_{c}\in\mathbb{C}\,\text{s.t.}\,\left|z_{c} \right|<
\left|\cfrac{z_{c}-\left(\bar t_a+\sum\limits _{b\neq c}z_{b}x_{ba}\right)}{ x_{ca}}\right|
\right\}
\,.
\end{align}
Just like $\K_a^+$ (eq. \eqref{2mtd constr}), this constraint will be of central importance when we will compute the integrals in \eqref{F}: the set $\K_a$, defined as
\begin{align}\label{cstr tot a}
\K_a&=\K^+_a\,\cap\,\K^-_a\nn
&=
\left\{z_{a}\in\mathbb{C}\,\text{s.t.}\,|z_{a}|>\left|
\cfrac{\bar t_a
+\sum\limits_{b\neq a}z_{b}x_{ba}}{1-x_{aa}}
\right|\,\right\}\nn
&\qquad\qquad\bigcap_{c\neq a}\,\,
\left\{z_{c}\in\mathbb{C}\,\text{s.t.}\,\left|z_{c} \right|<
\left|\cfrac{z_{c}-\left(\bar t_a+\sum\limits _{b\neq c}z_{b}x_{ba}\right)}{ x_{ca}}\right|
\right\}\,,
\end{align}
will in fact determine which poles are to be included by the contour \(\C_{a}\).


\section{Residues and constraints}\label{r&c}
In this appendix we will present the rule for including/excluding poles when calculating the contour integrals in eq. \eqref{F}, that is
\begin{align}\label{F copy}
 F^{[n]}(\{x_{ab}\},\{t_a\},\{\bar t_a\})=
\left(\prod_{a}\oint_{\mathcal{C}_{a}}\frac{dz_{a}}{2\pi i}\right)
\prod_a\,
 I_a(\vec z;\vec x_{a},t_a,\bar t_a)\,.
\end{align}
We recall that the integrands \(I_a\) are defined by
\begin{align}\label{I in  hat F copy}
I_a(\vec z;\vec x_{a},t_a,\bar t_a)=\cfrac{\exp\left(z_{a}\,t_{a}\right)}
{ z_{a}-\left(\bar t_{a}+\sum_{b}z_{b}\,x_{ba}\right)}\,.
\end{align}
The prescription is that we have to pick only the \(z_a\) pole coming from the \(I_a\) factor in the integrand of \eqref{F copy}, for each \(a\). Let us show how this rule arises.

If the quiver under study has $n$ nodes, each \(I_a\) will have $n$ poles, one for each $z$ variable. Explicitly
\begin{align}\label{defofpo}
 z_{a}^*=\cfrac{\bar t_a
+\sum\limits_{b\neq a}z_{b}x_{ba}}{1- x_{aa}}\,,\qquad\quad
 z_{c}^*=\cfrac{z_{a}-\left(\bar t_a+\sum\limits _{b\neq c}z_{b}x_{ba}\right)}{ x_{ca}}\,,
\qquad \forall c\neq a
\,.
\end{align}
From appendix \ref{sum over ab,alpha} we know however that we have to restrict to the set of \(\cup_a\{z_a\}\) that belongs to the intersection of the set \eqref{cstr tot a}
\begin{align}\label{cstr tot}
\K_a&=
\left\{z_{a}\in\mathbb{C}\,\text{s.t.}\,|z_{a}|>\left|
\cfrac{\bar t_a
+\sum\limits_{b\neq a}z_{b}x_{ba}}{1-x_{aa}}
\right|\,\right\}\nn
&\qquad\qquad\bigcap_{c\neq a}\,\,
\left\{z_{c}\in\mathbb{C}\,\text{s.t.}\,\left|z_{c} \right|<
\left|\cfrac{z_{a}-\left(\bar t_a+\sum\limits _{b\neq c}z_{b}x_{ba}\right)}{ x_{ca}}\right|
\right\}\nn
&=\left\{z_{a}\in\mathbb{C}\,\text{s.t.}\,|z_{a}|>\left|
z_{a}^*
\right|\,\right\}\bigcap_{c\neq a}\,\,
\left\{z_{c}\in\mathbb{C}\,\text{s.t.}\,\left|z_{c} \right|<
\left|z_{c}^*\right|
\right\}
\end{align}
with the set of \(\cup_a\{z_a\}\) satisfying the condition of absolute convergence \eqref{fub0-bif}:
\begin{align}\label{fub0-bif copy}
\left(\prod_b\sum_{p_{ba}}\right)&
\left(\sum_{b }p_{b a}\right)!\prod_{b}\cfrac{1}{p_{ba}!}\left|\frac{z_{b}x_{ba}z_{a}^{-1}}{1-z_{a}^{-1}\,\bar t_a}\right|^{p_{ba}}
<\infty
\,.
\end{align}
In the same appendix, we also argued that the former constraint \eqref{cstr tot} includes the latter \eqref{fub0-bif copy}: this means that if we impose \eqref{fub0-bif copy}, then \eqref{cstr tot} is also valid. But this is telling us that for \emph{any} $ I_{a}$ we only have to pick up the pole relative to the $z_{a}$ variable, and discard all the others. However this is a prescription which holds only before we perform any integration: after we do so, the poles for each of the remaining $z$ variables will have a different equation. This problem is anyway easily overcome: the constraint in \eqref{cstr tot} comes from the sums in \eqref{midstep3} that contribute to the $I_{a}$ piece of the integrand \eqref{F copy} alone. So in principle we could have chosen any \(a\) in \eqref{midstep3}, performed the sums over \(\cup_{b}p_{ba}\) only, got the $ I_{a}$ term together with the constraint above, inferred from the previous discussion that only the $z_{a}$ pole has to be picked up and finally compute the $z_{a}$ integration (all the other $z_{a}$ appearing in \eqref{midstep3} are regular and have no pole). Let us then imagine to be in such a situation, and for concreteness say that we have chosen to integrate over $z_{1}$. After the $z_{1}$ integration has been done, we are left with $n-1$ 
sums ($n$ being the number of nodes in the quiver) of the form already discussed in appendix \ref{sum over ab,alpha}, that is
\begin{align}\label{sum over bif copyX2}
\left(\prod_b\sum_{p_{ba}}\right)&
\left(\sum_{b }p_{b a}\right)!\prod_{b}\cfrac{1}{p_{ba}!}\left(\frac{z_{b}x_{ba}z_{a}^{-1}}{1-z_{a}^{-1}\,\bar t_a}\right)^{p_{ba}}
=\frac{1-z_{a}^{-1}\,\bar t_a}{1-z_{a}^{-1}\left(\bar t_a+\sum_{b}z_{b}\,x_{ba}\right)}\,,\qquad a\neq 1\,,
\end{align}
where now every $z_{1}$ has to be substituted with its pole equation, which will be of the form
\begin{align}
z_{1}\rightarrow z_{1}^*(z_{2},z_{3},...,z_{n};\vec{x})=\sum_{c>1}z_{c}\,a_{c}
\,,
\end{align}
for some coefficients $a_{c}$. As usual, we impose absolute convergence of the sums on the LHS of \eqref{sum over bif copyX2}. Adapting the notation of appendix \ref{sum over ab,alpha} to the present case, let us work with the simpler identity
\begin{align}\label{sum over bif copy simple tr}
\left(\prod_{b}\sum_{p_{b}=0}^\infty\right)\,\left(\sum_{b }p_{b }\right)!
\,\left(\prod_{b> 1}\frac{z_{b}^{p_{b}}}{p_{b}!}\right) \frac{(z_{1}^{*})^{p_{1}}}{p_{1}!}=
\frac{1}{1-\sum\limits _{b>1}z_{b}- z_{1}^*}\,,
\end{align}
which becomes \eqref{sum over bif copyX2} through the substitutions 
\begin{align}\label{not for sum bifs copy}
 z_{b}\rightarrow\frac{z_{b}x_{ba}\,z_{a}^{-1}}{1-z_{a}^{-1}\left(\bar t_a\right)}\,,\qquad\qquad   p_{b}\rightarrow p_{ba}\,.
\end{align}
Note that now we have
\begin{align}
z_{1}^{*}&\rightarrow\frac{z_{1}^*\,x_{1a}\,z_{a}^{-1}}{1-z_{a}^{-1}\,\bar t_a}=
\sum_{c>1}\frac{z_{c}\,a_{c}\,x_{1a}\,z_{a}^{-1}}{1-z_{a}^{-1}\,\bar t_a}=
\sum_{c>1}\frac{z_{c}\,\tilde x_{ca}\,z_{a}^{-1}}{1-z_{a}^{-1}\,\bar t_a}=
\sum_{c>1}\,\tilde z_{c}
\,,
\end{align}
in which we defined
\begin{align}\label{tilded xs in app}
 \tilde x_{ca}=a_{c}\,x_{1a}\,,\qquad  \tilde z_{c}=\frac{z_{c}\,\tilde x_{ca}\,z_{a}^{-1}}{1-z_{a}^{-1}\,\bar t_a}
\,.
\end{align}
Consider now the LHS of \eqref{sum over bif copy simple tr} and write it as 
\begin{align}
&\left(\prod_{b}\sum_{p_{b}=0}^\infty\right)\,\left(\sum_{b }p_{b }\right)!
\,\left(\prod_{b> 1}\frac{z_{b}^{p_{b}}}{p_{b}!}\right) \frac{(z_{1}^{*})^{p_{1}}}{p_{1}!}=
\left(\prod_{b}\sum_{p_{b}=0}^\infty\right)\,\left(\sum_{b }p_{b }\right)!
\,\left(\prod_{b> 1}\frac{z_{b}^{p_{b}}}{p_{b}!}\right)\, \frac{1}{p_{1}!}\left(\sum_{c>1}\,\tilde z_{c}\right)^{p_{1}}\nn
&\qquad=
\left(\prod_{b}\sum_{p_{b}=0}^\infty\right)\,\left(\sum_{b }p_{b }\right)!
\,\left(\prod_{b> 1}\frac{z_{b}^{p_{b}}}{p_{b}!}\right)\left(
\prod_{c>1}\,\sum_{q_{c}=0}^\infty
\right)
\,\delta\left(p_{1}-\sum_{c >1}q_{c }\right)\
\prod_{c>1}\,\frac{\tilde z_{c}^{q_{c}}}{q_{c}!}\,.
\end{align}
After summing over $p_{1}$ we obtain
\begin{align}
\left(\prod_{b}\sum_{p_{b}=0}^\infty\right)\,&\left(\sum_{b }p_{b }\right)!
\,\left(\prod_{b> 1}\frac{z_{b}^{p_{b}}}{p_{b}!}\right) \frac{(z_{1}^{*})^{p_{1}}}{p_{1}!}\nn
&=
\left(\prod_{b>1}\sum_{p_{b}=0\atop q_{b}=0}^\infty\right)\,\left(\sum_{b >1}p_{b }+ \sum_{c >1}q_{c } \right)!
\,\left(\prod_{b>1}\frac{z_{b}^{p_{b}}}{p_{b}!}\right)
\,\prod_{c>1}\,\frac{\tilde z_{c}^{q_{c}}}{q_{c}!}\nn
&=
\left(\prod_{b>1}\sum_{p_{b}=0\atop q_{b}=0}^\infty\right)\,\left(\sum_{b >1}(p_{b }+q_{b }) \right)!
\,\left(\prod_{b>1}\frac{z_{b}^{p_{b}}}{p_{b}!}\,\frac{\tilde z_{b}^{q_{b}}}{q_{b}!}\right)\,,
\end{align}
where the last equality follows from noticing that the $b$ and $c$ labels in the products and sums run over the same set of variables. Now multiplying the far right hand side of the above equation by $\prod_{b> 1}\frac{(p_{b}+q_{b})!}{(p_{b}+q_{b})!}=1$ 
and inserting the identity 
\begin{align}
\prod_{b> 1}\sum_{\lambda_{b}=0}^\infty \delta\left(\lambda_{b}-p_{b}-q_{b}\right)=1\,,
\end{align}
we get, exploiting the support of the delta function
\begin{align}
&\left(\prod_{b}\sum_{p_{b}=0}^\infty\right)\,\left(\sum_{b }p_{b }\right)!
\,\left(\prod_{b> 1}\frac{z_{b}^{p_{b}}}{p_{b}!}\right) \frac{(z_{1}^{*})^{p_{1}}}{p_{1}!}\nn
&\quad=
\left(\prod_{b>1}\sum_{p_{b}=0\atop q_{b}=0}^\infty\,\sum_{\lambda_{b}=0}^\infty\right)
\,\left(\sum_{b >1}\lambda_{b }\right)!
\left(\prod_{b>1}\frac{1}{\lambda_{b}!}\right)\,\prod_{b>1}\,\delta\left(\lambda_{b}-p_{b}-q_{b}\right)\frac{\lambda_{b}!}{p_{b}!\,q_{b}!}\,z_{b}^{p_{b}}\,\tilde z_{b}^{q_{b}}\nn
&\quad=
\left(\prod_{b>1}\,\sum_{\lambda_{b}=0}^\infty\right)
\,\left(\sum_{b >1}\lambda_{b }\right)!\,\prod_{b>1}\,\frac{1}{\lambda_{b}!}\,\left[\sum_{p_{b}=0\atop q_{b}=0}^\infty\,\delta\left(\lambda_{b}-p_{b}-q_{b}\right)\frac{\lambda_{b}!}{p_{b}!\,q_{b}!}\,z_{b}^{p_{b}}\,\tilde z_{b}^{q_{b}}\right]\,.
\end{align}
The quantity inside the square bracket is of the form
\begin{align}
\sum_{k1,k2=0}^\infty\,\delta\left( n-k_1-k_2\right)\,\frac{n!}{k_1!\,k_2!}\,a^{k_1}\,b^{k_2}=(a+b)^n\,,
\end{align}
so that we eventually have, relabelling $\lambda_{b}\rightarrow p_{b}$
\begin{align}
\left(\prod_{b}\sum_{p_{b}=0}^\infty\right)&\,\left(\sum_{b }p_{b }\right)!
\,\left(\prod_{b> 1}\frac{z_{b}^{p_{b}}}{p_{b}!}\right) \frac{(z_{1}^{*})^{p_{1}}}{p_{1}!}
=
\left(\prod_{b>1}\,\sum_{p_{b}=0}^\infty\right)
\,\left(\sum_{b >1}p_{b }\right)!\,\prod_{b>1}\,\frac{\left(z_{b}+\tilde z_{b}\right)^{p_{b}}}{p_{b}!}
\end{align}
for the LHS of eq. \eqref{sum over bif copy simple tr}. 

Consider now the RHS of the same formula: it reads
\begin{align}
\frac{1}{1-\sum\limits _{b>1}z_{b}- z_{1}^*}
=
\frac{1}{1-\sum\limits _{b>1}z_{b}-\sum\limits_{c>1}\,\tilde z_{c}}=
\frac{1}{1-\sum\limits _{b>1}(z_{b}+\tilde z_{b})}\,.
\end{align}
Equating the right hand sides of the last two equations we then get
\begin{align}\label{equating app}
\left(\prod_{b>1}\,\sum_{p_{b}=0}^\infty\right)
\,\left(\sum_{b >1}p_{b }\right)!\,\prod_{b>1}\,\frac{\left(z_{b}+\tilde z_{b}\right)^{p_{b}}}{p_{b}!}=
\frac{1}{1-\sum\limits _{b>1}(z_{b}+\tilde z_{b})}\,.
\end{align}
Using the substitutions in \eqref{not for sum bifs copy} and defining the new quantity $\hat x_{ba}\equiv x_{ba}+\tilde x_{ba}$ we immediately obtain
\begin{align}\label{hatted xs in app}
z_{b}+\tilde z_{b}\rightarrow\frac{z_{b}x_{ba}\,z_{a}^{-1}}{1-z_{a}^{-1}\,\bar t_a}+\frac{z_{b}\,\tilde x_{ba}\,z_{a}^{-1}}{1-z_{a}^{-1}\,\bar t_a}
=
\frac{z_{b}\,(x_{ba}+\tilde x_{ba})\,z_{a}^{-1}}{1-z_{a}^{-1}\,\bar t_a}
\equiv
\frac{z_{b}\,\hat x_{ba}\,z_{a}^{-1}}{1-z_{a}^{-1}\,\bar t_a}\,,
\end{align}
%
%
%
so that eq. \eqref{equating app} becomes
\begin{align}\label{sums after one integration}
\left(\prod_{b>1}\,\sum_{p_{ba}=0}^\infty\right)
\,\left(\sum_{b >1}p_{b a}\right)!\,&\prod_{b>1}\,
\frac{1}{p_{ba}!}
\left(\frac{z_{b}\,\hat x_{ba}\,z_{a}^{-1}}{1-z_{a}^{-1}\,\bar t_a}\right)^{p_{ba}}\nn
&=\frac{1-z_{a}^{-1}\,\bar t_a}{1-z_{a}^{-1}\left(\bar t_a+\sum\limits_{b>1}z_{b}\,\hat x_{ba}\right)}\,,\qquad a\neq1\,.
\end{align}
This is exactly the equation in \eqref{sum over bif copy} with the substitution $x_{ba}\rightarrow \hat x_{ba}$ and the removal of the first node. We have already proven such an equality in appendix \ref{sum over ab,alpha}, where we have also obtained the set of constraints in \eqref{cstr tot}. This means that the constraints coming from the convergence of the sums on the LHS of \eqref{sums after one integration} can be described by the intersection of the set
\begin{align}\label{cstr tot a after first integration}
\hat \K_a&
=
\left\{z_{a}\in\mathbb{C}\,\text{s.t.}\,|z_{a}|>\left|
\cfrac{\bar t_a
+\sum\limits_{b\neq a,1}z_{b}\hat  x_{ba}}{1-\hat  x_{aa}}
\right|\,\right\}\nn
&\qquad\qquad\bigcap_{c\neq a}\,\,
\left\{z_{c}\in\mathbb{C}\,\text{s.t.}\,\left|z_{c} \right|<
\left|\cfrac{z_{c}-\left(\bar t_a+\sum\limits _{b\neq c,1}z_{b}\hat  x_{ba}\right)}{ \hat  x_{ca}}\right|
\right\}\,,\qquad a\neq 1
\end{align}
with the region in $ \mathbb{ C}^{n-1}$ parametrized by $ \{ z_2 , \cdots , z_n  \}$
satisfying the absolute convergence condition
\begin{align}\label{abs convergence after one integration}
\left(\prod_{b>1}\,\sum_{p_{ba}=0}^\infty\right)
\,\left(\sum_{b >1}p_{b a}\right)!\,&\prod_{b>1}\,
\frac{1}{p_{ba}!}
\left|\frac{z_{b}\,\hat x_{ba}\,z_{a}^{-1}}{1-z_{a}^{-1}\,\bar t_a}\right|^{p_{ba}}
<\infty\,,\qquad a\neq1\,.
\end{align}
%
%
We stress once again that the former includes the latter. Such an intersection gives us a prescription on which poles to include/exclude after one integration has been done: in complete analogy to the situation discussed at the beginning of this section, we find that only the $z_{a}$ pole coming from the $ I_{a}$ term in the integrand of \eqref{F copy} has to be picked up, \(\forall a\, \neq 1\). The steps presented here are trivially generalisable, and they can be redone in the exact same way integration after integration. We can then say that, \emph{at any level of integration}, only the $z_{a}$ pole in the $ I_{a}$ factor has to be enclosed by \(\C_a\) in \eqref{F copy}. This is our pole prescription to perform integrals.


\section{Three node unflavoured quiver example}\label{Three Nodes Matter-Free Example residue way}

In this section we will provide an explicit example of application of the formulae presented in section \ref{The Matter-Free Case} to the three node unflavoured case. Let us start by writing $z_{1}^*$, $z_{2}^*$ and $z_{3}^*$. According to eq. \eqref{def pole j}, the equation for $z_1^*(z_2,z_3;\vec x)$ is obtained by solving for $z_1$ the equation
\begin{align}
I_1^{-1}(z_1,z_2,z_3;\vec x)=z_1-\sum_{b=1}^3z_b\,x_{b,1}=0\,,
\end{align}
that gives
\begin{align}\label{EX3 zi*}
z_1^*(z_2,z_3;\vec x)=\sum_{i>1}z_i\,\frac{x_{i,1}}{1-x_{1,1}}\,.
\end{align}
From \eqref{allpoles} we then have
\begin{align}\label{a_1 ex 3}
a_{i,1}=\frac{x_{i,1}}{1-x_{1,1}}\,.
\end{align}
We now turn to $z_2^*(z_3;\vec x)$, which is obtained by solving for $z_2$ the equation
\begin{align}
I_2^{-1}(z_1^*,z_2,z_3)=z_2-\sum_{b>1}^3z_b\,x_{b,2}-z_1^* x_{1,2}=0\,.
\end{align}
Using \eqref{EX3 zi*} we get
\begin{align}
z_2^*(z_3;\vec x)=\sum_{i>2}z_i\,\frac{\left(x_{i,1}x_{1,2}+x_{i,2}(1-x_{1,1})\right)}{(1-x_{1,1})(1-x_{2,2})-x_{1,2}x_{2,1}}\,,
\end{align}
so that
\begin{align}\label{a_2 ex 3}
a_{i,2}=\frac{\left(x_{i,1}x_{1,2}+x_{i,2}(1-x_{1,1})\right)}{(1-x_{1,1})(1-x_{2,2})-x_{1,2}x_{2,1}}\,.
\end{align}
Finally, $I_3^{-1}(z_1^*,z_2^*,z_3)=0$ is solved by $z_3^*=0$. We can now write down the pole coefficients $\hat a_{i,p}^{[r]}$, which we will need in computing $F_0^{[3]}$. Following the definition given in \eqref{hatted a}, we have
\begin{align}
& \hat a_{i,p}^{[0]}=0\,,\\[3mm]
& \hat a_{i,p}^{[1]}=a_{i,p}+\sum_{\lambda=p+1}^1a_{i,\lambda}^{[1]}\,a_{\lambda,p}=a_{i,p}\,,
\end{align}
and 
\begin{equation}
 \hat a_{i,p}^{[2]}=a_{i,p}+\sum_{\lambda=p+1}^2a_{i,\lambda}^{[2]}\,a_{\lambda,p}=
\left\{
\begin{array}{ll}
a_{i,1}+\hat a_{i,2}^{[2]}a_{2,1}=a_{i,1}+a_{i,2}\,a_{2,1}\quad&\text{if }p=1\\[3mm]
a_{i,p}\quad&\text{if }p>1
\end{array}
\right.\,.
\end{equation}
Using eqs. \eqref{a_1 ex 3} and \eqref{a_2 ex 3}, and noting that 
\begin{align}
1-x_{1,1}=G_{[1]}\,,\qquad\quad(1-x_{1,1})(1-x_{2,2})-x_{1,2}x_{2,1}=G_{[2]}\,,
\end{align}
we can also write
\begin{equation}
 \hat a_{i,p}^{[2]}=
\left\{
\begin{array}{ll}
\displaystyle a_{i,1}+\frac{a_{i,2}\,x_{2,1}}{G_{[1]}}=\frac{x_{i,2}x_{2,1}+x_{i,1}(1-x_{2,2})}{G_{[2]}}\quad&\text{if }p=1\\[3mm]
a_{i,p}\quad&\text{if }p>1
\end{array}
\right.\,.
\end{equation}
By using eq. \eqref{allpoles same z}, 
\begin{align}\label{allpoles same z app EX3}
z_j^{*[r]}=z_j^*(z_{r+1},...,z_n;\vec{x})=\sum_{i>r}z_i\, \hat a_{i,j}^{[r]}\,,\qquad 1\leq j\leq r\,,
\end{align}
we then obtain
\begin{align}
z_{1}^{*[1]}(z_2,z_3,\vec x)=\sum_{i>1}^3z_i\,\hat a_{i,1}^{[1]}=\sum_{i=2,3}\frac{z_i\,x_{i,1}}{G_{[1]}}\,,
\end{align}
and similarly
\begin{align}
&z_{1}^{*[2]}(z_3,\vec x)=\sum_{i>2}^3z_i\,\hat a_{i,1}^{[2]}=z_3\,\frac{x_{3,2}x_{2,1}+x_{3,1}(1-x_{2,2})}{G_{[2]}}\,,\\[3mm]
&z_{2}^{*[2]}(z_3,\vec x)=\sum_{i>2}^3z_i\,\hat a_{i,2}^{[2]}=z_3\,\frac{x_{3,1}x_{1,2}+x_{3,2}(1-x_{1,1})}{G_{[2]}}\,.
\end{align}
Finally, we can compute $F_0^{[3]}$ using formula \eqref{final}. We have
\begin{align}\label{final for 3}
F_0^{[3]}&=
\prod_{i=1}^3H_i(\vec x)
=\prod_{i=1}^{3}\left(1-x_{{{i}},{{i}}}-\sum_{q=1}^{i-1} \hat a_{{i},q}^{[i-1]}\,x_{q,{{i}}}\right)^{-1}\\[3mm]
&=
(1-x_{1,1})^{-1}\left(1-x_{2,2}-\hat a_{2,1}^{[1]}x_{1,2}\right)^{-1}\left(
1-x_{3,3}-\hat a_{3,1}^{[2]}x_{1,3}-\hat a_{3,2}^{[2]}x_{2,3}
\right)^{-1}\nn
&=
(1-x_{1,1})^{-1}\left(1-x_{2,2}-a_{2,1}x_{1,2}\right)^{-1}\left(
1-x_{3,3}-a_{3,1}x_{1,3}-a_{3,2}\left(\frac{x_{2,1}x_{1,3}}{1-x_{1,1}}+x_{2,3}\right)
\right)^{-1}\,.\nonumber
\end{align}
Using the equations for $a_{2,1}$ and $a_{3,1}$ defined in \eqref{a_1 ex 3} and $a_{3,2}$ defined in \eqref{a_2 ex 3}, we get
\begin{align}\label{final for 3 adv}
F_0^{[3]}&=\left((1-x_{1,1})(1-x_{2,2})-x_{1,2}x_{2,1}\right)^{-1}
\left(
1-x_{3,3}-\frac{x_{1,3}x_{3,1}}{1-x_{1,1}}\right.\nn
&\qquad\qquad\qquad\qquad\qquad\qquad
\left.-\frac{x_{3,1}x_{1,2}+x_{3,2}(1-x_{1,1})}{(1-x_{1,1})(1-x_{2,2})-x_{1,2}x_{2,1}}\left(\frac{x_{2,1}x_{1,3}}{1-x_{11}}+x_{23}\right)
\right)^{-1}\nn
&=
\left(
((1-x_{1,1})(1-x_{2,2})-x_{1,2}x_{2,1})\left(1-x_{3,3}-\frac{x_{1,3}x_{3,1}}{1-x_{1,1}}\right)\right.\nn
&\qquad\qquad\qquad\qquad\qquad\qquad
\left.-(x_{3,1}x_{1,2}+x_{3,2}(1-x_{1,1}))\left(\frac{x_{2,1}x_{1,3}}{1-x_{1,1}}+x_{2,3}\right)
\right)^{-1}\nn
&=
\left(
1-x_{1,1}-x_{2,2}-x_{3,3}-x_{1,2}x_{2,1}+x_{1,1}x_{2,2}-x_{1,3}x_{3,1}+x_{1,1}x_{3,3}-x_{2,3}x_{3,2}+x_{2,2}x_{3,3}\right.\nn
&
\left.-x_{1,1}x_{2,2}x_{3,3}+x_{1,1}x_{2,3}x_{3,2}+x_{2,2}x_{1,3}x_{3,1}+x_{3,3}x_{1,2}x_{2,1}-x_{1,2}x_{2,3}x_{3,1}-x_{1,3}x_{3,2}x_{2,1}
\right)^{-1}\,,
\end{align}
which concludes our computation.

\subsection{Permutation formula}\label{three nodes example}
Let us now give an example of the application of formula \eqref{sj formula} in this simple case of a three node unflavoured quiver. We have already computed the correct answer $F_0^{[3]}$ in the previous section, so we can explicitly check that \eqref{sj formula} indeed reproduces the correct result. Let us call the three nodes of the quiver simply 1, 2 and 3.
We can immediately write the simple loops $y_{\sigma^{(i)}}(\{x_{ab}\})$ using eq. \eqref{single cycle perm}:
\begin{equation}
\begin{array}{lll}
 y_{(1)}(\{x_{ab}\})=x_{11}\,,&  y_{(2)}(\{x_{ab}\})=x_{22}\,,&  y_{(3)}(\{x_{ab}\})=x_{33}\,,\\[3mm]
 y_{(12)}(\{x_{ab}\})=x_{12}x_{21}\,,& y_{(13)}(\{x_{ab}\})=x_{13}x_{31}\,,& y_{(23)}(\{x_{ab}\})=x_{23}x_{32}\,,\\[3mm]
 y_{(123)}(\{x_{ab}\})=x_{12}x_{23}x_{31}\qquad\quad &  y_{(132)}(\{x_{ab}\})=x_{13}x_{32}x_{21}\,.
\end{array}
\end{equation}
From these quantities we can construct $y_\sigma(\{x_{ab}\})$, for every \(\sigma\), by using the definition in eq. \eqref{cycle f}:
\begin{align}\label{cycle f copy other}
y_\sigma(\{x_{ab}\})&=(-1)^{c_\sigma}\prod_iy_{\sigma^{(i)}}(\{x_{ab}\})\,.
\end{align}
For example, if we had $\sigma=(12)(3)$, then 
\begin{align}
y_{(12)(3)}(\{x_{ab}\})=(-1)^2\,y_{(12)}(\{x_{ab}\})\,\,y_{(3)}(\{x_{ab}\})=x_{12}x_{21}\,x_{33}\,,
\end{align}
the power 2 in the $-1$ comes from the fact that $\sigma=(12)(3)$ is a product of two cycles. Getting back to our three node quiver example, there are 7 non empty subsets that we can form out of the set $\{1,2,3\}$, namely $\{1\}$, $\{2\}$, $\{3\}$, $\{12\}$, $\{13\}$, $\{23\}$, $\{123\}$. According to eq. \eqref{cycle f} we then have
\begin{subequations}
\begin{align}
&\sum_{\sigma\in\text{Sym}(\{1\})}y_\sigma(\{x_{ab}\})=y_{(1)}(\{x_{ab}\})=-x_{1,1}\,,\\[3mm]
&\sum_{\sigma\in\text{Sym}(\{2\})}y_\sigma(\{x_{ab}\})=y_{(2)}(\{x_{ab}\})=-x_{2,2}\,,\\[3mm]
&\sum_{\sigma\in\text{Sym}(\{3\})}y_\sigma(\{x_{ab}\})=y_{(3)}(\{x_{ab}\})=-x_{3,3}\,,\\[3mm]
&\sum_{\sigma\in\text{Sym}(\{12\})}y_\sigma(\{x_{ab}\})=y_{(1)(2)}(\{x_{ab}\})+y_{(12)}(\{x_{ab}\})=x_{1,1}x_{2,2}-x_{1,2}x_{2,1}\,,\\[3mm]
&\sum_{\sigma\in\text{Sym}(\{13\})}y_\sigma(\{x_{ab}\}))=y_{(1)(3)}(\{x_{ab}\})+y_{(13)}(\{x_{ab}\})=x_{1,1}x_{3,3}-x_{1,3}x_{3,1}\,,\\[3mm]
&\sum_{\sigma\in\text{Sym}(\{23\})}y_\sigma(\{x_{ab}\}))=y_{(2)(3)}(\{x_{ab}\})+y_{(23)}(\{x_{ab}\})=x_{2,2}x_{3,3}-x_{2,3}x_{3,2}\,,\\[3mm]
&\sum_{\sigma\in\text{Sym}(\{123\})}y_\sigma(\{x_{ab}\})=
y_{(1)(2)(3)}(\{x_{ab}\})+y_{(12)(3)}(\{x_{ab}\})+y_{(13)(2)}(\{x_{ab}\})\nn
&\qquad\qquad\qquad\qquad\qquad\qquad\qquad
+y_{(23)(1)}(\{x_{ab}\})+y_{(123)}(\{x_{ab}\})+y_{(132)}(\{x_{ab}\})\nn
&\qquad\qquad\qquad\qquad\quad\,
=
-x_{1,1}x_{2,2}x_{3,3}+x_{1,2}x_{2,1}x_{3,3}+x_{1,3}x_{3,1}x_{2,2}\nn
&\qquad\qquad\qquad\qquad\qquad\qquad
+x_{2,3}x_{3,2}x_{1,1}-x_{1,2}x_{2,3}x_{3,1}-x_{1,3}x_{3,2}x_{2,1}\,.
\end{align}
\end{subequations}
Summing all of the terms above we get
\begin{align}\label{3nodes ex by perms}
F_0^{[3]}&=\left(1-x_{1,1}-x_{2,2}-x_{3,3}-x_{1,2}x_{2,1}+x_{1,1}x_{2,2}-x_{1,3}x_{3,1}+x_{1,1}x_{3,3}-x_{2,3}x_{3,2}+x_{2,2}x_{3,3}\right.\nn
&
\left.-x_{1,1}x_{2,2}x_{3,3}+x_{1,1}x_{2,3}x_{3,2}+x_{2,2}x_{1,3}x_{3,1}+x_{3,3}x_{1,2}x_{2,1}-x_{1,2}x_{2,3}x_{3,1}-x_{1,3}x_{3,2}x_{2,1}
\right)^{-1}\,,
\end{align}
in perfect agreement with \eqref{final for 3 adv}.

\subsection{Determinant formula}\label{det three nodes ex}
To conclude this section we now calculate \(F_0^{[3]}\) yet another time, using the determinant formula: 
\begin{equation}\label{det(1-L) in text copy}
F^{[n]}_0=\frac{1}{\det\left(\mathbb 1_n- X_n\right)}\,,\qquad\quad \left. X_n\right|_{ij}=
x_{ij}\,,\quad
1\leq(i,j)\leq n\,.
\end{equation}
This is the simplest way to calculate \(F_0^{[3]}\). Since
\begin{equation}
X_{3}=
\left(
\begin{array}{ccc}
x_{11}&x_{12}&x_{13}\\
x_{21}&x_{22}&x_{23}\\
x_{31}&x_{32}&x_{33}
\end{array}
\right)\,,
\end{equation}
we have
\begin{equation}
F_0^{[3]}=
\det\,^{-1}(\mathbb 1_3-X_3)
=\det\,^{-1}
\left(
\begin{array}{ccc}
1-x_{11}&-x_{12}&-x_{13}\\
-x_{21}&1-x_{22}&-x_{23}\\
-x_{31}&-x_{32}&1-x_{33}
\end{array}
\right)\,,
\end{equation}
and so we immediately get
\begin{align}
F_0^{[3]}&=\left(1-x_{1,1}-x_{2,2}-x_{3,3}-x_{1,2}x_{2,1}+x_{1,1}x_{2,2}-x_{1,3}x_{3,1}+x_{1,1}x_{3,3}-x_{2,3}x_{3,2}+x_{2,2}x_{3,3}\right.\nn
&
\left.-x_{1,1}x_{2,2}x_{3,3}+x_{1,1}x_{2,3}x_{3,2}+x_{2,2}x_{1,3}x_{3,1}+x_{3,3}x_{1,2}x_{2,1}-x_{1,2}x_{2,3}x_{3,1}-x_{1,3}x_{3,2}x_{2,1}
\right)^{-1}\,.
\end{align}
This is the same result we obtained using other computational methods earlier in this section.


\section{An equation for the pole coefficients in term of paths}\label{attempt}

In this section we will prove eq. \eqref{guess}:
\begin{align}\label{guess copy}
G_{[r]}\,\hat a_{p,q}^{[r]}=\sum_{t=0}^{r-1}\left(\sum^r_{i_1,i_2,..,i_t=1\atop i_1\neq i_2 \neq...\neq i_t\neq q}G_{[r]\setminus \{q,\cup_{h=1}^t i_h\} }x_{p,i_1}x_{i_1,i_2}x_{i_2,i_3}\cdots x_{i_{t-1},i_t}x_{i_t,q}\right)\,.
\end{align}
In the case $q=r$ this identity becomes particularly easy to prove, so let us start with this one.

From the definitions \eqref{initial formula recursion} and \eqref{def of a} we get
\begin{align}\label{comp1 cp}
\hat a_{i,r}^{[r]}=a_{i,r}=\frac{x_{i,r}+\sum\limits_{\lambda=1}^{r-1}\,\hat a_{i,\lambda}^{[r-1]}\,x_{\lambda,r}}
{1-\left(x_{r,r}+\sum\limits_{\lambda=1}^{r-1}\,\hat a_{r,\lambda}^{[r-1]}\,x_{\lambda,r}\right)}\,.
\end{align}
Now let us multiply and divide the far RHS above by $G_{[r-1]}$: recalling eqs. \eqref{jnoj} and \eqref{res form} we have
\begin{align}
G_{[r-1]}\left[1-\left(x_{r,r}+\sum\limits_{\lambda=1}^{r-1}\,\hat a_{r,\lambda}^{[r-1]}\,x_{\lambda,r}\right)\right]=G_{[r-1]}\,\frac{G_{[r]}}{G_{[r-1]}}=G_{[r]}\,,
\end{align}
so that we can write eq. \eqref{comp1 cp} as
\begin{align}\label{ready for induction g re}
G_{[r]}\,\hat a_{i,r}^{[r]}=G_{[r-1]}\,x_{i,r}+\sum\limits_{\lambda\in [r-1]}\,G_{[r-1]}\,\hat a_{i,\lambda}^{[r-1]}\,x_{\lambda,r}\,.
\end{align}
Using the last equation we can prove eq. \eqref{guess copy}, for the $q=r$ case, by induction. The identity is trivial for 1 point: it just reads
\begin{align}\label{1pt Gs and as2}
G_{[1]}\,\hat a_{i,1}^{[1]}=G_{[1]\setminus \{1\} }\,x_{i,1}=G_{[0] }\,x_{i,1}
\end{align}
for any $i>1$, and since
\begin{align}
G_{[0]}=1\,,\qquad G_{[1]}=1-x_{1,1}\,,\qquad \hat a_{i,1}^{[1]}= a_{i,1}=\frac{x_{i,1}}{1-x_{1,1}}\,,
\end{align}
eq. \eqref{1pt Gs and as2} is trivially satisfied. Let us now assume \eqref{guess copy} is true for $r-1$ points and let us show that it holds for $r$ points too. We can then use \eqref{guess copy} in the terms $G_{[r-1]}\,\hat a_{i,\lambda}^{[r-1]}$ of \eqref{ready for induction g re}, to obtain
\begin{align}\label{ins guess g}
G_{[r]}\,\hat a_{i,r}^{[r]}&=G_{[r-1]}\,x_{i,r}\nn
&\quad+\sum\limits_{\lambda=1}^{r-1}\,
\sum_{t=0}^{r-2}
\sum_{i_1,i_2,..,i_t=1\atop i_1\neq i_2 \neq...\neq i_t\neq \lambda}^{r-1}G_{[r-1]\setminus \{\lambda,\cup_{h=1}^t i_h\} }x_{i,i_1}x_{i_1,i_2}x_{i_2,i_3}\cdots x_{i_{t-1},i_t}x_{i_t,\lambda}
\,x_{\lambda,r}\,.
\end{align}
The next step is just a relabelling of the summation variables: first relabel $\lambda\rightarrow i_{t+1}$ and then $t\rightarrow t'=t+1$ to get (dropping the prime symbol on $t$)
\begin{align}\label{ins guess g almost}
G_{[r]}\,\hat a_{i,r}^{[r]}&=G_{[r-1]}\,x_{i,r}\nn
&\quad+
\sum_{t=1}^{r-1}
\sum_{i_1,i_2,..,i_t=1\atop i_1\neq i_2 \neq...\neq i_t}^{r-1}G_{[r-1]\setminus \{\cup_{h=1}^t i_h\} }x_{i,i_1}x_{i_1,i_2}x_{i_2,i_3}\cdots x_{i_{t-1},i_t}
\,x_{i_t,r}\,.
\end{align}
Note that the first term on the RHS of the above equation is just the $t=0$ component of the sum following it, so that 
\begin{align}\label{ins guess g almost2}
G_{[r]}\,\hat a_{i,r}^{[r]}&=
\sum_{t=0}^{r-1}
\sum_{i_1,i_2,..,i_t=1\atop i_1\neq i_2 \neq...\neq i_t}^{r-1}G_{[r-1]\setminus \{\cup_{h=1}^t i_h\} }x_{i,i_1}x_{i_1,i_2}x_{i_2,i_3}\cdots x_{i_{t-1},i_t}
\,x_{i_t,r}\,,
\end{align}
which using $G_{[r-1]}=G_{[r]\setminus \{r\}}$ we can write as
\begin{align}\label{ins guess g done}
G_{[r]}\,\hat a_{i,r}^{[r]}&=
\sum_{t=0}^{r-1}\left(
\sum_{i_1,i_2,..,i_t=1\atop i_1\neq i_2 \neq...\neq i_t\neq r}^rG_{[r]\setminus \{r,\cup_{h=1}^t i_h\} }x_{i,i_1}x_{i_1,i_2}x_{i_2,i_3}\cdots x_{i_{t-1},i_t}
\,x_{i_t,r}\right)\,,
\end{align}
which is exactly eq. \eqref{guess copy} for the case $q=r$. This observation concludes the first part of our proof.

The case $q\neq r$ could be potentially difficult to analyse, but we can overcome this complication using a trick: loosely speaking we will change the order of integration in \eqref{F0}, in such a way that the $z_q$ variable, corresponding to the $q$ node, will be integrated last. This will allow us to use the same induction process mentioned above, with trivial modifications. To begin with, we will argue that the order of integration does not affect the expression for the $ \hat a_{i,j}^{[r]}$ coefficients defined in \eqref{allpoles same z} and \eqref{hatted a}. 

Consider again eq. \eqref{allpoles same z}:
\begin{align}\label{allpoles same z copy}
z_j^{*[r]}=z_j^*(z_{r+1},...,z_n;\vec{x})=\sum_{i>r}z_i\, \hat a_{i,j}^{[r]}\,.
\end{align}
These are the equations for the poles of the $z_j$ ($1\leq j\leq r$) variables after we have integrated over $z_{1},z_{2},...,z_{r}$ \emph{in this order}, which in section \ref{The Matter-Free Case} we called `natural ordering'. We labelled this \emph{ordered} set as $\{z_{1},z_{2},...,z_{r}\}\equiv[r]$. Now consider integrating over the same set of variables $z_{1},z_{2},...,z_{r}$, but in a \emph{different order}, which we call $\{z_{\sigma(1)},z_{\sigma(2)},...,z_{\sigma(r)}\}\equiv[r]_\sigma$. We then have, analogously to eq. \eqref{allpoles same z copy},
\begin{align}\label{allpoles same z different order}
z_{\sigma(j)}^{*[r]_\sigma}=z_{\sigma(j)}^*(z_{r+1},...,z_n;\vec{x})=\sum_{i>r}z_i\, \hat a_{i,{\sigma(j)}}^{[r]_\sigma}\,.
\end{align}
The key observation is that equations \eqref{allpoles same z copy} and \eqref{allpoles same z different order} have to contain the same set of equations. 
To see this, suppose that we want to calculate the $z_{r+1}$ pole equation. Following section \ref{The Matter-Free Case} we would have
\begin{align}
(1-x_{r+1,r+1})z_{r+1}&=\sum_{b>{r+1}}z_b\,x_{b,{r+1}}+\sum_{i=1,..,r}z_i^{*[r]}\,x_{i,{r+1}}\,,
\end{align}
if we use the $[r]$ set (the natural ordering), and
\begin{align}
(1-x_{r+1,r+1})z_{r+1}&=\sum_{b>{r+1}}z_b\,x_{b,{r+1}}+\sum_{i=1,..,r}z_{\sigma(i)}^{*[r]_\sigma}\,x_{\sigma(i),{r+1}}\nn
&\equiv
\sum_{b>{r+1}}z_b\,x_{b,{r+1}}+\sum_{i=1,..,r}z_{i}^{*[r]_\sigma}\,x_{i,{r+1}}
\,.
\end{align}
if we use the $[r]_\sigma$ set. Now take the difference of the two equations above to get
\begin{align}\label{zero?}
0
=\sum_{i=1,..,r}\left(z_i^{*[r]}-z_{i}^{*[r]_\sigma}\right)\,x_{i,{r+1}}\,.
\end{align}
Since $x_{i,r+1}$ does not appear inside $z_i^{*[r]}$ or $z_{i}^{*[r]_\sigma}$, for any $i$, the only way that the RHS of \eqref{zero?} can be zero is that each term in the sum vanish on its own, so that
\begin{align}\label{order does not matter}
z_i^{*[r]}=z_{i}^{*[r]_\sigma}\qquad\qquad\forall i\in \{1,2,...,r\}\,.
\end{align}
This indeed shows that \eqref{allpoles same z copy} and \eqref{allpoles same z different order} do contain the same set of equations. More precisely, since eq. \eqref{order does not matter} does not depend on a particular $\sigma$, we see that the order in which we compute integrals does not matter: after $r$ integrations, whatever the order, the pole equations will be described by \eqref{allpoles same z copy}. Eq. \eqref{order does not matter} also implies that
\begin{align}\label{order does not matter on hat a}
\hat a_{i,q}^{[r]}
=\hat a_{i,q}^{[r]_{\sigma}}\,,
\end{align}
if $[r]$ and $[r]_{\sigma}$ differ only by the order of their elements. This is what we need to prove the identity \eqref{guess copy} for generic $q$. The proof will be based upon a comparison between $\hat a$ coefficients computed in two different orderings.

Let us then choose the ordering $[r]_{\sigma_q}=\{z_1,z_2,...,z_{q-1},z_{q+1},...,z_r,z_q\}$, which we will just call $[r]_{q}$ for notational purposes. From \eqref{order does not matter on hat a} we have then 
\begin{align}\label{comp1}
\hat a_{i,q}^{[r]}=\hat a_{i,q}^{[r]_q}=\frac{x_{i,q}+\sum\limits_{\lambda\in [r-1]_q}\,\hat a_{i,\lambda}^{[r-1]_q}\,x_{\lambda,q}}
{1-\left(x_{q,q}+\sum\limits_{\lambda\in [r-1]_q}\,\hat a_{q,\lambda}^{[r-1]_q}\,x_{\lambda,q}\right)}\,,
\end{align}
in which the last equality follows from \eqref{def of a}: with the ordering $[r]_q$, $z_q$ is in fact the last variable to be integrated over, so that it plays the role of the starting point \eqref{initial formula recursion} in the recursion relation \eqref{formula recursion}. We are therefore in the same configuration discussed at the beginning of this section, where the right lower index of $\hat a$ corresponds to the last one in the ordering $[r]_q$: we can therefore redo the steps \eqref{comp1 cp} - \eqref{ins guess g done}, with trivial modifications, to obtain
\begin{align}
G_{ [r]} \hat a_{i,q}^{[r]}= G_{ [r]} \hat a_{i,q}^{[r]_q}=
\sum_{t=0}^{r-1}\left(\sum^r_{i_1,i_2,..,i_t=1\atop i_1\neq i_2 \neq...\neq i_t\neq q}G_{[r]\setminus \{q,\cup_{h=1}^t i_h\} }x_{i,i_1}x_{i_1,i_2}x_{i_2,i_3}\cdots x_{i_{t-1},i_t}x_{i_t,q}\right)\,.
\end{align}
Eq. \eqref{guess copy} is then proved.

\section{The building block $ F_0^{[n]} $ and closed string word counting: Examples }\label{sec:FandCSW} 

Let us consider the 2-node case. We will verify that the coefficients in the expansion of $ F_0^{[2]} $ count words made from letters corresponding to simple loops in the 2-node quiver, with one edge for every specified start and end point. Thus there are letters $\hat y_1 , \hat y_2 , \hat y_{12}$. We require that letters corresponding to loops which do not share a node commute. Thus we here have $ \hat y_1 \hat y_2 = \hat y_2 \hat y_1 $. 
It is useful to define 
\begin{align}
 y_1 \equiv x_{11} \,,\qquad \quad
 y_2 \equiv x_{22} \,,\qquad\quad
 y_{12} \equiv x_{12} x_{21}\,, 
\end{align}
together with
\begin{align}
F_0^{[2]}  & = { 1 \over  1 - y_{1} - y_{2} - y_{12} + y_1 y_2   }  = { 1 \over ( 1 - y_1) ( 1 - y_2 ) - y_{12}  }  \nn
   & = { 1 \over ( 1 - y_1) ( 1-  y_2 ) } { 1 \over  1 - {\displaystyle y_{12} \over { \displaystyle(  1 - y_1  ) ( 1- y_2   ) } }}\,.
\end{align}
Expanding this we get
\bea 
F_0^{[2]} &= &  \sum_{ n_1 , n_2  =0}^{ \infty } y_1^{n_1} y_2^{n_2}  \sum_{  m = 0 }^{ \infty } \left  ( {  y_{12} \over ( 1 - y_1 ) ( 1 - y_2  ) } \right )^m \nn  
 & = & \sum_{ n_1 , n_2 = 0 }^{ \infty } \sum_{ m=0} \sum_{ k_1 , k_2 =0 } y_1^{n_1} y_2^{n_2} y^m y_1^{ k_1 } y_2^{k_2}  { (  m+k_1 -1)!\over k_1 !  (m- 1)  ! } { ( m + k_2 -1 ) ! \over k_2 ! ( m-1 ) ! } \,,
\eea
and defining $ N_1 = n_1 + k_1 $ and $ N_2 = n_2 + k_2 $ we can write 
\bea 
F_0^{[2]} = \sum_{ N_1 , N_2 , m = 0 }^{ \infty } y_1^{  N_1  } y_2^{ N_2  } y^m   \sum_{ k_1 = 0 }^{ N_1 } \sum_{ k_2 = 0 }^{ N_2 }  { (  m+k_1 -1)!\over k_1 !  (m- 1)  ! } { ( m + k_2 -1 ) ! \over k_2 ! ( m-1 ) ! } \,.
\eea
Finally, using the identity 
\bea\label{binomialID}  
 \sum_{ k_1 = 0 }^{ N_1 }{ (  m+k_1 -1)!\over k_1 !  (m- 1)  ! } =  {  ( m + N_1 ) ! \over m! N_1 ! }   = \sum_{ k_1 =0}^{ N_1 } { (m)_{k_1} \over k_1 ! }\,,
\eea
the expansion of $F$ reads
\bea\label{2nodeexp} 
F_0^{[2]} = \sum_{ N_1 , N_2 , m = 0 }^{ \infty }  y_1^{  N_1  } y_2^{ N_2  } y_{12}^m   { ( m + N_1 )! \over m! N_1 ! } { ( m + N_2 )! \over m! N_2 ! } \,.
\eea

The coefficient counts the number of words made from letters $\hat y_{1} , \hat y_2 , \hat y_{12} $, with the condition that $ \hat y_1 \hat y_2  = \hat y_2 \hat y_1 $. The words containing $m$ copies of $\hat y_{12} $ can be built, by writing the $\hat y_{12}$ letters out in a line, with spaces between them, and then inserting the $N_1$  $\hat y_{1}$ letters in any of the $m+1$ slots. Now build a sequence of $N_1$ numbers, recording which slot the first $ \hat y_1 $ goes into, which the second goes into and so on. 
Each number in the sequence is something between $ 1$ and $ m+1$. Each such sequence can be mapped to a state $ e_{ a_1 } \otimes e_{a_2} \dots ... e_{a_{N_1} } $. Sequences related by the symmetrization procedure of shuffling around the $N_1$ factors 
correspond to same word, because what matters is what goes in the $m+1$ slots, not the order in which the $N_1$ copies of $x_{11}$ were put there. 
Thus the sequences are in one-one correspondence with a basis for the symmetric tensors $Sym ( V_{ m+1 }^{ \otimes N_1 } ) $.
The dimension of this space is precisely 
\bea 
\text{dim}\left(Sym ( V_{ m+1 }^{ \otimes N_1 } )  \right)={ ( m + N_1 )! \over m! N_1 ! } \,.
\eea 
Then we can insert the $\hat y_2  $ in the $m+1$ slots and we get the other factor. 
This proves that, in the 2-node case, the words in the language we defined are counted by the $F_0^{[2]} $-function.

Let us now turn to the three node case. Let us define 
\begin{align}
  y_{i} \equiv x_{ii} \,, \qquad\quad
  y_{ij} \equiv x_{ij} x_{ji}\,, \qquad  \quad
 y_{ijk} \equiv x_{ij} x_{jk} x_{ki} \,.
\end{align}
In this case 
\begin{align} 
 F^{[3]}_0  &=  ( 1- y_1 )^{-1}  ( 1- y_2 )^{-1}  ( 1- y_3 )^{-1}   \left [ 1 - { y_{12} \over ( 1 - y_1 ) ( 1- y_2 ) } - { y_{13} \over ( 1 - y_1 ) ( 1- y_3 ) } \right. \nn
&\qquad\qquad \left.-{ y_{23} \over ( 1- y_2 ) ( 1- y_3 ) }  - { y_{123 }   \over ( 1- y_1) ( 1- y_2) ( 1 - y_3 ) } - { y_{132}  \over ( 1- y_1) ( 1- y_2) ( 1- y_3) } \right ]^{-1}  \nn 
& = \sum_{ m_1 , m_2 , m_3 =0 }^{ \infty }  ~ \sum_{ p_1 , \cdots , p_5 =0}^{ \infty } 
 y_1^{m_1} \,y_2^{m_2} \,y_3^{m_3} \,y_{12}^{p_1} \,\,y_{13}^{p_2}\, \,y_{23}^{p_3}\,\, y_{123}^{p_4}\, \,y_{132}^{p_5} \nn 
& \qquad \times\, { ( p_1 + p_2 + \cdots + p_5 ) ! \over p_1! p_2! p_3! p_4! p_5 ! } { 1 \over ( 1 - y_1)^{ p_1 + p_2 +p_4 + p_5 } ( 1- y_2 )^{ p_1 + p_3+ p_4 + p_5 } ( 1- y_3 )^{ p_2 + p_3 + p_4 + p_5 } }  \nn
& = \sum_{ m_1 , m_2 , m_3 =0 }^{ \infty }  ~ \sum_{ p_1 , \cdots , p_5 =0}^{ \infty } 
 y_1^{m_1} \,y_2^{m_2}\, y_3^{m_3}\, y_{12}^{p_1}\, y_{13}^{p_2}\, y_{23}^{p_3}\, y_{123}^{p_4}\, y_{132}^{p_5} \,  { ( p_1 + p_2 + \cdots + p_5 ) ! \over p_1! p_2! p_3! p_4! p_5 ! }  \nn 
 & \qquad\times\sum_{ l_1 , l_2 , l_3 =0 }^{ \infty } 
 {  ( p_1 + p_2 + p_4 + p_5 )_{ l_1 } \over l_1 ! }  {  ( p_1 + p_3 + p_4 + p_5 )_{ l_2}    \over l_2 ! } {  ( p_2 + p_3 + p_4 + p_5 )_{ l_3 } \over l_3 ! }  y_1^{ l_1 } y_2^{ l_2 } y_3^{ l_3 } \nn
&=  \sum_{ n_1 , n_2 , n_3 =0 }^{ \infty }  ~ \sum_{ p_1 , \cdots , p_5 =0}^{ \infty } 
 y_1^{n_1}\, y_2^{n_2}\,y_3^{n_3} \,y_{12}^{p_1} \,\,y_{13}^{p_2} \,\,y_{23}^{p_3} \,\,y_{123}^{p_4}\,\, y_{132}^{p_5}   { ( p_1 + p_2 + \cdots + p_5 ) ! \over p_1! p_2! p_3! p_4! p_5 ! }  \\[3mm]
&\qquad \times \sum_{ l_1  =0 }^{ n_1} \sum_{ l_2 =0}^{ n_2 }  \sum_{ l_3 =0 }^{ n_3  } 
 {  ( p_1 + p_2 + p_4 + p_5 )_{ l_1 } \over l_1 ! }  {  ( p_1 + p_3 + p_4 + p_5 )_{ l_2}    \over l_2 ! } {  ( p_2 + p_3 + p_4 + p_5 )_{ l_3 } \over  l_3 ! }  y_1^{ l_1 } y_2^{ l_2 } y_3^{ l_3 }\,.\nonumber
\end{align}
Finally we use the identity \eqref{binomialID} above three times, to get
\bea 
&& F_0^{ [3]} = \sum_{ n_1 , n_2 , n_3 =0 }^{ \infty }  ~ \sum_{ p_1 , \cdots , p_5 =0}^{ \infty } 
 y_1^{n_1}\, y_2^{n_2}\, y_3^{n_3}\, y_{12}^{p_1} \,\,y_{13}^{p_2}\,\, y_{23}^{p_3}\,\, y_{123}^{p_4} \,\,y_{132}^{p_5}  \, { ( p_1 + p_2 + \cdots + p_5 ) ! \over p_1! p_2! p_3! p_4! p_5 ! }  \\[3mm]
&& \qquad\qquad\qquad \times\,
{ p_1 + p_2 + p_4 + p_5  + n_1 \choose n_1 } { p_1 + p_3 + p_4 + p_5 + n_2 \choose n_2 } { p_2 + p_3 + p_4 + p_5 \choose n_3 }\,. \nonumber
\eea
For the closed string words in this case, there are letters $ \hat y_i , \hat y_{ij}  , \hat y_{ ijk} $. 
The five letters $ \hat y_{ 12} , \hat y_{ 13} , \hat y_{ 23} ,$ $ \hat y_{ 123} , \hat y_{ 132} $ do not commute with each other. 
$\hat y_1 , \hat y_2 , \hat y_3 $ commute with each other. $ \hat y_1$ commutes with $ \hat y_{23} $. $ \hat y_2 $ commutes with $ \hat y_{13} $, and $ \hat y_3$ commutes with $ \hat y_{12}$. We can build an arbitrary word by first fixing the numbers $ p_1 ,\, p_2, ... , p_5 $ of the letters from the set $ \{ \hat y_{ij } , \hat y_{ ijk } \}$. Then choose an order of these. The first multinomial factor
\bea 
{  ( p_1 + \cdots + p_5 ) ! \over p_1 !  \cdots p_5  ! } 
\eea
gives the number of choices of this order. For each fixed order of these, we can insert the $ \hat y_i $. Consider the insertion of the $ \hat y_1 $ and choose the number $n_1$ of them. We have $ ( p_1 + p_2 + p_4 + p_5 +1 ) $ slots which specify where, relative to $ \hat y_{ 12} , \hat y_{ 13} , \hat y_{123} , \hat y_{ 132 }$, we are inserting these. As in the 2-node case, this is the dimension of $ Sym^{ n_1}  ( V_{ p_1 + p_2 + p_4 + p_5  +1 } ) $ which is given by 
\bea 
{ p_1 + p_2 + p_4 + p_5  + n_1 \choose n_1 } \,.
\eea
The position relative to $ \hat y_{23} $ is immaterial in the word counting because $ \hat y_1 $ commutes with this. Hence $p_3$ does not appear in the above formula. In the same way, the insertion of the $ \hat y_{2} $ and $ \hat y_3 $ account for the additional binomial factors. Since the $ \hat y_i $ commute with each other, the insertion of the $ \hat y_2$ is insensitive to the previous insertion of the $ \hat y_1 $. Likewise the insertion of the $ \hat y_3 $ is insensitive to the positions of the $ \hat y_1 , \hat y_2$. Hence the word counting for specified $ p_1 , \cdots , p_5 ,  n_1 , n_2 , n_3 $ has separate factors corresponding to insertions of $ \hat y_1 , \hat y_2 , \hat y_3 $ among the mutually non-commuting set $ \{ \hat y_{ ij} , \hat y_{ ijk } \} $. 

These examples illustrate the general fact that the function $ F_0^{[n]} ( x_{ ab} ) $ counts 
words made from letters corresponding to simple loops in the complete $n$-node quiver, with the condition that letters corresponding to loops without a shared node commute.

\section{Deriving the flavoured $F^{[n]}$ function}\label{using guess in matter}
In this section we will prove eq. \eqref{final2}:
\begin{align}\label{final2 copy}
F^{[n]}=F^{[n]}_0\,\exp\left( t_p\bar t_q\,\partial^{p,q}\,\log F_0^{[n]}\vphantom{\sum}\right)\,.
\end{align}
We will start from eq. \eqref{mat}:
\begin{align}\label{mat copy}
F^{[n]}=\prod_{j=1}^{n}\left(\frac
{\displaystyle
\exp{\left(\hat a_{0,j}^{[n]}\, t_j\right)}}
{\displaystyle 
1-x_{{j},{j}}-\sum_{i=1}^{j-1} \hat a_{j,i}^{[j-1]}\,x_{i,{j}}
}\right)\,.
\end{align}
We already know that the denominatorof \eqref{mat copy} is
\begin{align}\label{denfwm2}
\prod_{l=1}^{n}&\left(1-x_{{{l}},{{l}}}-\sum_{q=1}^{l-1} \hat a_{{l},q}^{[l-1]}\,x_{q,{{l}}}\right)=
\prod_{l=1}^{n}\,\frac{G_{[l]}}{G_{[l-1]}}=
G_{[n]}\,,
\end{align}
so that we only need to work on its numerator, which is the exponentiation of the sum $\sum\limits_{k=1}^n \hat a_{0,k}^{[n]}\, t_k$. As we did in section \ref{full gen fun sec}, let us now set $\bar t_p=x_{0,p}$ and $t_p=x_{p,0}$. We can multiply and divide \eqref{denfwm2} by $G_{[n]}$ to get
\begin{align}
\sum\limits_{k=1}^n \hat a_{0,k}^{[n]}\,x_{k,0}=
\frac{1}{G_{[n]}}\sum\limits_{k=1}^n G_{[n]}\,\hat a_{0,k}^{[n]}\,x_{k,0}\,.
\end{align}
Using eq. \eqref{guess} on each of the terms $G_{[n]}\,\hat a_{0,k}^{[n]}$ in the sum above gives
\begin{align}\label{bftrk}
\sum\limits_{k=1}^n \hat a_{0,k}^{[n]}\,x_{k,0}&=
\frac{1}{G_{[n]}}\sum\limits_{k=1}^n x_{k,0}
\sum_{t=0}^{n-1}\sum^{n}_{i_1,i_2,..,i_t=1\atop  i_1\neq i_2 \neq...\neq i_t\neq k}G_{[n]\setminus \{k,\cup_{h=1}^t i_h\} }x_{0,i_1}x_{i_1,i_2}x_{i_2,i_3}\cdots x_{i_{t-1},i_t}x_{i_t,k}\nn
&=
\frac{1}{G_{[n]}}\sum\limits_{k=1}^n
\sum_{t=0}^{n-1}\sum^{n}_{i_1,i_2,..,i_t=1\atop  i_1\neq i_2 \neq...\neq i_t\neq k}G_{[n]\setminus \{k,\cup_{h=1}^t i_h\} }x_{0,i_1}x_{i_1,i_2}x_{i_2,i_3}\cdots x_{i_{t-1},i_t}x_{i_t,k} x_{k,0}\,.
\end{align}
Consider the product of $x_{ab}$ coefficients in the equation above, 
\begin{align}\label{string of x}
x_{0,i_1}x_{i_1,i_2}x_{i_2,i_3}\cdots x_{i_{t-1},i_t}x_{i_t,k} x_{k,0}\,.
\end{align}
This can be interpreted as a path on the quiver starting from node 0, passing through \(t\) intermediate nodes \(i_h\), \(1\leq h\leq t\), reaching node \(k\) and returning back at node 0. Crucially, since all the $i_h$ nodes in this term do not ever take the value $k$ (because of the summation ranges in \eqref{bftrk}), such a path never intersects itself. Our aim now is to factor out the 0 node from such a term, rewriting it as a path starting from node \(k\), passing through the same \(t\) intermediate nodes \(i_h\) and ending at node \(k\) again. We can achieve this goal by letting an appropriate derivative act on the string of \(x
_{ab}\) coefficients in \eqref{string of x}. Consider the identity
\begin{align}\label{trick2}
x_{0,i_1}&x_{i_1,i_2}x_{i_2,i_3}\cdots x_{i_{t-1},i_t}x_{i_t,k} x_{k,0}\nn
&=
x_{k,0}\left(x_{0,i_1}\,\frac{\partial}{\partial x_{k,i_1}}+x_{0,k}\,\frac{\partial}{\partial x_{k,k}}\right)
x_{k,i_1}x_{i_1,i_2}x_{i_2,i_3}\cdots x_{i_{t-1},i_t}x_{i_t,k}
\end{align}
(no sum on $k$ or $i_1$), where we added the term
\begin{align}
\left(x_{0,i_1}\,\frac{\partial}{\partial x_{k,i_1}}+x_{0,k}\,\frac{\partial}{\partial x_{k,k}}\right)x_{k,i_1}=x_{0,i_1}\,,\qquad i_1\neq k\,.
\end{align}
The $\partial / \partial x_{k,k}$ derivative has been added in order to account for the $t=0$ case (the one in which there are no intermediate steps in the path \eqref{string of x}, which would just read $x_{0,k}x_{k,0}$): in this situation we would trivially get
\begin{align}
x_{0,k}x_{k,0}=
x_{k,0}\left(0+x_{0,k}\,\frac{\partial}{\partial x_{k,k}}\right)
x_{k,k}=x_{0,k}x_{k,0}\,,
\end{align}
so that the identity \eqref{trick2} holds for any $t\geq0$. Note also that we can rewrite the same equation as
\begin{align}\label{trick 3}
x_{0,i_1}&x_{i_1,i_2}x_{i_2,i_3}\cdots x_{i_{t-1},i_t}x_{i_t,k} x_{k,0}=\nn
&=
x_{k,0}\left(\sum_{p=1\atop p\neq k}^{n}x_{0,p}\,\frac{\partial}{\partial x_{k,p}}+x_{0,k}\,\frac{\partial}{\partial x_{k,k}}\right)
x_{k,i_1}x_{i_1,i_2}x_{i_2,i_3}\cdots x_{i_{t-1},i_t}x_{i_t,k}\,,
\end{align}
where the values that $p$ can take ($\{1,2,...,n\}\setminus \{k\}$) are the same ones on which $i_1$ runs in the sum in \eqref{bftrk}: all the $i_1,i_2,...,i_t$ indices never take the value $k$, leaving $x_{k,i_1}$ as the only variable on which the $\partial / \partial x_{k,p}$ derivative can act with nonzero result.
%
%
%
We can then rewrite the identity \eqref{trick 3} as
\begin{align}\label{trick4}
x_{0,i_1}&x_{i_1,i_2}x_{i_2,i_3}\cdots x_{i_{t-1},i_t}x_{i_t,k} x_{k,0}\nn
&=
x_{k,0}\left(\sum_{p=1}^{n}x_{0,p}\,\frac{\partial}{\partial x_{k,p}}\right)
x_{k,i_1}x_{i_1,i_2}x_{i_2,i_3}\cdots x_{i_{t-1},i_t}x_{i_t,k}\nn
&=
\sum_{p=1}^{n} t_k \bar t_p\,\partial^{k,p}\,
x_{k,i_1}x_{i_1,i_2}x_{i_2,i_3}\cdots x_{i_{t-1},i_t}x_{i_t,k}\,,
\end{align}
where in the last line we set $\frac{\partial}{\partial x_{k,q}}=\partial^{k,q}$ and used the original notation $x_{k,0}= t_k$, $x_{0,k}=\bar t_k$. At this stage, we successfully rewrote a our initial path \((0,i_1,i_2,...,i_t,k,0)\) in terms of a suitable differential operator acting on a new path \((k,i_1,i_2,...,i_t,k)\).

Inserting eq. \eqref{trick4} into \eqref{bftrk} gives
\begin{align}\label{aftrk}
\sum\limits_{k=1}^n& \hat a_{0,k}^{[n]}\,x_{k,0}=\\[3mm]
&=
\frac{1}{G_{[n]}}\sum\limits_{k=1}^n
\sum_{t=0}^{n-1}\sum^{n}_{i_1,i_2,..,i_t=1\atop  i_1\neq i_2 \neq...\neq i_t\neq k}G_{[n]\setminus \{k,\cup_{h=1}^t i_h\} }
\sum_{p=1}^{n}y_k\bar y_p
\,\partial^{k,p}\,
x_{k,i_1}x_{i_1,i_2}x_{i_2,i_3}\cdots x_{i_{t-1},i_t}x_{i_t,k}\nonumber\,.
\end{align}
Note that $\partial_{k,p}$ can pass through $G_{[n]\setminus \{k,\cup_{h=1}^t i_h\} }$, since the latter does not contain the $k$-th point (by construction). We can then write
\begin{align}\label{aftrk2}
\sum\limits_{k=1}^n& \hat a_{0,k}^{[n]}\,x_{k,0}\nn
&=
\frac{1}{G_{[n]}}
\sum\limits_{k,p=1}^n
 t_k \bar t_p\,\partial^{k,p}
\sum_{t=0}^{n-1}\sum^{n}_{i_1,i_2,..,i_t=1\atop  i_1\neq i_2 \neq...\neq i_t\neq k}
G_{[n]\setminus \{k,\cup_{h=1}^t i_h\} }
x_{k,i_1}x_{i_1,i_2}x_{i_2,i_3}\cdots x_{i_{t-1},i_t}x_{i_t,k}\\[3mm]
&=
\frac{1}{G_{[n]}}
\sum\limits_{k,p=1}^n
 t_k \bar t_p\,\partial^{k,p}\left(
\sum_{t=0}^{n-1}\sum^{n}_{i_1,i_2,..,i_t=1\atop  i_1\neq i_2 \neq...\neq i_t\neq k}
G_{[n]\setminus \{k,\cup_{h=1}^t i_h\} }
x_{k,i_1}x_{i_1,i_2}x_{i_2,i_3}\cdots x_{i_{t-1},i_t}x_{i_t,k}-G_{[n]\setminus \{k\}}\right)\,,\nonumber
\end{align}
where in the last line we added $G_{[n]\setminus \{k\}}$ under the derivative action: indeed $\partial_{k,p}G_{[n]\setminus \{k\}}=0$, since $G_{[n]\setminus \{k\}}$ does not contain the $k$-th point, and thus any $x_{k,p}$ \(\forall p\). Note that the term in the round brackets of the equation above is just $-G_{[n]}$. The definition of $G_{[n]}$ we gave in eq. \eqref{good G} reads
\begin{align}\label{good G n}
G_{[n]}&=G_{[n-1]}-\sum_{t=0}^{n-1}\sum^{n-1}_{i_1,i_2,..,i_t=1\atop  i_1\neq i_2 \neq...\neq i_t}G_{[n-1]\setminus \{\cup_{h=1}^t i_h\} }x_{n,i_1}x_{i_1,i_2}x_{i_2,i_3}\cdots x_{i_{t-1},i_t}x_{i_t,n}\nn
&=
G_{[n]\setminus\{n\}}-\sum_{t=0}^{n-1}\sum^{n}_{i_1,i_2,..,i_t=1\atop  i_1\neq i_2 \neq...\neq i_t\neq n}G_{[n]\setminus \{n,\cup_{h=1}^t i_h\} }x_{n,i_1}x_{i_1,i_2}x_{i_2,i_3}\cdots x_{i_{t-1},i_t}x_{i_t,n}\,,
\end{align}
but the same equation holds if, instead of \(n\), we remove any integer $1\leq k\leq n$ from the set $ [n] $:
\begin{align}\label{good G n->k}
G_{[n]}=G_{[n]\setminus \{k\}}-\sum_{t=0}^{n-1}\sum^{n}_{i_1,i_2,..,i_t=1\atop  i_1\neq i_2 \neq...\neq i_t\neq k}G_{[n]\setminus \{k,\cup_{h=1}^t i_h\} }x_{k,i_1}x_{i_1,i_2}x_{i_2,i_3}\cdots x_{i_{t-1},i_t}x_{i_t,k}\,.
\end{align}

Using \eqref{good G n->k} in \eqref{aftrk2} gives then
\begin{align}\label{aftrk4}
\sum\limits_{k=1}^n x_{k,0}\,\hat a_{0,k}^{[n]}&=
-\frac{1}{G_{[n]}}
\sum\limits_{k,p=1}^n
 t_k \bar t_p\,\partial^{k,p}\, G_{[n]}
 =-\sum\limits_{k,p=1}^n
 t_k \bar t_p\,\partial^{k,p}\,\log G_{[n]}\,,
\end{align}
so that we can write, for the numerator of $F$ in \eqref{mat copy}
\begin{align}
\prod_{k=1}^n\exp\left(\hat a_{0,k}^{[n]}\,x_{k,0}\right)
=\exp\left(\sum\limits_{k=1}^n \hat a_{0,k}^{[n]}\,x_{k,0}\right)
=\exp\left(-\sum\limits_{k,p=1}^n
 t_k \bar t_p\,\partial^{k,p}\,\log G_{[n]}\right)\,.
\end{align}
This means that \(F^{[n]}\) can be written as
\begin{align}\label{final!!}
F^{[n]}=\cfrac{\exp\left(-\sum\limits_{k,p=1}^n
 t_k \bar t_p\,\partial^{k,p}\,\log G_{[n]}\right)}
{G_{[n]}}\,,
\end{align}
or, using Einstein summation
\begin{align}\label{final!! ein sum}
F^{[n]}=\cfrac{\exp\left(- t_k \bar t_p\,\partial^{k,p}\,\log G_{[n]}\vphantom{\sum}\right)}
{G_{[n]}}\,.
\end{align}
Recalling that $F^{[n]}_0={G_{[n]}}^{-1}$, where $F_0$ is the generating function for the unflavoured case, we also have
\begin{align}
F^{[n]}=
F^{[n]}_0
\exp\left( t_k \bar t_p\,\partial^{k,p}\,\log F^{[n]}_0\vphantom{\sum}\right)\,.
\end{align}
Furthermore, considering the chain of equalities
\begin{align}
\frac{(-1)^{p+q}\,M_{p,q}}{\det(\mathbb 1_n- X_n)}=
-\frac{1}{\det(\mathbb 1_n- X_n)}\partial^{p,q}\det(\mathbb 1_n- X_n)=\partial^{p,q}\log \left(\frac{1}{\det(\mathbb 1_n- X_n)}\right)
\end{align}
we finally get to 
\begin{align}\label{final2 mod minors copy}
F^{[n]}=F^{[n]}_0\,\exp\left(\sum_{p,q=1}^n\, t_p\bar t_q\,\,\frac{(-1)^{p+q}\,M_{p,q}}{\det(\mathbb 1_n- X_n)} \right)\,.
\end{align}
The latter is exactly \eqref{final2 mod minors}.

\pagebreak

\bibliographystyle{utphys}
\bibliography{biblio}

\end{document}